\documentclass[epsfig,12pt,onecolumn]{article}
\usepackage{amsfonts}
\usepackage{amssymb}
\usepackage{amsmath}
\usepackage{multicol}
\usepackage{graphicx}
\usepackage{float}
\usepackage{caption}
\usepackage{xcolor}
\usepackage{hyperref}

\hypersetup{colorlinks,
citecolor=blue,
filecolor=blue,
linkcolor=blue,
urlcolor=black}
\makeatletter \@addtoreset{equation}{section}
\renewcommand{\theequation}{\thesection.\arabic{equation}}
\def \be{\begin{equation}}
\def \ee{\end{equation}}
\def \bea{\begin{eqnarray}}
\def \eea{\end{eqnarray}}

\newcommand{\nc}{\newcommand}
\nc{\al}{\alpha} \nc{\bib}{\bibitem} \nc{\la}{\lambda}
\nc{\C}{\mbox{\hspace{1.24mm}\rule{0.2mm}{2.5mm}\hspace{-2.7mm} C}}
\nc{\R}{\mbox{\hspace{.04mm}\rule{0.2mm}{2.8mm}\hspace{-1.5mm} R}}
\input{tcilatex}
\begin{document}

\title{\textbf{Lax Operator and superspin chains from}\\
\textbf{\ 4D CS gauge theory}}
\author{Y. Boujakhrout, E.H Saidi, R. Ahl Laamara, L.B Drissi \\
{\small 1. LPHE-MS, Science Faculty}, {\small Mohammed V University in
Rabat, Morocco}\\
{\small 2. Centre of Physics and Mathematics, CPM- Morocco}}
\maketitle

\begin{abstract}
We study the properties of interacting line defects in the four-dimensional
Chern Simons (CS) gauge theory with invariance given by the $SL\left(
m|n\right) $ super-group family. From this theory, we derive the oscillator
realisation of the Lax operator for superspin chains with $SL(m|n)$
symmetry. To this end, we investigate the holomorphic property of the
bosonic Lax operator $\mathcal{L}$ and build a differential equation $%
\mathfrak{D}\mathcal{L}=0$ solved by the Costello-Gaioto-Yagi realisation of 
$\mathcal{L}$ in the framework of\ the CS theory. We generalize this
construction to the case of gauge super-groups, and develop a Dynkin
super-diagram algorithm to\ deal with the decomposition of the Lie
superalgebras. We obtain the generalisation of the Lax operator describing
the interaction between the electric Wilson super-lines and the magnetic 't
Hooft super-defects. This coupling is given in terms of a mixture of bosonic
and fermionic oscillator degrees of freedom in the phase space of
magnetically charged 't Hooft super-lines. The purely fermionic realisation
of the superspin chain Lax operator is also investigated and it is found to
coincide exactly with the $\mathbb{Z}_{2}$- gradation of Lie superalgebras.\ 
\newline
Keywords: 4D Chern-Simons theory, Super-gauge symmetry, Lie superalgebras
and Dynkin super-diagrams, Superspin chains and integrability, Super- Lax
operator.
\end{abstract}


\section{Introduction}

Four-dimensional Chern-Simons theory living on\textrm{\ }$\mathbb{R}%
^{2}\times C$\textrm{\ }is a topological gauge field theory with a
complexified gauge symmetry $G$ \cite{1A}. Its basic observables are given
by line and surface defects such as the electrically charged Wilson lines
and the magnetically charged 't Hooft lines \cite{1A,1B,2B,2BA,3B,4B,5B}.
These lines expand in the topological plane $\mathbb{R}^{2}$ and are located
at a point $z$ of the complex holomorphic line $C$. The Chern-Simons (CS)
gauge theory offers a powerful framework to study the Yang-Baxter equation
(YBE) of integrable 2D systems \cite{1A,1C,2C,3C,4C,5C} and statistical
mechanics of quantum spin chains \cite{1D,1DA,1DB,1DC,1DD,1DE,2D,2DA}. This
connection between the two research areas is sometimes termed as the
Gauge/YBE correspondence \cite{1E,1EA}. In these regards, the topological
invariance of the crossings of three Wilson lines in the 4D theory, which
can be interpreted as interactions between three particle states, yields a
beautiful graphic realisation of the YBE. Meanwhile, the R-matrix is
represented by the crossing of two Wilson lines and is nicely calculated in
the 4D CS\ gauge theory using the Feynman diagram method \cite{1A,1C,1EB,1EC}%
.\textrm{\ }

In the same spirit, a quantum integrable XXX spin chain of $N$ nodes can be
generated in the CS gauge theory by taking $N$ electrically charged Wilson
lines located at a point $z$ of\textrm{\ }$C$ \cite{1DA,1D}\textrm{. }These
parallel lines are aligned along a direction of $\mathbb{R}^{2}$ and are
simultaneously crossed by a perpendicular magnetic 't Hooft line at $%
z^{\prime }\neq z$. The 't Hooft line defect plays an important role in this
modeling as it was interpreted in terms of the transfer (monodromy) matrix 
\cite{2B,3B,1ED} and the Q-operators of the spin chain \cite{1D,1EE,1EF}. In
this setup, the nodes' spin states of the quantum chain are identified with
the weight states of the Wilson lines, which in addition to the spectral
parameter $z$, are characterised by\emph{\ }highest weight representations $%
\boldsymbol{R}$ of the gauge symmetry $G$\textrm{\ }\cite{5B}. Moreover, to
every crossing vertex, corresponding to a node of the spin chain, is
associated a Lax operator (L-operator) describing the Wilson-'t Hooft lines'
coupling. Thus, the RLL\ equations of the spin chain integrability can be
graphically represented following the YBE/Gauge correspondence by the
crossings of two Wilson lines\textrm{\ }with a 't Hooft line and with each
other.

In this paper, we investigate the integrability of superspin chains in the
framework of the 4D CS theory with gauge super-groups while focussing on the 
$SL(m|n)$ family. On the chain side, the superspin states are as formulated
in \cite{1G} with values in the fundamental representation of $SL(m|n)$. On
the gauge theory side, the superspin chain is described by $N$ Wilson
super-lines\textrm{\ }crossed by a 't Hooft super-line\emph{\ }charged under%
\emph{\ }$SL(m|n)$. These super-lines are graded extensions of the bosonic
ones of the CS theory; they are described in sub-subsection \emph{5.1.2; }%
\textrm{in particular} eqs(\ref{524})-(\ref{526}) and the Figures \textbf{%
\ref{sl}}, \textbf{\ref{fig14}}.\textrm{\ }To that purpose, we develop the
study of the extension of the standard CS theory to the case of classical
gauge super-groups as well as the implementation of the super-line defects
and their interactions.\textrm{\ }We begin by revisiting the construction of
the L-operator in the CS theory with bosonic gauge symmetry and explicitize
the derivation of the parallel transport of the gauge fields in the presence
of 't Hooft line defects with Dirac-like singularity following \cite{1D}.%
\textrm{\ }We also build the differential Lax equation, solved by the
oscillator realisation of the L-operator, and use it to motivate its
extension to supergroups.\textrm{\ }Then, we describe useful aspects
concerning Lie superalgebras and their representations; and propose a
diagrammatic algorithm to approach the construction of the degenerate
L-operators for every node of the $sl(m|n)$ spin chain.\ This description
has been dictated by: $\left( i\right) $ the absence of a generalised
Levi-theorem for superalgebras' decomposition \textrm{\cite{1L,2L}} and $%
\left( ii\right) $ the multiplicity of\textrm{\ }Dynkin Super-Diagrams (DSD)
associated to a given superalgebra underlying the gauge supergroup symmetry.
Next, we describe the basics of the CS theory with $SL(m|n)$ gauge
invariance. We focus on the distinguished $sl(m|n)$ superalgebras
characterized by a minimal number of fermionic nodes in the DSDs and provide
new results concerning the explicit calculation of the super- Lax operators
from the gauge theory in consideration. These super L-operators are given in
terms of a mixed system of bosonic and fermionic oscillators that we study
in details. On one hand, these results contribute to understand better the
behaviour of the super- gauge fields in the presence of 't Hooft lines
acting like magnetic Dirac monopoles. On the other hand, we recover explicit
results from the literature of integrable superspin chains. This finding
extends the consistency of the Gauge/YBE correspondence to supergroups\emph{%
\ }and opens the door for other links to supersymmetric quiver gauge
theories and D-brane systems of type II string theory and M2/M5-brane
systems of M-theory\  \cite{1H,2H,3H}.

The organisation is as follows. In section 2, we recall basic features of
the topological 4D Chern Simons theory with bosonic gauge symmetry $G$. We
describe the moduli space of solutions to the equations of motion in the
presence of interacting Wilson and 't Hooft lines and show how the Dirac
singularity properties of the magnetic 't Hooft line lead to an exact
description of the oscillator Lax operator for XXX spin chains with bosonic
symmetry.\textrm{\ }In section 3, we derive the differential equation $%
\mathcal{D}L=0$ verified by the CGY formula $e^{X}z^{\mu }e^{Y}$ \cite{1D}
for the oscillator realisation of the L-operator. We rely on the fact that
this formula is based on the Levi- decomposition of Lie algebras which means
that $L(z)$ is a function of the three quantities $\left( \mu ,X,Y\right) $
obeying an $sl\left( 2\right) $ algebra. We also link this equation to the
usual time evolution equation of the Lax operator. Then, we assume that this
characterizing behaviour of $L(z)$ described by the differential equation is
also valid for superalgebras and use this assumption to treat CS theory with
a\ super- gauge invariance. For illustration, we study the example of the $%
GL\left( 1|1\right) $ theory as a simple graded extension of the $GL(2)$
case.\textrm{\ }In section 4, we investigate the CS theory for the case of
gauge supergroups and describe the useful mathematical tools needed for this
study, in particular, the issue regarding the non uniqueness of the DSDs. In
section 5, we provide the fundamental building blocks of the CS with $%
SL\left( m|n\right) $ gauge invariance ($m\neq n$) and define its basic
elements that we will need to generalise the expression of the oscillator
Lax operator in the super- gauge theory. Here, the distinguished
superalgebra is decomposed by cutting a node of the DSD in analogy to the
Levi- decomposition of Lie algebras. In section 6, we build the super
L-operator associated to $SL\left( m|n\right) $ and explicit the associated
bosonic and fermionic oscillator degrees of freedom. We also specify\ the
special pure fermionic case where the Lax operator of the superspin chain is
described by fermionic harmonic oscillators. Section 7 is devoted to
conclusions and comments. \textrm{Three} appendices A, B and C including
details are reported in section 8.

\section{Lax operator from 4D CS theory}

In this section, we recall the field action of the 4D Chern-Simons theory on 
$\mathbb{R}^{2}\times C$ with a simply connected gauge symmetry $G$. Then,
we investigate the presence of a 't Hooft line defect with magnetic charge $%
\mu $ (tH$_{\mathrm{\gamma }_{0}}^{\mu }$ for short), interacting with an
electrically charged Wilson line W$_{\mathrm{\xi }_{z}}^{\boldsymbol{R}}$.
For this coupled system, we show that the Lax operator $L_{\mathrm{\gamma }%
_{0},\mathrm{\xi }_{z}}^{\boldsymbol{R}},$ encoding the coupling tH$_{%
\mathrm{\gamma }_{0}}^{\mu }$-W$_{\mathrm{\xi }_{z}}^{\boldsymbol{R}}$, is
holomorphic in $z$ and can be put into the following factorised form \textrm{%
\cite{1D}}%
\begin{equation}
L_{\mathrm{\gamma }_{0},\mathrm{\xi }_{z}}^{\boldsymbol{R}}\left( z\right)
=e^{X_{\boldsymbol{R}}}z^{\mathbf{\mu }}e^{Y_{\boldsymbol{R}}}\qquad ,\qquad
L_{\mathrm{\gamma }_{0},\mathrm{\xi }_{z}}^{\boldsymbol{R}}\in G_{\left[ %
\left[ z\right] \right] }  \label{l}
\end{equation}%
In this relation, $X_{\boldsymbol{R}}=\sum b^{m}X_{m}^{\boldsymbol{R}}$ and $%
Y_{\boldsymbol{R}}=\sum c_{m}Y_{\boldsymbol{R}}^{m}$ where $b^{m}$ and $%
c_{m} $ are the coordinates of the phase space underlying the RLL
integrability\ equation. The $X_{m}^{\boldsymbol{R}}$ and $Y_{\boldsymbol{R}%
}^{m}$ are generators of nilpotent algebras $\boldsymbol{n}_{\pm }$
descendant from the Levi- decomposition of the Lie algebra $\boldsymbol{g}$
of the gauge symmetry $G$. They play an important role in the study; they
will be described in details later.

\subsection{Topological 4D CS field action}

Here, we describe the field action of the 4D Chern-Simons gauge theory with
a bosonic-like gauge symmetry $G$ and give useful tools in order to derive
the general expression (\ref{l}) of the Lax operator (L-operator).\newline
The 4D Chern-Simons theory built\ in \textrm{\cite{1A}} is a topological
theory living on the typical 4- manifold $\boldsymbol{M}_{4}=\  \mathbb{R}%
^{2}\times C$ parameterised by $\left( x,y,z\right) .$ The real $\left(
x,y\right) $ are the local coordinates of $\mathbb{R}^{2}$ and the complex $%
z $ is the usual local coordinate of the complex plane $\mathbb{C}$. It can
be also viewed as a local coordinate $Z_{1}/Z_{2}$ parameterising an open
patch in the complex projective line $\mathbb{CP}^{1}.$ This theory is
characterized by the complexified gauge symmetry $G$ that will be here as $%
SL\left( m\right) $ and later as the supergroup $SL\left( m|n\right) .$ The
gauge field connection given by%
\begin{equation}
\mathcal{A}=dx\mathcal{A}_{x}+dy\mathcal{A}_{y}+d\bar{z}\mathcal{A}_{\bar{z}}
\label{1f}
\end{equation}%
This is a complex 1-form gauge potential valued in the Lie \textrm{algebra} $%
g$ of the gauge symmetry $G$. So, we have the expansion $\mathcal{A}=t_{a}%
\mathcal{A}^{a}$ with $t_{a}$ standing for the generators of $G$.\newline
The field action $S\left[ \mathcal{A}\right] $ describing the space dynamics
of the gauge field $\mathcal{A}$ reads in the p-form language as follows 
\begin{equation}
S_{4dCS}=\int_{\mathbb{R}^{2}\times \mathbb{CP}^{1}}dz\wedge tr\left[ 
\mathcal{A}\wedge d\mathcal{A}+\frac{2}{3}\mathcal{A}\wedge \mathcal{A}%
\wedge \mathcal{A}\right]  \label{ac}
\end{equation}%
The field equation of the gauge connection $\mathcal{A}$ without external
objects like line defects is given by $\delta S_{4dCS}/\delta \mathcal{A}=0$
and reads as 
\begin{equation}
\mathcal{F}=d\mathcal{A}+\mathcal{A}\wedge \mathcal{A}=0  \label{f}
\end{equation}%
The solution of this flat 2-form field strength is given by the topological
gauge connection $\mathcal{A}=\mathfrak{g}^{-1}d\mathfrak{g}$ with $%
\mathfrak{g}$ being an \textrm{element} of the gauge symmetry group $(%
\mathfrak{g}\in G)$. \newline
Using the covariant derivatives $\mathcal{D}_{\mathrm{m}}=\partial _{\mathrm{%
m}}+\mathcal{A}_{\mathrm{m}}$ with label $\mathrm{m}=x,y,\bar{z},$ we can
express the equation of motion like $\left[ \mathcal{D}_{\mathrm{m}},%
\mathcal{D}_{\mathrm{n}}\right] =0$ reading explicitly as 
\begin{equation}
\begin{tabular}{lll}
$\partial _{x}\mathcal{A}_{y}-\partial _{y}\mathcal{A}_{x}+\left[ \mathcal{A}%
_{x},\mathcal{A}_{y}\right] $ & $=$ & $0$ \\ 
$\partial _{x}\mathcal{A}_{\bar{z}}-\partial _{\bar{z}}\mathcal{A}_{x}+\left[
\mathcal{A}_{x},\mathcal{A}_{\bar{z}}\right] $ & $=$ & $0$ \\ 
$\partial _{y}\mathcal{A}_{\bar{z}}-\partial _{\bar{z}}\mathcal{A}_{y}+\left[
\mathcal{A}_{y},\mathcal{A}_{\bar{z}}\right] $ & $=$ & $0$%
\end{tabular}
\label{gt}
\end{equation}%
If we assume that $\mathcal{A}_{x}=0$ and $\mathcal{A}_{\bar{z}}=0$ (the
conditions for tH$_{\mathrm{\gamma }_{0}}^{\mu }$), the above relations
reduce to $\partial _{x}\mathcal{A}_{y}=0$ and $\partial _{\bar{z}}\mathcal{A%
}_{y}=0$; they show that the component $\mathcal{A}_{y}$ is analytic in $z$
with no dependence in $x$; 
\begin{equation}
\mathcal{A}_{y}=\mathcal{A}_{y}\left( y,z\right)
\end{equation}

\subsection{Implementing the 't Hooft line in CS theory}

In the case where the 4D CS theory is equipped with\textrm{\ }a\textrm{\ }%
magnetically charged 't Hooft line defect tH$_{\mathrm{\gamma }_{0}}^{\mu }$
that couples to the CS field; the field action (\ref{ac}) is deformed like $%
S_{4dCS}+S_{int}$[{\small tH}$_{{\small \gamma }_{{\small 0}}}^{{\small \mu }%
}$]. The new field equation of motion of the gauge potential is no longer
trivial \textrm{\cite{1D}}; the 2-form field strength $\mathcal{F}$ is not
flat ($\mathcal{F}\neq 0$). This non flatness deformation can be imagined in
terms of a Dirac monopole with non trivial first Chern class $c_{1}=k$
(magnetic charge) that\textrm{\ }we write as follows 
\begin{equation}
c_{1}=\int_{\mathbb{S}^{2}}\mathcal{F}_{_{U\left( 1\right) }}
\end{equation}%
where $\mathbb{S}^{2}$ is a sphere surrounding the 't Hooft line. In these
regards, recall that for a hermitian non abelian Yang-Mills theory with
gauge symmetry $\mathcal{G}$, the magnetic Dirac monopole is implemented in
the gauge group by a coweight $\mathbf{\mu }:U\left( 1\right) \rightarrow 
\mathcal{G}$. As a consequence, one has a Dirac monopole such that the gauge
field $\mathcal{A}$ defines on $\mathbb{S}^{2}$ a $\mathcal{G}$-bundle
related to the $U\left( 1\right) $ monopole line bundle by the coweight $%
\mathbf{\mu }=k_{i}\omega _{i}$ with integers $k_{i}$ and fundamental
coweights $\omega _{i}$ of $\mathcal{G}$. Further details on this matter are
reported in the appendix A where we also explain how underlying constraint
relations lead to the following expression the L-operator%
\begin{equation}
L(z;\mu )=e^{X}z^{\mu }e^{Y}  \label{lm}
\end{equation}%
In this relation first obtained by Costello-Gaiotto-Yagi (CGY) in \textrm{%
\cite{1D},} the operators $X$ and $Y$ are valued in the nilpotent algebras $%
\boldsymbol{n}_{\pm }$ of the Levi-decomposition of the gauge symmetry $G$.
As such, they can be expanded as follows 
\begin{equation}
X=\sum_{i=1}^{\dim \boldsymbol{n}_{+}}b^{i}X_{i}\qquad ,\qquad
Y=\sum_{i=1}^{\dim \boldsymbol{n}_{-}}c_{i}Y^{i}
\end{equation}%
where the $X_{i}$'s and $Y^{i}$'s are respectively the generators of $%
\boldsymbol{n}_{+}$ and $\boldsymbol{n}_{-}$. The coefficients $b^{i}$ and $%
c_{i}$ are the Darboux coordinates of the phase space of the L-operator.
Notice that these $b^{i}$'s and $c_{i}$'s are classical variables. At the
quantum level, these phase space variables are promoted to creation $\hat{b}%
^{i}$ and annihilation $\hat{c}_{i}$ operators satisfying the canonical
commutation relations of the bosonic harmonic oscillators namely%
\begin{equation}
\lbrack \hat{c}_{k},\hat{b}^{i}]=\delta _{k}^{i}\qquad ,\qquad \lbrack \hat{b%
}^{i},\hat{b}^{k}]=\left[ \hat{c}_{i},\hat{c}_{k}\right] =0
\end{equation}%
These quantum relations will be used later when studying the quantum Lax
operator; see section 6.

\section{Lax equation in 4D CS theory}

In sub-section 3.1, we revisit the construction of the Costello-Gaiotto-Yagi
(CGY) Lax operator $e^{X}z^{\mu }e^{Y}$ for the bosonic gauge symmetry $%
sl\left( 2\right) $ (for short $\mathcal{L}_{{\small sl}_{2}}$); and use
this result to show that $\mathcal{L}_{{\small sl}_{2}}=e^{X}z^{\mathbf{\mu }%
}e^{Y}$ extends also to the supergauge invariance $sl\left( 1|1\right) $
that we denote like $\mathcal{L}_{{\small sl}_{1|1}}=e^{\Psi }z^{\mathbf{\mu 
}}e^{\Phi }$. In subsection 3.2, we consider the $\mathcal{L}_{{\small sl}%
_{2}}$ and $\mathcal{L}_{{\small sl}_{1|1}};$ and show that both obey the
typical Lax equations $\partial _{t}\mathcal{L}=\left[ A_{t},\mathcal{L}%
\right] $ with pair $\left( \mathcal{L},A_{t}\right) $ to be constructed.

\subsection{From Lax operator to super-Lax operator}

\subsubsection{L-operator for $\mathcal{L}_{{\protect \small sl}_{2}}$ theory}

We start with the CGY Lax operator $e^{X}z^{\mu }e^{Y}$ and think about the
triplet $\left( \mu ,X,Y\right) $ in terms of the three $sl\left( 2\right) $
generators $\left( h,E_{\pm \alpha }\right) $ as follows 
\begin{equation}
X=bE_{+\alpha },\qquad Y=cE_{-\alpha }\qquad ,\qquad z^{\mathbf{\mu }}=z^{h}
\end{equation}%
where $b$ and $c$ are complex parameters. The $h,E_{\pm \alpha }$ obey the
commutation relations 
\begin{equation}
\left[ \mathcal{E},\mathcal{F}\right] =h,\qquad \left[ h,\mathcal{E}\right]
=+\mathcal{E},\qquad \left[ h,\mathcal{F}\right] =-\mathcal{F}  \label{ef}
\end{equation}%
where we have set $\mathcal{E}=E_{+\alpha }$ and $\mathcal{F}=E_{-\alpha }$.
From these relations, we deduce the algebra $\left[ \mu ,X\right] =+X$ and $%
\left[ \mu ,Y\right] =-Y.$ By using the vector basis $\left \{
e_{1},e_{2}\right \} \equiv \left \{ \left \vert 1\right \rangle
,\left
\vert 2\right \rangle \right \} $ of the fundamental representation
of $sl\left( 2\right) ,$ we can solve these relations like 
\begin{equation}
X=b\left \vert 1\right \rangle \left \langle 2\right \vert ,\qquad Y=c\left
\vert 2\right \rangle \left \langle 1\right \vert \qquad ,\qquad \mu =\frac{1%
}{2}\left( P_{1}-P_{2}\right)
\end{equation}%
with $P_{1}=\left \vert 1\right \rangle \left \langle 1\right \vert $ and $%
P_{2}=\left \vert 2\right \rangle \left \langle 2\right \vert .$ By
substituting these expressions into $e^{X}z^{\mu }e^{Y}$, we end up with the
well known expression of $\mathcal{L}_{{\small sl}_{2}}$. As these
calculations are interesting, let us give some details. First, we find that $%
\mathcal{L}_{{\small sl}_{2}}$ is expressed in terms of $X,Y$ and the
projectors as%
\begin{equation}
\begin{tabular}{lll}
$\mathcal{L}_{{\small sl}_{2}}$ & $=$ & $z^{\frac{1}{2}}P_{1}+z^{-\frac{1}{2}%
}P_{2}+z^{\frac{1}{2}}XP_{1}Y+z^{-\frac{1}{2}}XP_{2}Y+$ \\ 
&  & $z^{\frac{1}{2}}XP_{1}+z^{-\frac{1}{2}}XP_{2}+z^{\frac{1}{2}}P_{1}Y+z^{-%
\frac{1}{2}}P_{2}Y$%
\end{tabular}%
\end{equation}%
Moreover, using the properties $XP_{1}=0$ and $P_{1}Y=0$ as well as 
\begin{equation}
P_{1}X=X,\qquad XP_{2}=X,\qquad YP_{1}=Y,\qquad P_{2}Y=Y
\end{equation}%
the L-operator takes the form%
\begin{equation}
\mathcal{L}_{{\small sl}_{2}}=P_{1}\left( z^{\frac{1}{2}}+z^{-\frac{1}{2}%
}XY\right) P_{1}+P_{2}\left( z^{-\frac{1}{2}}\right) P_{2}+P_{1}\left( z^{-%
\frac{1}{2}}X\right) P_{2}+P_{2}\left( z^{-\frac{1}{2}}Y\right) P_{1}
\end{equation}%
It reads in the matrix language $\left( \mathcal{L}_{{\small sl}_{2}}\right)
_{ij}=P_{i}\mathcal{L}_{{\small sl}_{2}}P_{j}$ as follows%
\begin{equation}
\mathcal{L}_{{\small sl}_{2}}=\left( 
\begin{array}{cc}
z^{\frac{1}{2}}+z^{-\frac{1}{2}}bc & z^{-\frac{1}{2}}b \\ 
z^{-\frac{1}{2}}c & z^{-\frac{1}{2}}%
\end{array}%
\right)
\end{equation}%
where one recognises the usual $bc$ term corresponding to the energy of the
free bosonic harmonic oscillator. By writing $bc$ as $\frac{1}{2}(bc+cb)$
and thinking of these $b,c$ parameters (Darboux-coordinates) in terms of
creation $\hat{b}=\hat{a}^{\dagger }$ and annihilation $\hat{c}=\hat{a}$
operators with commutator $\hat{a}\hat{a}^{\dagger }-\hat{a}^{\dagger }\hat{a%
}=1,$ we get $\hat{b}\hat{c}=\hat{a}^{\dagger }\hat{a}+\frac{1}{2}$ and then
the following quantum L-operator 
\begin{equation}
\mathcal{\hat{L}}_{{\small sl}_{2}}=\left( 
\begin{array}{cc}
z^{\frac{1}{2}}+z^{-\frac{1}{2}}\left( a^{\dagger }a+\frac{1}{2}\right) & 
z^{-\frac{1}{2}\hat{a}\dagger } \\ 
z^{-\frac{1}{2}}\hat{a} & z^{-\frac{1}{2}}%
\end{array}%
\right)
\end{equation}%
Multiplying by $z^{\frac{1}{2}}$, we discover the expression of $\mathcal{%
\hat{L}}_{{\small sl}_{2}}$ obtained by algebraic methods.

\subsubsection{Super L-operator for $sl\left( 1|1\right) $ theory}

Here, we extend the analysis done for $\mathcal{L}_{{\small sl}_{2}}$ to the
super $sl\left( 1|1\right) $. For that, we begin by recalling some useful
features. $\left( \mathbf{1}\right) $ the $sl\left( 1|1\right) $ is a sub-
superalgebra of $gl\left( 1|1\right) $ with vanishing supertrace \textrm{%
\cite{2J,3J}}. $\left( \mathbf{2}\right) $ The $gl\left( 1|1\right)
=gl\left( 1|1\right) _{\bar{0}}\oplus gl\left( 1|1\right) _{\bar{1}}$ has
even and odd sectors with $gl\left( 1|1\right) _{\bar{0}}=gl\left( 1\right)
\oplus gl\left( 1\right) .$ It has rank 2 and four dimensions generated by: $%
\left( i\right) $ two bosonic generators $K$ and $J;$ and $\left( ii\right) $
two fermionic $\Psi ^{+}$ and $\Phi ^{-}$ satisfying%
\begin{equation}
\begin{tabular}{lllllll}
$\left \{ \Psi ^{+},\Phi ^{-}\right \} $ & $=$ & $K$ & $\quad ,\quad $ & $%
\left[ K,J\right] $ & $=$ & $0$ \\ 
$\  \left[ J,\Psi ^{+}\right] $ & $=$ & $+\Psi ^{+}$ & $\quad ,\quad $ & $%
\left[ K,\Psi ^{+}\right] $ & $=$ & $0$ \\ 
$\  \left[ J,\Phi ^{-}\right] $ & $=$ & $-\Phi ^{-}$ & $\quad ,\quad $ & $%
\left[ K,\Phi ^{-}\right] $ & $=$ & $0$%
\end{tabular}
\label{311}
\end{equation}%
as well as $\left( \Psi ^{+}\right) ^{2}=\left( \Phi ^{-}\right) ^{2}=0.$
The Casimir $C$ of $gl\left( 1|1\right) $ is given by $C=\left( 2J-1\right)
E+2\Phi ^{-}\Psi ^{+}.$ To determine the super $\mathcal{L}_{{\small sl}%
_{1|1}},$ we assume that it is given by the same formula as $\mathcal{L}_{%
{\small sl}_{2}}$ namely 
\begin{equation}
\mathcal{L}_{{\small sl}_{1|1}}=e^{\Psi }z^{\mu }e^{\Phi }  \label{PS}
\end{equation}%
but with triplet $\left( \Psi ,\Phi ,\mathbf{\mu }\right) $ as follows%
\begin{equation}
\Psi =\beta ^{-}\Psi ^{+}\qquad ,\qquad \Phi =\gamma ^{+}\Phi ^{-}\qquad
,\qquad z^{\mathbf{\mu }}=z^{J}
\end{equation}%
where $\beta ^{-},\gamma ^{+}$ are now fermions satisfying $\left( \beta
^{-}\right) ^{2}=\left( \gamma ^{+}\right) ^{2}=0.$ Repeating the analysis
done for the bosonic $\mathcal{L}_{{\small sl}_{2}},$ we end up with the
following super L-operator,%
\begin{equation}
\mathcal{L}_{{\small sl}_{1|1}}=\left( 
\begin{array}{cc}
z^{\frac{1}{2}}+z^{-\frac{1}{2}}\beta ^{-}\gamma ^{+} & z^{-\frac{1}{2}%
}\beta ^{-} \\ 
z^{-\frac{1}{2}}\gamma ^{+} & z^{-\frac{1}{2}}%
\end{array}%
\right)  \label{lg}
\end{equation}%
In this expression, we recognise the typical $\beta ^{-}\gamma ^{+}$ term
corresponding to the energy of a free classical fermionic oscillator. By
writing it as $\frac{1}{2}(\beta ^{-}\gamma ^{+}-\gamma ^{+}\beta ^{-})$ and
promoting $\left( \beta ^{-},\gamma ^{+}\right) $ to operators $(\hat{\beta}%
^{-},\hat{\gamma}^{+})$, we obtain the quantum version of (\ref{lg}).
Indeed, thinking of $(\hat{\beta}^{-},\hat{\gamma}^{+})$ as creation ($\hat{%
\beta}^{-}=\hat{\xi}^{\dagger })$ and annihilation ($\hat{\gamma}^{+}=\hat{%
\xi})$ operators with canonical anti-commutator 
\begin{equation}
\hat{\xi}\hat{\xi}^{\dagger }+\hat{\xi}^{\dagger }\hat{\xi}=1\qquad ,\qquad 
\hat{\xi}^{2}=0\qquad ,\qquad (\hat{\xi}^{\dagger })^{2}=0,
\end{equation}%
it follows that $\hat{\beta}^{-}\hat{\gamma}^{+}=\hat{\xi}^{\dagger }\hat{\xi%
}-\frac{1}{2}.$ Therefore, the quantum $\mathcal{\hat{L}}_{{\small sl}%
_{1|1}} $ reads as,%
\begin{equation}
\mathcal{\hat{L}}_{{\small sl}_{1|1}}=\left( 
\begin{array}{cc}
z^{\frac{1}{2}}+z^{-\frac{1}{2}}(\hat{\xi}^{\dagger }\hat{\xi}-\frac{1}{2})
& z^{-\frac{1}{2}}\hat{\xi}^{\dagger } \\ 
z^{-\frac{1}{2}}\hat{\xi} & z^{-\frac{1}{2}}%
\end{array}%
\right)
\end{equation}%
It agrees with the one obtained in \textrm{\cite{1G} using algebraic methods
and indicates the consistency of the CS formalism for supergroup symmetries}%
. Notice that $\mathcal{\hat{L}}_{{\small sl}_{1|1}}$ has only one fermionic
oscillator $(\hat{\beta}^{-},\hat{\gamma}^{+}).$ This feature will be
explained when we consider DSDs.

\subsection{CGY operator as solution of $\mathfrak{D}L=0$}

Here we show that the $L_{{\small CGY}}$, derived from parallel transport of
gauge configuration as revisited in the appendix A, can be also viewed as a
solution of a differential equation $\mathfrak{D}L_{{\small CGY}}=0.$ First,
we consider the $sl\left( m\right) $ bosonic $\mathcal{L}_{{\small sl}%
_{m}}=e^{X}z^{\mu }e^{Y}$ by zooming on the $sl\left( 2\right) $ theory.
Then, we generalise this equation to $\mathcal{L}_{{\small sl}_{m|n}}$ while
focussing on the leading $sl(1|1)$.

\subsubsection{Determining $\mathfrak{D}\mathcal{L}_{{\protect \small sl}%
_{2}}=0$}

As a foreword to the $SL(m|n)$ case, we consider at first the CS theory with
gauge symmetry $G=SL\left( m\right) $ and look for the algebraic equation 
\begin{equation}
\mathfrak{D}\mathcal{L}_{{\small sl}_{m}}=0  \label{dl}
\end{equation}%
whose solution is given by the parallel transport eq(\ref{2}) detailed in
appendix A. To that purpose, we recall the Levi- decomposition $sl_{m}=%
\boldsymbol{n}_{+}\oplus \boldsymbol{l}_{\mu }\oplus \boldsymbol{n}_{-}$ 
\textrm{\cite{1D,1J,2J,3J,4J}}, 
\begin{equation*}
\boldsymbol{l}_{\mu }=sl\left( 1\right) \oplus sl\left( m-1\right) ,%
\boldsymbol{\qquad n}_{\pm }=\left( \boldsymbol{m-1}\right) _{\pm },\qquad %
\left[ \mathbf{\mu },\boldsymbol{n}_{\pm }\right] =\pm \boldsymbol{n}_{\pm }
\end{equation*}%
with $\mathbf{\mu }$ refering to the adjoint action of the minuscule
coweight $\mu $. For $\boldsymbol{sl}_{2}$ generated by $\left \{ h,E_{\pm
}\right \} ,$ we have $\boldsymbol{sl}_{2}=\boldsymbol{1}_{+}\oplus 
\boldsymbol{sl}_{1}\oplus \boldsymbol{1}_{-}$ with $\left[ h,E_{\pm }\right]
=\pm E_{\pm }$ as in (\ref{ef}). \newline
To determine the differential eq(\ref{dl}), we start from the oscillator
realisation of the L-operator (\ref{lm}) with nilpotent matrix operators as $%
X=b\mathcal{E}$ and $Y=c\mathcal{F}$. Then, we compute the commutator $%
ad_{\mu }\left( L\right) =\left[ \mu ,L\right] .$ The $ad_{\mu }$ is just
the derivation in the Lie algebra obeying $ad_{\mu }\left( AB\right) =\left[
ad_{\mu }\left( A\right) \right] B+A\left[ ad_{\mu }\left( B\right) \right] $%
. Applying this property to the L-operator, we find%
\begin{equation}
ad_{\mu }\left( L\right) =\left[ ad_{\mu }\left( X\right) \right] L+L\left[
ad_{\mu }\left( Y\right) \right]  \label{ad}
\end{equation}%
where we have used $ad_{\mu }\left( z^{\mu }\right) =0$ and $ad_{\mu }\left(
Y\right) e^{Y}=e^{Y}ad_{\mu }\left( Y\right) .$ Then, using $ad_{\mu }\left(
X\right) =X,$ $ad_{\mu }\left( Y\right) =-Y$ and putting back into (\ref{ad}%
), we obtain%
\begin{equation}
ad_{\mu }\left( L\right) =XL-LY  \label{39}
\end{equation}%
By thinking of $XL$ and $LY$ in terms of the left $l_{X}$ and the right $%
r_{Y}$ multiplications acting like $l_{X}\left( L\right) =XL$ and $%
r_{Y}\left( L\right) =LY$, we can put (\ref{39}) into the form $\mathfrak{D}%
\mathcal{L}_{{\small sl}_{2}}=0$ with%
\begin{equation}
\mathfrak{D}=ad_{\mu }-\left( l_{X}-r_{Y}\right)
\end{equation}%
This operator involves the triplet $\left( \mu ,X,Y\right) $; as such it can
be imagined as $\mathfrak{D}=\mathfrak{D}_{\left( \mu ,X,Y\right) }.$ To
interpret to this abstract operator in classical physics, we use the
following correspondence with Hamiltonian systems living on a phase space $%
\mathcal{E}_{ph}$ parameterized by $\left( q,p\right) $. We have 
\begin{equation}
\begin{tabular}{lll}
$ad_{\mu }L$ & : & $\frac{\partial L}{\partial t}$ \\ 
$l_{X}\left( L\right) $ & : & $\frac{\partial H}{\partial p}\frac{\partial L%
}{\partial q}$ \\ 
$r_{Y}\left( L\right) $ & : & $\frac{\partial L}{\partial p}\frac{\partial H%
}{\partial q}$%
\end{tabular}%
\   \label{lxy}
\end{equation}%
with Hamiltonian $H\left( q,p\right) $ governing the dynamics. Putting these
relations back into (\ref{39}), we obtain the familiar evolution equation $%
\frac{\partial L}{\partial t}=\left \{ H,L\right \} _{PB}.$ At the quantum
level, it is equivalent to the Heisenberg equation of motion $\frac{\partial
L}{\partial t}=\left[ iH,L\right] $ (Lax equation with $A_{t}=iH$). From the
correspondence (\ref{lxy}), we learn that the $X$ and $Y$ operators used in
the CGY construction are nothing but the Hamiltonian vector fields $\frac{%
\partial H}{\partial p}\frac{\partial }{\partial q}$ and $\frac{\partial H}{%
\partial q}\frac{\partial }{\partial p}.$ Moreover, writing the Hamiltonian
as $(bp^{2}+cq^{2})/2,$ we end up with $X=b\mathcal{E}$ and $Y=c\mathcal{F}$
as well as%
\begin{equation}
\mathcal{E}=p\frac{\partial }{\partial q},\qquad \mathcal{F}=q\frac{\partial 
}{\partial p},\qquad \mathbf{\mu }=p\frac{\partial }{\partial p}-q\frac{%
\partial }{\partial q}
\end{equation}

\subsubsection{Extension to super $\mathcal{L}_{{\protect \small sl}_{1|1}}$}

First, recall that the Lie superalgebra $gl\left( 1|1\right) $ is four
dimensional and obeys eq(\ref{311}). It has two fermionic generators $\Psi
^{+},\Phi ^{-}$ and two bosonic $J,K$. In the graded phase space $\mathcal{E}%
_{ph}^{1|1}$\ with super coordinates $\left( q,p;\chi ^{+},\chi ^{-}\right) $%
, the fermionic generator are realised as 
\begin{equation}
\begin{tabular}{lll}
$\Psi ^{+}$ & $=$ & $\chi ^{+}\frac{\partial }{\partial p}+q\frac{\partial }{%
\partial \chi ^{-}}$ \\ 
$\Phi ^{-}$ & $=$ & $\chi ^{-}\frac{\partial }{\partial q}+p\frac{\partial }{%
\partial \chi ^{+}}$%
\end{tabular}
\label{fop}
\end{equation}%
and the bosonic ones like 
\begin{equation}
\begin{tabular}{lll}
$J$ & $=$ & $\left( \chi ^{+}\frac{\partial }{\partial \chi ^{+}}-\chi ^{-}%
\frac{\partial }{\partial \chi ^{-}}\right) +\left( q\frac{\partial }{%
\partial q}-p\frac{\partial }{\partial p}\right) $ \\ 
$K$ & $=$ & $\left( \chi ^{+}\frac{\partial }{\partial \chi ^{+}}+\chi ^{-}%
\frac{\partial }{\partial \chi ^{-}}\right) +\left( q\frac{\partial }{%
\partial q}+p\frac{\partial }{\partial p}\right) $%
\end{tabular}%
\end{equation}%
To determine the differential equation whose solution is given by eq(\ref{39}%
) namely $\mathcal{L}_{{\small sl}_{1|1}}=e^{\Psi }z^{\mu }e^{\Phi },$ we
repeat the same calculations done for $\mathcal{L}_{{\small sl}_{2}}$ to
obtain%
\begin{equation}
ad_{\mu }\left( \mathcal{L}_{{\small sl}_{1|1}}\right) -\left( \Psi \mathcal{%
L}_{{\small sl}_{1|1}}-\mathcal{L}_{{\small sl}_{1|1}}\Phi \right) =0
\label{312}
\end{equation}%
where $\Psi $ and $\Phi $ have two contributions like%
\begin{equation}
\Psi =\mathrm{\beta }^{-}\Psi ^{+}\qquad ,\qquad \Phi =\mathrm{\gamma }%
^{+}\Phi ^{-}  \label{313}
\end{equation}%
where $\left( \mathrm{\beta }^{-},\mathrm{\gamma }^{+}\right) $ are
fermionic-like Darboux coordinates. The interpretation of eqs(\ref{312}-\ref%
{313}) is given by the extension of (\ref{lxy}) to the graded phase space
with bosonic $\left( q,p\right) $ and fermionic $\left( \chi ^{+},\chi
^{-}\right) $ coordinates\footnote{%
\ Notice that for supersymmetric oscillator of SUSY quantum mechanics , the
supercharges $\hat{Q}^{+},\hat{Q}^{-}$ read in terms of bosonic ($b$%
)/fermionic ($f$) operators as $\sqrt{\omega }\hat{b}^{\dagger }\hat{f}$ and 
$\sqrt{\omega }\hat{b}\hat{f}^{\dagger }$ \cite{5J}.}. The homologue of (\ref%
{lxy}) reads for ${\small sl}\left( 1|1\right) $ as 
\begin{equation}
\begin{tabular}{lll}
$ad_{\mu }\mathcal{L}$ & : & $\frac{\partial \mathcal{L}}{\partial t}$ \\ 
$l_{\Psi }\left( \mathcal{L}\right) $ & : & $\frac{\partial H}{\partial \chi
^{+}}\frac{\partial \mathcal{L}}{\partial \chi ^{-}}+\frac{\partial H}{%
\partial q}\frac{\partial \mathcal{L}}{\partial p}$ \\ 
$r_{\Phi }\left( \mathcal{L}\right) $ & : & $\frac{\partial H}{\partial \chi
^{-}}\frac{\partial \mathcal{L}}{\partial \chi ^{+}}+\frac{\partial H}{%
\partial p}\frac{\partial \mathcal{L}}{\partial q}$%
\end{tabular}%
\end{equation}%
where $H=\mathrm{\beta }^{-}\left( \chi ^{+}q\right) +\mathrm{\gamma }%
^{+}\left( \chi ^{-}p\right) .$ Putting these relations back into (\ref{39}%
), we obtain the familiar evolution equation $\frac{\partial \mathcal{L}}{%
\partial t}=\left \{ H,\mathcal{L}\right \} _{PB}.$ From this
correspondence, we identify the $\Psi $ and $\Phi $ operators used in the
super- $\mathcal{L}_{{\small sl}_{1|1}}$ with the vector fields $\frac{%
\partial H}{\partial \chi ^{+}}\frac{\partial }{\partial \chi ^{-}}+\frac{%
\partial H}{\partial q}\frac{\partial }{\partial p}$ and $\frac{\partial H}{%
\partial \chi ^{-}}\frac{\partial }{\partial \chi ^{+}}+\frac{\partial H}{%
\partial p}\frac{\partial }{\partial q}.$ By substituting, we obtain $\Psi =%
\mathrm{\beta }^{-}\Psi ^{+}$ and $\Phi =\mathrm{\gamma }^{+}\Phi ^{-}$ with
fermionic operators $\Psi ^{+}$ and $\Phi ^{-}$\ as in (\ref{fop}).

\section{Chern-Simons with gauge supergroups}

In this section, we give basic tools needed for the study of 4D CS theory
with gauge symmetry given by classical \emph{super-groups\ }$G;$\emph{\ }and
for\emph{\ }\textrm{the construction of the }super-Lax operators $L\left(
z\right) $. Other elements like Verma modules of $G$ are given in Appendix B
as they are necessary for the investigation of superspin chains
characterized the graded RLL equation \textrm{\cite{0K}.} 
\begin{equation*}
R\left( z_{1}-z_{2}\right) L\left( z_{1}\right) L\left( z_{2}\right)
=L\left( z_{2}\right) L\left( z_{1}\right) R\left( z_{1}-z_{2}\right)
\end{equation*}%
\textrm{Generally speaking}, classical super-groups $G$ and their Lie
superalgebras $\boldsymbol{g=g}_{\bar{0}}\oplus \boldsymbol{g}_{\bar{1}}$
are made of two building blocks; they are classified in literature as
sketched here below \textrm{\cite{1K},}%
\begin{equation}
\begin{tabular}{l|l|l}
$\  \  \  \  \  \  \  \  \  \  \boldsymbol{g}$ & $\  \  \  \  \  \  \  \  \  \  \  \  \ 
\boldsymbol{g}_{\bar{0}}$ & $\  \  \  \  \  \  \  \  \  \boldsymbol{g}_{\bar{1}}$ \\ 
\hline
$\  \ A\left( m-1,n-1\right) $ $\  \ $ & $\  \ A_{m-1}\oplus A_{n-1}\oplus
gl\left( 1\right) $ $\  \ $ & $\  \  \left( m,\bar{n}\right) \oplus \left( \bar{%
m},n\right) $ \\ 
$\  \ A\left( m-1,m-1\right) $ $\  \ $ & $\  \ A_{m-1}\oplus A_{m-1}$ & $\  \
\left( m,\bar{m}\right) \oplus \left( \bar{m},m\right) $ $\  \  \ $ \\ 
$\  \ C\left( m+1\right) $ & $\  \ C_{m}\oplus gl\left( 1\right) $ & $\  \
\left( 2m\right) \oplus \left( 2m\right) $ \\ 
$\  \ B\left( m,n\right) $ & $\  \ B_{m}\oplus C_{n}$ & $\  \  \left(
2m+1,2n\right) $ \\ 
$\  \ D\left( m,n\right) $ & $\  \ D_{m}\oplus C_{n}$ & $\  \  \left(
2m,2n\right) $ \\ 
$\  \ F\left( 4\right) $ & $\  \ A_{1}\oplus B_{3}$ & $\  \  \left( 2,8\right) $
\\ 
$\  \ G\left( 3\right) $ & $\  \ A_{1}\oplus G_{2}$ & $\  \  \left( 2,7\right) $
\\ 
$\  \ D\left( 2,1;\alpha \right) $ & $\  \ A_{1}\oplus A_{1}\oplus A_{1}$ & $\
\  \left( 2,2,2\right) $ \\ \hline
\end{tabular}
\label{tabg}
\end{equation}%
\begin{equation*}
\end{equation*}%
where $A\left( m-1,n-1\right) $ designates $sl\left( m|n\right) $ which we
will focus on here. Several results about these graded algebras and their
quantization were obtained in the Lie superalgebra literature; they
generalise the bosonic-like ones; some of them will be commented in this
study, \textrm{related others are described in literature; see for instance 
\cite{1KA,1KB,1KC,1KD,1KE}}. \newline
\textrm{In the first subsection, we will introduce the }$gl\left( m|n\right) 
$\textrm{\ algebra and describe useful mathematical tools for the present
study. In the second one, we study some illustrating examples and the
associated \textquotedblleft Dynkin diagrams\textquotedblright \ to manifest
the \emph{non-uniqueness} of the DSDs (Dynkin Super-Diagrams) of Lie
superalgebras in contrast to the bosonic Lie algebras. As such, a given 4D
CS theory with }$GL\left( m|n\right) $\textrm{\ invariance may have several
DSDs and consequently lead to different super L-operators.}

\subsection{Lie superalgebras: $gl\left( m|n\right) $ and $sl\left(
m|n\right) $ family}

As $sl\left( m|n\right) $ is a Lie sub-superalgebra of $gl\left( m|n\right)
, $ it is interesting to work with $gl\left( m|n\right) .$ The restriction
to $sl\left( m|n\right) $ can be obtained by imposing the super-traceless
(str) condition leading to 
\begin{equation}
\begin{tabular}{l|l|l}
\ superalgebra \  & \ dimension & \  \ rank \  \  \  \  \  \  \  \  \\ \hline
$\  \  \ gl\left( m|n\right) $ & $\left( m+n\right) ^{2}$ & $m+n$ \\ \hline
$\  \  \ sl\left( m|n\right) $ & $\left( m+n\right) ^{2}-1$ & $m+n-1$ \\ \hline
\end{tabular}%
\end{equation}

\subsubsection{The $gl\left( m|n\right) $\ superalgebra}

The Lie superalgebra $gl\left( m|n\right) $ is a $\mathbb{Z}_{2}$- graded
vector space with two particular subspaces: $\left( 1\right) $ an even
subspace $gl\left( m|n\right) _{\bar{0}}$ given by $gl\left( m\right) \oplus
gl\left( n\right) .$ $\left( 2\right) $ an odd subspace $gl\left( m|n\right)
_{\bar{1}}$ given by a module of $gl\left( m|n\right) _{\bar{0}}$. The super 
$gl\left( m|n\right) $ is endowed by a $\mathbb{Z}_{2}$- graded commutator
often termed as super-bracket given by \textrm{\cite{1K}}%
\begin{equation}
\left[ X,Y\right \} =XY-\left( -\right) ^{\left \vert A\right \vert \left
\vert B\right \vert }YX  \label{41}
\end{equation}%
In this relation, the degree $\left \vert Z\right \vert $ refers to the two
classes of the $\mathbb{Z}_{2}$-gradation namely $\left \vert Z\right \vert =%
\bar{0},$ for the bosonic generators, and $\left \vert Z\right \vert =\bar{1}
$ for the fermionic ones. To fix the ideas, we have for the bosonic
generators the usual Lie bracket $\left[ B_{1},B_{2}\right] $ while for the
fermionic ones we have the anticommutator $\left \{ F_{1},F_{2}\right \} .$
For the mixture, we have the commutators $[B,F\}$.\newline
\textrm{In this context,} a natural way to think of $gl\left( m|n\right) $
is in terms of $End\left( \mathbb{C}^{m|n}\right) $ acting on the graded
vector space $\mathbb{C}^{m|n}$. As such, the super-matrices of $End\left( 
\mathbb{C}^{m|n}\right) $ have the form\textrm{\footnote{%
The form of the supermatrix presented in eq(\ref{44}) corresponds to the
minimal fermionic node situation.}}%
\begin{equation}
M_{\left( m|n\right) \times \left( m|n\right) }=\left( 
\begin{array}{cc}
A_{m\times m} & B_{m\times n} \\ 
C_{n\times m} & D_{n\times n}%
\end{array}%
\right)  \label{44}
\end{equation}%
For the even subalgebra $g_{\bar{0}}=gl\left( m|n\right) _{\bar{0}},$ we
have $B_{m\times n}=C_{n\times m}=0.$ For the odd subspace $gl\left(
m|n\right) _{\bar{1}}$, we have $A_{m\times m}=0$ and $D_{n\times n}=0$. 
\textrm{Notice as well} that the odd space $gl\left( m|n\right) _{\bar{1}}$
can be also splitted like $g_{+1}\oplus g_{-1}$ where $g_{\pm 1}$ are
nilpotent subalgebras corresponding to triangular super-matrices. In the
representation language of the even part $gl\left( m\right) \oplus gl\left(
n\right) ,$ the $g_{\pm 1}$ can be interpreted in terms of bi-fundamentals
like 
\begin{equation}
g_{+1}\sim \left( \boldsymbol{m},\boldsymbol{\bar{n}}\right) \qquad ,\qquad
g_{-1}\sim \left( \boldsymbol{\bar{m}},\boldsymbol{n}\right)
\end{equation}%
The complex vector space $\mathbb{C}^{m|n}$ is generated by m bosonic basis
vector $\left( b_{1},...,b_{m}\right) $ and n fermionic-like partners $%
\left( f_{1},...,f_{n}\right) .$ Generally speaking, these basis vectors of $%
\mathbb{C}^{m|n}$ can be collectively denoted like $e_{\text{\textsc{a}}%
}=\left( e_{1},...,e_{m+n}\right) $ with the $\mathbb{Z}_{2}$-grading
property 
\begin{equation}
\begin{tabular}{l|l|l}
$\deg \text{\textsc{a}}=\left \vert \text{\textsc{a}}\right \vert $ & $\  \  \ 
\bar{0}$ \  \  \  & $\  \  \bar{1}$ \  \  \\ \hline
$\  \  \  \  \  \  \ e_{\text{\textsc{a}}}$ & $\  \  \mathbb{C}^{m}$ \  \  & $\ 
\mathbb{C}^{n}$%
\end{tabular}%
\end{equation}%
It turns out that the ordering of the vectors in the set $\left \{ e_{\text{%
\textsc{a}}}\right \} $ is important in the study of Lie superalgebras and
their representations. Different orderings of the $e_{\text{\textsc{a}}}$'s
lead to different DSDs for the same Lie superalgebra. In other words, a
given Lie superalgebra has many representative DSDs\textrm{.}\newline
To get more insight into the super-algebraic structure of $gl\left(
m|n\right) ,$ we denote its $\left( n+m\right) ^{2}$ generators as $\mathcal{%
E}_{\text{\textsc{ab}}}$ with labels \textsc{a,b}$=1,...,m+n$, and express
its graded commutations as%
\begin{equation}
\left[ \mathcal{E}_{\text{\textsc{ab}}},\mathcal{E}_{\text{\textsc{cd}}%
}\right \} =\delta _{\text{\textsc{bc}}}\mathcal{E}_{\text{\textsc{ad}}%
}-\left( -\right) ^{\left \vert \mathcal{E}_{\text{\textsc{ab}}}\right \vert
\left \vert \mathcal{E}_{\text{\textsc{cd}}}\right \vert }\delta _{\text{%
\textsc{da}}}\mathcal{E}_{\text{\textsc{cb}}}  \label{gc}
\end{equation}%
with 
\begin{equation}
\left \vert \mathcal{E}_{\text{\textsc{ab}}}\right \vert \equiv \deg \left
\vert \mathcal{E}_{\text{\textsc{ab}}}\right \vert =\left \vert \text{%
\textsc{a}}\right \vert +\left \vert \text{\textsc{b}}\right \vert
\end{equation}%
For the degrees $\left \vert \mathcal{E}_{\text{\textsc{ab}}}\right \vert
=0, $ the labels \textsc{a} and \textsc{b} are either both bosonic or both
fermionic. For $\left \vert \mathcal{E}_{\text{\textsc{ab}}}\right \vert =1,$
the labels \textsc{a} and \textsc{b} have opposite degrees. Using the
convention notation $e_{\text{\textsc{a}}}=b_{a},f_{i}$ with the label $a\in
J_{1}$ for bosons and the label $i\in J_{2}$ for fermions such that $%
J_{1}\cup J_{2}=\left \{ 1,2,...,m+n\right \} $, we can split the
super-generators $\mathcal{E}_{\text{\textsc{ab}}}$ into four types as%
\begin{equation}
\mathcal{E}_{\text{ab}}\qquad \mathcal{E}_{\text{ij}}^{\prime }\qquad 
\mathcal{\tilde{E}}_{\text{ai}}\qquad \mathcal{\tilde{E}}_{\text{ia}%
}^{\prime }  \label{ep}
\end{equation}%
So, we have: $\left( i\right) $ $m^{2}+n^{2}$ bosonic generators; $m^{2}$
operators $\mathcal{E}_{\text{ab}}$ and $n^{2}$ operators $\mathcal{E}_{%
\text{ij}}^{\prime }$. $\left( ii\right) $ $2mn$ fermionic generators; $mn$
operators $\mathcal{\tilde{E}}_{\text{ai}}$ and $nm$ operators $\mathcal{%
\tilde{E}}_{\text{ia}}^{\prime }$.\newline
The Cartan subalgebra of $sl\left( m|n\right) ,$ giving the quantum numbers
of the physical states, is generated by r diagonal operators $H_{\text{%
\textsc{a}}}.$ They read in terms of the diagonal $\mathcal{E}_{\text{%
\textsc{aa}}}$ as follows%
\begin{equation}
H_{\text{\textsc{a}}}=\left( -\right) ^{\left \vert \text{\textsc{a}}\right
\vert }\mathcal{E}_{\text{\textsc{aa}}}-\left( -\right) ^{\left \vert \text{%
\textsc{a}}{\small +1}\right \vert }\mathcal{E}_{\left( \text{\textsc{a}}%
{\small +1}\right) \left( \text{\textsc{a}}{\small +1}\right) }  \label{r}
\end{equation}%
Because of the $\mathbb{Z}_{2}$-gradation, we have four writings of the
generators $H_{\text{\textsc{a}}}$, they are as follows%
\begin{equation}
\begin{tabular}{l|l||l|l}
\multicolumn{2}{l}{\small \  \  \  \  \  \  \  \  \ bosonic sector} & 
\multicolumn{2}{||l}{\small \  \  \  \  \  \  \ fermionic sector} \\ \hline
$\  \left \vert \text{\textsc{a}}\right \vert ,\left \vert \text{\textsc{a}}%
{\tiny +1}\right \vert $ \  & $\  \  \  \  \  \  \  \ H_{\text{\textsc{a}}}$ & $\
\left \vert \text{\textsc{a}}\right \vert ,\left \vert \text{\textsc{a}}%
{\tiny +1}\right \vert $ \  & $\  \  \  \  \  \  \ H_{\text{\textsc{a}}}$ \\ \hline
$\  \  \  \bar{0},\bar{0}$ & $\mathcal{E}_{aa}-\mathcal{E}_{\left( a+1\right)
\left( a+1\right) }$ & $\  \  \  \bar{0},\bar{1}$ & $+\mathcal{E}_{ai}^{\prime
}+\mathcal{E}_{\left( a+1\right) \left( i+1\right) }^{\prime }$ \\ 
$\  \  \  \bar{1},\bar{1}$ & $\mathcal{\tilde{E}}_{\left( i+1\right) \left(
i+1\right) }-\mathcal{\tilde{E}}_{ii}$ & $\  \  \  \bar{1},\bar{0}$ & $-%
\mathcal{\tilde{E}}_{ia}^{\prime }-\mathcal{\tilde{E}}_{\left( i+1\right)
\left( a+1\right) }^{\prime }$ \\ \hline
\end{tabular}%
\end{equation}

\subsubsection{Root super-system and generalized Cartan matrix}

The roots $\alpha _{\text{\textsc{ab}}}$ of the $sl\left( m|n\right) $
(super- roots) are of two kinds: bosonic roots and fermionic ones. They are
expressed in terms of the unit weight vectors $\epsilon _{\text{\textsc{a}}}=%
\mathcal{E}_{\text{\textsc{aa}}}^{\vee }$ (the dual of $\mathcal{E}_{\text{%
\textsc{aa}}}$ ).\  \  \ 

$\bullet $ \emph{Root super-system }$\Phi _{sl_{m|n}}$\newline
The root system $\Phi _{sl_{m|n}}$ has $\left( m+n\right) \left(
m+n-1\right) $ roots $\alpha _{\text{\textsc{ab}}}$ realised as $\epsilon _{%
\text{\textsc{a}}}-\epsilon _{\text{\textsc{b}}}$ with \textsc{a}$\neq $%
\textsc{b}$.$ Half of these super- roots are positive (\textsc{a}$<$\textsc{b%
}) and the other half are negative (\textsc{a}$>$\textsc{b}). The positive
roots are generated by r simple roots $\alpha _{\text{\textsc{a}}}$ given by%
\begin{equation}
\alpha _{\text{\textsc{a}}}=\epsilon _{\text{\textsc{a}}}-\epsilon _{\text{%
\textsc{a}}+1}
\end{equation}%
The degree of these simple roots depend on the ordering of the $\epsilon _{%
\text{\textsc{a}}}$'s. The step operators $\mathcal{E}_{\alpha _{\text{%
\textsc{a}}}}\equiv E_{\text{\textsc{a}}}$ and $\mathcal{E}_{-\alpha _{\text{%
\textsc{a}}}}\equiv F_{\text{\textsc{a}}}$ together with $H_{\text{\textsc{a}%
}},$ defining the Chevalley basis, obey%
\begin{equation}
\begin{tabular}{lll}
$\left[ H_{\text{\textsc{a}}},E_{\text{\textsc{b}}}\right \} $ & $=$ & $+K_{%
\text{\textsc{ab}}}E_{\text{\textsc{b}}}$ \\ 
$\left[ H_{\text{\textsc{a}}},F_{\text{\textsc{b}}}\right \} $ & $=$ & $-K_{%
\text{\textsc{ab}}}F_{\text{\textsc{b}}}$ \\ 
$\left[ E_{\text{\textsc{a}}},E_{\text{\textsc{b}}}\right \} $ & $=$ & $%
\delta _{\text{\textsc{ab}}}H_{\text{\textsc{a}}}\left( -\right) ^{\left
\vert \text{\textsc{a}}\right \vert }$%
\end{tabular}%
\end{equation}%
where $K_{\text{\textsc{ab}}}=\alpha _{\text{\textsc{a}}}\left( H_{\text{%
\textsc{b}}}\right) $ is the super- Cartan matrix of $sl\left( m|n\right) $
given by%
\begin{equation}
K_{\text{\textsc{ab}}}=\delta _{\text{\textsc{ab}}}\left[ \left( -\right)
^{\left \vert \text{\textsc{a}}\right \vert }+\left( -\right) ^{\left \vert 
\text{\textsc{a}}+1\right \vert }\right] -\left( -\right) ^{\left \vert 
\text{\textsc{a}}+1\right \vert }\delta _{\left( \text{\textsc{a+1}}\right) 
\text{\textsc{b}}}-\left( -\right) ^{\left \vert \text{\textsc{a}}\right
\vert }\delta _{\text{\textsc{a}}\left( \text{\textsc{b+1}}\right) }
\end{equation}%
The matrix $K_{\text{\textsc{ab}}}$ extends the usual $sl\left( m\right) $
algebra namely $K_{ab}=2\delta _{ab}-\delta _{\left( a+1\right) b}-\delta
_{a\left( b+1\right) }.$ It allows to encode the structure of $sl\left(
m|n\right) $ into a generalised Dynkin diagram. This Dynkin super- diagram
has r nodes labeled by the simple roots $\alpha _{\text{\textsc{A}}}.$
Because of the degrees of the $\alpha _{\text{\textsc{A}}}$'s, the nodes are
of two kinds: $\left( \mathbf{1}\right) $ bosonic (\textrm{blank)} nodes
associated with $K_{\text{\textsc{aa}}}=\pm 2$. $\left( \mathbf{2}\right) $
fermionic (\textrm{grey)} nodes associated with $K_{\text{\textsc{aa}}}=0$.
As noticed before, the DSD of Lie superalgebras depend on the ordering of
the vector basis $\left( e_{1},...,e_{m}\right) $ of $\mathbb{C}^{m|n}$ and
the associated $\left( \epsilon _{1},...,\epsilon _{m}\right) $. This
feature is illustrated on the following example.\  \ 

$\bullet $ \emph{Distinguished root system of }$sl\left( m|n\right) $ 
\newline
Here, we give the root system\emph{\ }of $sl\left( m|n\right) $ in the
distinguished weight basis where the $n+m$ unit weight vectors $\epsilon _{%
\text{\textsc{a}}}$ are ordered like $\left( \varepsilon _{a}|\delta
_{i}\right) $ with $\epsilon _{m+i}=\delta _{i}$ and 
\begin{equation}
\varepsilon _{a}=\left( \varepsilon _{1},...,\varepsilon _{m}\right) \qquad
,\qquad \delta _{i}=\left( \delta _{1},...,\delta _{n}\right)
\end{equation}%
As such, the set of distinguished roots $\beta _{\text{\textsc{ab}}%
}=\epsilon _{\text{\textsc{a}}}-\epsilon _{\text{\textsc{b}}}$ split into
thee subsets as follows%
\begin{equation}
\begin{tabular}{|l|l|l|l|l|}
\hline
{\small root} & $\  \  \  \  \  \  \alpha _{ab}$ & $\  \  \  \  \alpha _{ij}^{\prime }$
& $\  \  \  \  \tilde{\alpha}_{ai}$ & $\  \ -\tilde{\alpha}_{ai}$ \\ \hline
{\small value} & $\  \  \  \varepsilon _{a}-\varepsilon _{b}$ & $\  \  \delta
_{i}-\delta _{j}$ & $\  \varepsilon _{a}-\delta _{i}$ \  \  \  & $\  \delta
_{i}-\varepsilon _{a}$ \  \  \  \\ \hline
{\small number} & $m\left( m-1\right) $ \  & $\  \ n\left( n-1\right) $ \  \ 
& $\  \  \  \ mn$ & $\  \  \  \ mn$ \\ \hline
{\small degree} & {\small \  \  \  \ even} & {\small \  \  \  \ even} & {\small \
\  \  \ odd \  \ } & {\small \  \  \  \  \ odd \  \ } \\ \hline
\end{tabular}%
\end{equation}%
where $\tilde{\alpha}_{ia}=-\tilde{\alpha}_{ai}$. Similarly, the set of the
simple roots $\alpha _{\text{\textsc{A}}}$ split into three kinds as shown
in the following table with $\left( \alpha _{a}\right) ^{2}=2$ and $\left(
\alpha _{i}^{\prime }\right) ^{2}=-2$ as well as $\left( \tilde{\alpha}%
\right) ^{2}=0$.%
\begin{equation}
\begin{tabular}{|l|l|l|l|}
\hline
{\small simple root } & $\  \  \  \  \alpha _{a}$ & $\  \  \  \  \alpha _{i}^{\prime
}$ & $\  \  \  \tilde{\alpha}$ \\ \hline
\  \ {\small value} & $\  \varepsilon _{a}-\varepsilon _{a+1}$ & $\  \delta
_{i}-\delta _{i+1}$ & $\  \varepsilon _{m}-\delta _{1}$ \  \  \  \\ \hline
\  \ {\small number} & $\  \  \ m-1$ \  & $\  \ n-1$ \  \  & $\  \  \  \ 1$ \\ \hline
\  \  \ {\small degree} & {\small \  \  \  \ even} & {\small \  \  \  \ even} & 
{\small \  \  \  \ odd \  \ } \\ \hline
\end{tabular}%
\end{equation}

\subsection{Lie superalgebras $gl\left( 2|1\right) $ and $gl\left(
3|2\right) $}

Here, we study two examples of Lie superalgebras aiming to illustrate how a
given Lie superalgebra has several DSDs.

\subsubsection{The superalgebra $gl\left( 2|1\right) $}

This is the simplest Lie superalgebra coming after the $gl\left( 1|1\right) $
considered before. The dimension of $gl\left( 2|1\right) $ is equal to 9 and
its rank is $r=3$. Its even part $gl\left( 2|1\right) _{\bar{0}}$ is given
by $gl\left( 2\right) \oplus gl\left( 1\right) $. The $sl\left( 2|1\right) $
sub-superalgebra of $gl\left( 2|1\right) $ is obtained by imposing the
super-traceless condition. The two Cartan generators $H_{1},H_{2}$ of $%
sl\left( 2|1\right) $ read in terms of the projectors $\mathcal{E}_{\text{%
\textsc{aa}}}=\left \vert \text{\textsc{a}}\right \rangle \left \langle 
\text{\textsc{a}}\right \vert $ as in eq(\ref{r}); they depend on the
grading of the vector basis $e_{\text{\textsc{a}}}\equiv \left \vert \text{%
\textsc{a}}\right \rangle $ of the superspace $\mathbb{C}^{2|1}$ and the
orderings of the vector basis $\left( e_{1},e_{2},e_{3}\right) $ as given in
table \textbf{\ref{table 1}}. 
\begin{table}[tbp]
\begin{center}
$%
\begin{tabular}{||l||l||l||l||l||l||}
\hline \hline
basis & $\left( e_{1},e_{2},e_{3}\right) $ & $\alpha _{\text{a}}^{2}=2$ & $%
\alpha _{\text{a}}^{2}=-2$ & $\alpha _{\text{a}}^{2}=0$ & \  \ Dynkin diagram
\\ \hline \hline
\  \ I & $\left( b_{1},b_{2},f\right) $ & 1 & 0 & 1 & %
\includegraphics[width=2.4cm]{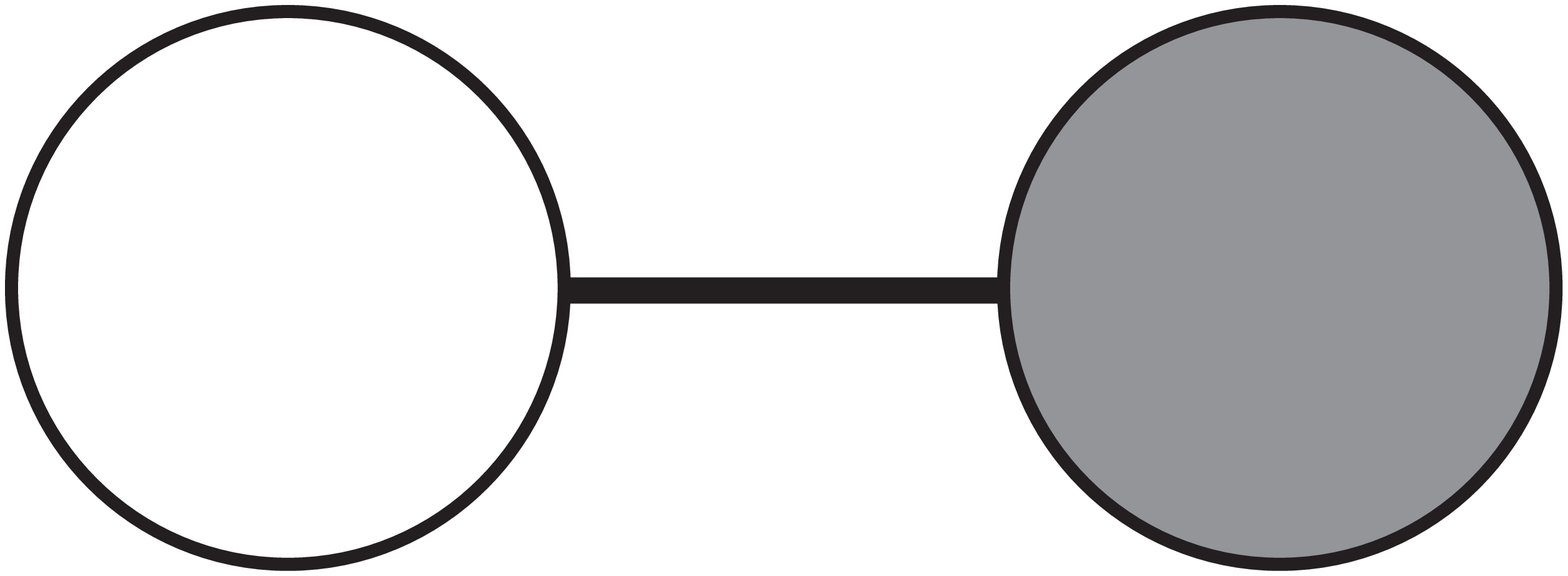} \\ \hline \hline
\ II & $\left( b_{1},f,b_{2}\right) $ & 0 & 0 & 2 & %
\includegraphics[width=2.5cm]{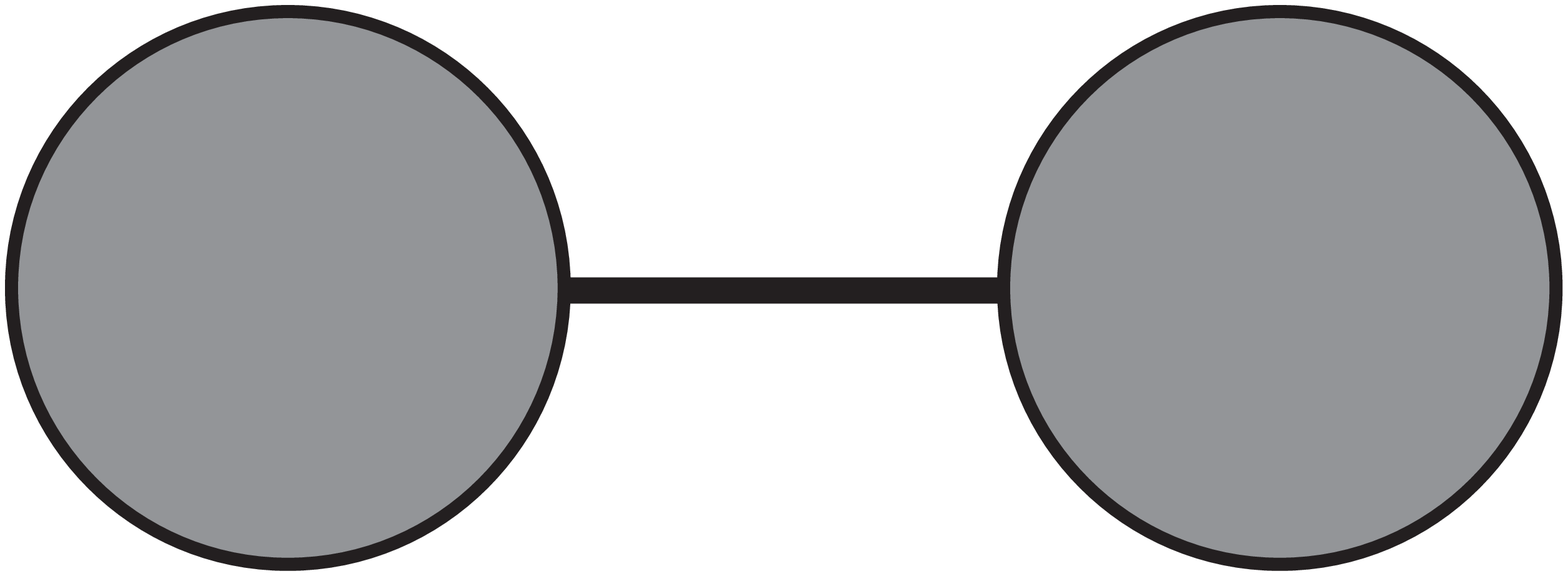} \\ \hline \hline
\end{tabular}%
$%
\end{center}
\caption{Two Dynkin super-diagrams for $sl\left( 2|1\right) $. They have two
nodes. The first has one bosonic node and one fermionic node. The second has
two fermionic nodes }
\label{table 1}
\end{table}
The other missing orderings in this table are equivalent to the given ones;
they are related by Weyl symmetry transformations. \  \ 

$\bullet $ \emph{DSD for the basis choice }I \emph{in table} \textbf{\ref%
{table 1}}\newline
In the case where the three vectors of the basis I are ordered like $\left(
b_{1},b_{2},f\right) $, the two Cartan generators $H_{\text{\textsc{a}}}$ of
the superalgebra \emph{sl}$\left( 2|1\right) $ are given by%
\begin{equation}
H_{1}=\mathcal{E}_{11}-\mathcal{E}_{22},\qquad H_{2}=\mathcal{E}_{22}+%
\mathcal{\tilde{E}}_{33}
\end{equation}%
They have vanishing supertrace. The two simple roots $\alpha _{\text{\textsc{%
a}}}=\epsilon _{\text{\textsc{a}}}-\epsilon _{\text{\textsc{a}}+1}$ read as
follows%
\begin{equation}
\alpha _{1}=\varepsilon _{1}-\varepsilon _{2}\quad ,\quad \alpha
_{2}=\varepsilon _{2}-\delta
\end{equation}%
with gradings as $\left \vert \alpha _{1}\right \vert =0$ and $\left \vert
\alpha _{2}\right \vert =1$. The associated super- Cartan matrix is given by%
\begin{equation}
K_{\text{\textsc{ab}}}=\left( 
\begin{array}{cc}
2 & -1 \\ 
-1 & 0%
\end{array}%
\right)  \label{k1}
\end{equation}%
The root system\emph{\ }$\Phi _{sl_{2|1}}$ has six roots; three positive and
three negative; they are given by $\pm \alpha _{1},\pm \alpha _{2}$ and $\pm
\alpha _{3}=\pm \left( \alpha _{1}+\alpha _{2}\right) $ with the grading $%
\left \vert \alpha _{3}\right \vert =1.$ This grading feature indicates that 
\emph{sl}$\left( 2|1\right) $ has four fermionic step operators $E_{\pm
\alpha _{2}},E_{\pm \alpha _{3}}$; and two bosonic ones $E_{\pm \alpha _{1}}$%
.

$\bullet $ \emph{DSD for the basis choice }II \emph{in table} \textbf{\ref%
{table 1}}\newline
Here, the vectors of the basis choice II are ordered like $\left(
b_{1},f,b_{2}\right) $. The two associated Cartan generators $H_{\text{%
\textsc{a}}}$ of the superalgebra \emph{sl}$\left( 2|1\right) $ in the basis
II are given by%
\begin{equation}
H_{1}=+\mathcal{E}_{11}+\mathcal{\tilde{E}}_{22},\qquad H_{2}=-\mathcal{%
\tilde{E}}_{22}-\mathcal{E}_{33}
\end{equation}%
The two simple roots $\alpha _{\text{\textsc{a}}}=\epsilon _{\text{\textsc{a}%
}}-\epsilon _{\text{\textsc{a}}+1}$ read as follows%
\begin{equation}
\alpha _{1}=\varepsilon _{1}-\delta \quad ,\quad \alpha _{2}=\delta
-\varepsilon _{2}
\end{equation}%
with the same grading $\left \vert \alpha _{1}\right \vert =\left \vert
\alpha _{2}\right \vert =1.$ The associated super- Cartan matrix is given by%
\begin{equation}
K_{\text{\textsc{ab}}}=\left( 
\begin{array}{cc}
0 & -1 \\ 
-1 & 0%
\end{array}%
\right)  \label{k2}
\end{equation}%
Notice that the Cartan matrices (\ref{k1}) and (\ref{k2}) are different;
they give two different DSDs for the same Lie superalgebra \emph{sl}$\left(
2|1\right) $ as depicted in table \textbf{\ref{table 1}}.

\subsubsection{The superalgebra $gl\left( 3|2\right) $}

The dimension of the $gl\left( 3|2\right) $ Lie superalgebra is equal to 25
and has rank $r=5$. Its even part $gl\left( 3|2\right) _{\bar{0}}$ is given
by $gl\left( 3\right) \oplus gl\left( 2\right) $. The four Cartan generators 
$H_{1},H_{2},H_{3},H_{4}$ of the $sl\left( 3|2\right) $ read in terms of the
projectors $\mathcal{E}_{\text{\textsc{aa}}}=\left \vert \text{\textsc{a}}%
\right \rangle \left \langle \text{\textsc{a}}\right \vert $ as in eq(\ref{r}%
). Their expression depend on the grading of the vector basis $e_{\text{%
\textsc{a}}}\equiv \left \vert \text{\textsc{a}}\right \rangle $ of the
superspace $\mathbb{C}^{3|2}$ and on the ordering of the three bosonic $%
\left( b_{1},b_{2},b_{3}\right) $ and the two fermionic $\left(
f_{1},f_{2}\right) $ within the basis $\left(
e_{1},e_{2},e_{3},e_{4},e_{5}\right) $. Up to Weyl transformations, we
distinguish five different orderings given in Table \textbf{\ref{table 2}}. 
\begin{table}[tbp]
\begin{center}
$%
\begin{tabular}{||l||l||l||l||l||l||}
\hline \hline
basis & $\left( e_{1},e_{2},e_{3},e_{4},e_{5}\right) $ & $\alpha _{\text{a}%
}^{2}=2$ & $\alpha _{\text{a}}^{2}=-2$ & $\alpha _{\text{a}}^{2}=0$ & Dynkin
diagram \\ \hline \hline
\  \ I & $\left. \left( b_{1},b_{2},b_{3},f_{1},f_{2}\right) \right. $ & 2 & 1
& 1 & \includegraphics[width=3cm]{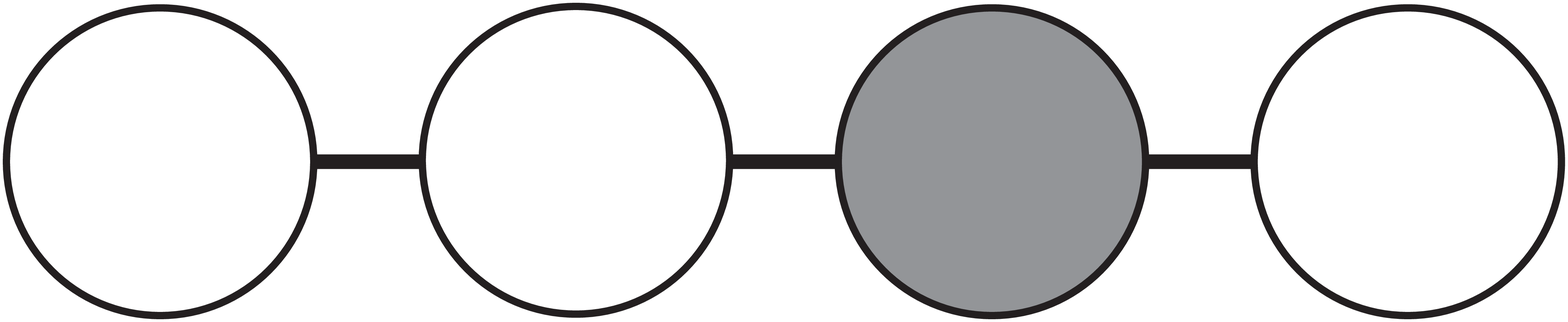} \\ \hline \hline
\ II & $\left( b_{1},b_{2},f_{1},b_{3},f_{2}\right) $ & 1 & 0 & 3 & %
\includegraphics[width=3cm]{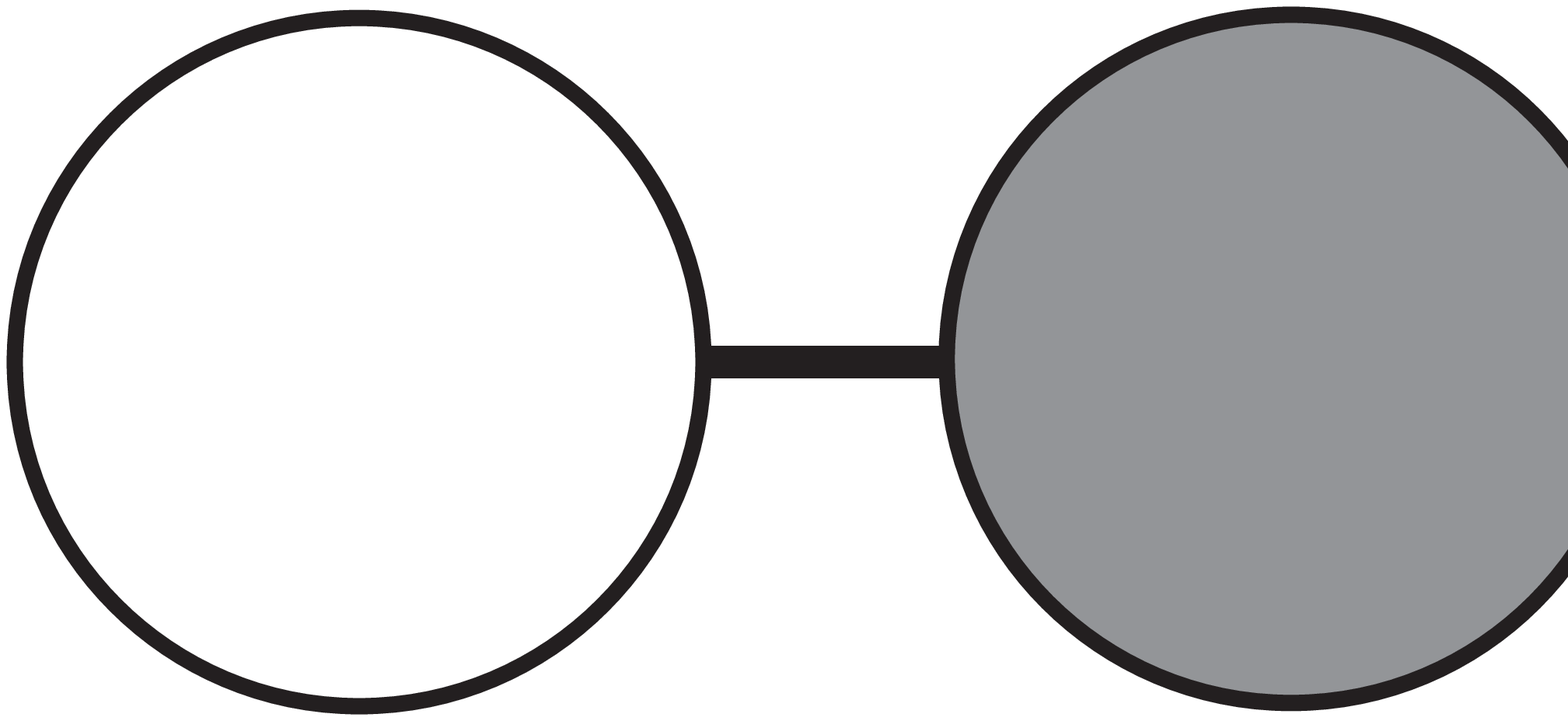} \\ \hline \hline
\ III & $\left( b_{1},b_{2},f_{1},f_{2},b_{3}\right) $ & 1 & 1 & 2 & %
\includegraphics[width=3cm]{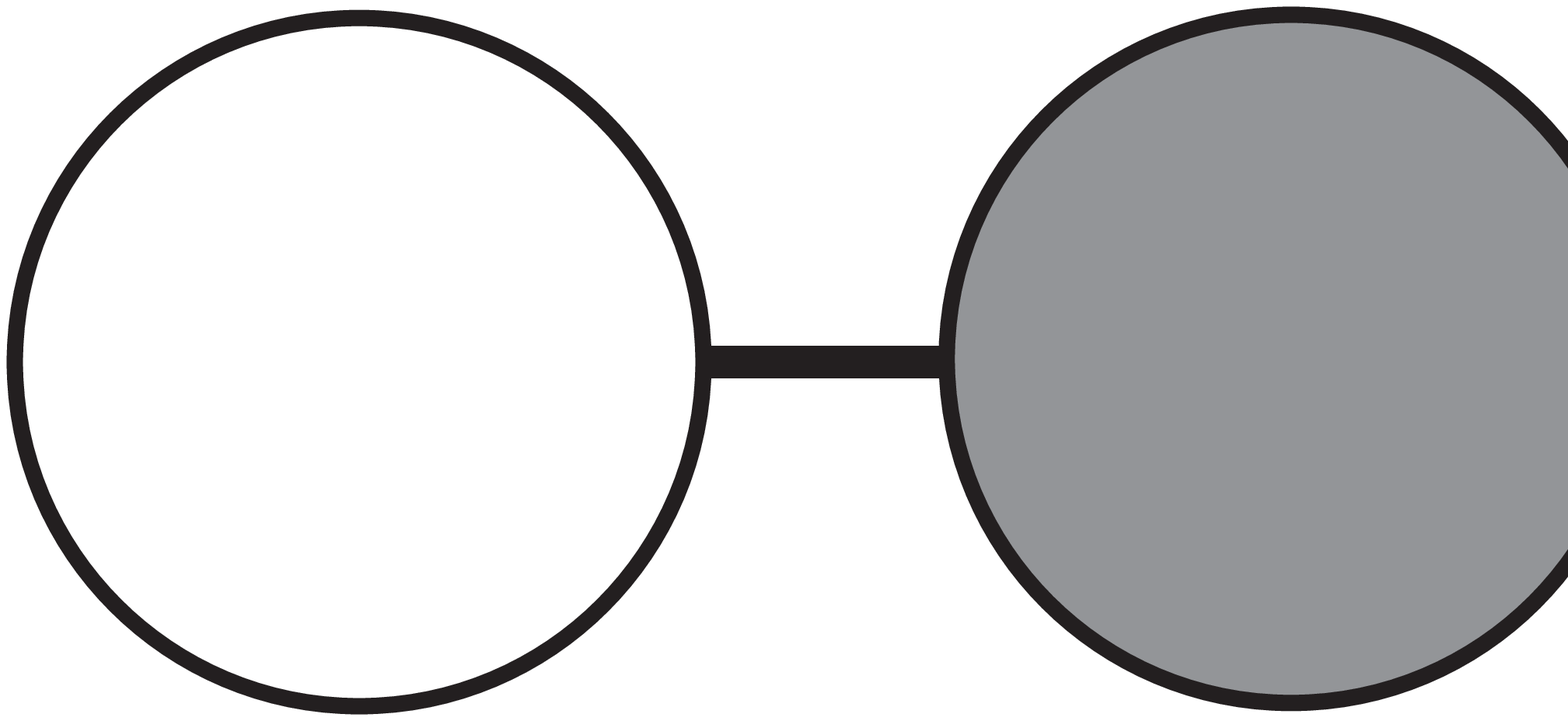} \\ \hline \hline
\ IV & $\left( b_{1},f_{1},b_{2},f_{2},b_{3}\right) $ & 0 & 0 & 4 & %
\includegraphics[width=3cm]{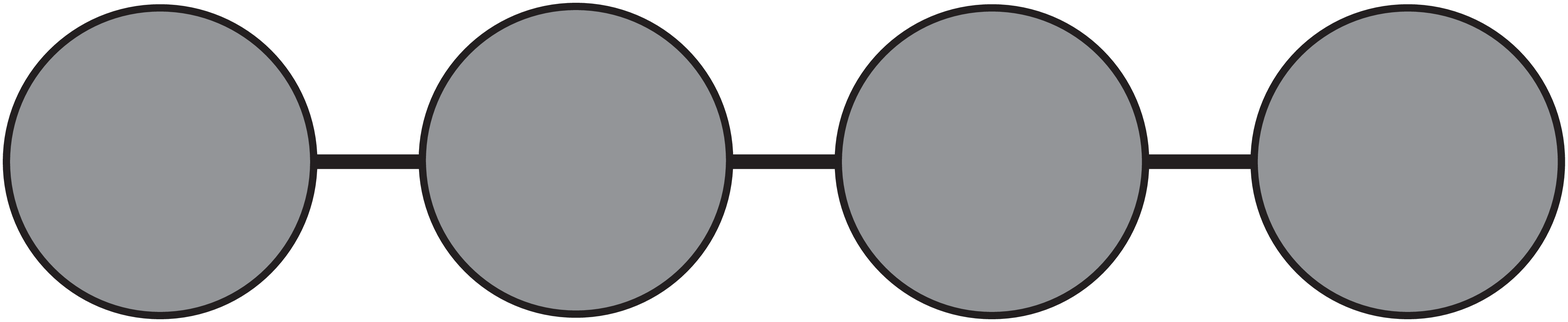} \\ \hline \hline
\ V & $\left( f_{1},b_{1},b_{2},b_{3},f_{2}\right) $ & 2 & 0 & 2 & %
\includegraphics[width=3cm]{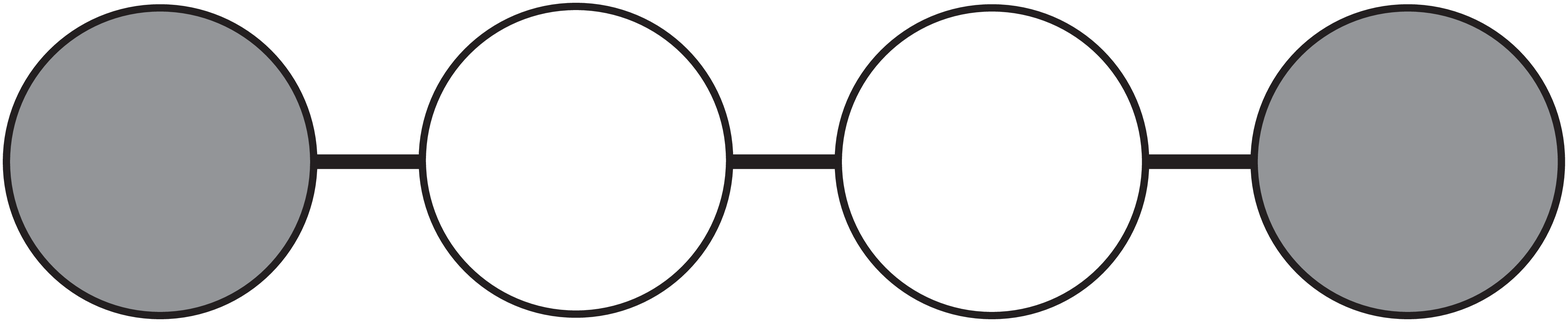} \\ \hline \hline
\end{tabular}%
$%
\end{center}
\caption{{}Five Dynkin super-diagrams for $sl\left( 3|2\right) $. They have
four nodes with various numbers of fermionic nodes. The first DSD has three
bosonic nodes and one fermionic node. These DSDs have at least one fermionic
node. Notice that the number of fermionic nodes is not the unique parameter
needed to classify the DSDs.}
\label{table 2}
\end{table}
Below, we describe the DSD associated with the basis III. A similar
treatment can be done for the other four basis. \newline
In the basis III of the Table \textbf{\ref{table 2}}, the vectors in $\left(
e_{1},e_{2},e_{3},e_{4},e_{5}\right) $ are ordered as $\left(
b_{1},b_{2},f_{1},f_{2},b_{3}\right) $. The four Cartan generators $H_{\text{%
\textsc{a}}}$ of the superalgebra \emph{sl}$\left( 3|2\right) $ in this
basis are therefore given by%
\begin{equation}
\begin{tabular}{lllllll}
$H_{1}$ & $=$ & $+\mathcal{E}_{11}-\mathcal{E}_{22}$ & , & $H_{3}$ & $=$ & $-%
\mathcal{\tilde{E}}_{33}+\mathcal{\tilde{E}}_{44}$ \\ 
$H_{2}$ & $=$ & $+\mathcal{E}_{22}+\mathcal{\tilde{E}}_{33}$ & , & $H_{4}$ & 
$=$ & $-\mathcal{\tilde{E}}_{44}-\mathcal{E}_{55}$%
\end{tabular}%
\end{equation}%
To construct the super- Dynkin diagram of the superalgebra $sl\left(
3|2\right) $ with the basis III, we use the following ordering%
\begin{equation}
\begin{tabular}{lll}
$\left( e_{1},e_{2},e_{3},e_{4},e_{5}\right) $ & $=$ & $\left(
b_{1},b_{2},f_{1},f_{2},b_{3}\right) $ \\ 
$\left( \epsilon _{1},\epsilon _{2},\epsilon _{3},\epsilon _{4},\epsilon
_{5}\right) $ & $=$ & $\left( \varepsilon _{1},\varepsilon _{2},\delta
_{1},\delta _{2},\varepsilon _{3}\right) $%
\end{tabular}%
\end{equation}%
with $\varepsilon _{i}^{2}=1$ and $\delta _{i}^{2}=-1.$ For this ordering,
the four simple roots $\alpha _{\text{\textsc{a}}}=\epsilon _{\text{\textsc{a%
}}}-\epsilon _{\text{\textsc{a}}+1}$ of $sl\left( 3|2\right) $ have the
grading $\left \vert \alpha _{1}\right \vert =\left \vert \alpha
_{2}\right
\vert =0$ and $\left \vert \alpha _{3}\right \vert =\left \vert
\alpha _{4}\right \vert =0$; and read as follows%
\begin{equation}
\begin{tabular}{lllll}
$\alpha _{1}$ & $=\varepsilon _{1}-\varepsilon _{2}$ & $\quad ,\quad $ & $%
\alpha _{3}$ & $=\delta _{1}-\delta _{2}$ \\ 
$\alpha _{2}$ & $=\varepsilon _{2}-\delta _{1}$ & $\quad ,\quad $ & $\alpha
_{4}$ & $=\delta _{2}-\varepsilon _{3}$%
\end{tabular}
\label{sa}
\end{equation}%
The other twelve roots of the super-system\emph{\ }$\Phi _{sl_{3|2}}$ are
given by 
\begin{equation}
\begin{tabular}{lllll}
$\pm \left( \alpha _{1}+\alpha _{2}\right) $ & $,$ & $\pm \left( \alpha
_{1}+\alpha _{2}+\alpha _{3}\right) $ & $,$ & $\pm \left( \alpha _{1}+\alpha
_{2}+\alpha _{3}+\alpha _{4}\right) $ \\ 
$\pm \left( \alpha _{2}+\alpha _{3}\right) $ & $,$ & $\pm \left( \alpha
_{2}+\alpha _{3}+\alpha _{4}\right) $ & $,$ & $\pm \left( \alpha _{3}+\alpha
_{4}\right) $%
\end{tabular}%
\end{equation}%
Six of these roots are bosonic; they correspond to $gl\left( 3\right) \oplus
gl\left( 2\right) $. The twelve others are fermionic. The super- Cartan
matrix associated with (\ref{sa}) reads as follows%
\begin{equation}
K_{\text{\textsc{ab}}}=\left( 
\begin{array}{cccc}
2 & -1 & 0 & 0 \\ 
-1 & 0 & +1 & 0 \\ 
0 & +1 & -2 & +1 \\ 
0 & 0 & +1 & 0%
\end{array}%
\right)
\end{equation}

\subsubsection{Dynkin super- diagrams: case $sl\left( 3|2\right) $}

\begin{figure}[tbph]
\begin{center}
\includegraphics[width=6cm]{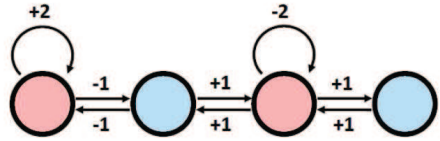}
\end{center}
\par
\vspace{-0.5cm}
\caption{Building the Dynkin diagram of Lie superalgebra $gl\left(
m|n\right) .$ Here we give the example the Dynkin diagram of $gl\left(
3|2\right) $ with weight basis ordered as $\left( \protect \varepsilon _{1},%
\protect \varepsilon _{2},\protect \delta _{1},\protect \delta _{2},\protect%
\varepsilon _{3}\right) .$ }
\label{gl32}
\end{figure}

The Dynkin super- diagrams of $sl\left( 3|2\right) $ have four nodes.
Because of the grading of the simple roots, we distinguish five types of
diagrams as in Table \textbf{\ref{table 2}}. These super-diagrams have a
nice interpretation in the study of integrable superspin chains; in
particular in the correspondence between Bethe equations and 2D $\mathcal{N}%
=\left( 2,2\right) $ quiver gauge theories \textrm{\cite{1H}}. To draw one
of the super-diagrams of $sl\left( 3|2\right) ,$ we start by fixing the
degrees of $\left( \epsilon _{1},\epsilon _{2},\epsilon _{3},\epsilon
_{4},\epsilon _{5}\right) ;$ that is a basis weight vectors of $sl\left(
3|2\right) $. As an example, we take this basis as $\left( \varepsilon
_{1},\varepsilon _{2},\delta _{1},\delta _{2},\varepsilon _{3}\right) $ and
represent it graphically as follows%
\begin{equation}
\begin{tabular}{lllll}
$\varepsilon _{1}$ & $\varepsilon _{2}$ & $\delta _{1}$ & $\delta _{2}$ & $%
\varepsilon _{3}$ \\ 
$\  \textcolor{red}{|}$ & $\textcolor{red}{|}$ & $\textcolor{blue}{|}$ & $%
\textcolor{blue}{|}$ & $\textcolor{red}{|}$%
\end{tabular}%
\end{equation}%
The simple roots $\alpha _{\text{\textsc{a}}}=\epsilon _{\text{\textsc{a}}%
}-\epsilon _{\text{\textsc{a+1}}}$ are represented by circle nodes $\bigcirc 
$ between each pair of adjacent vertical lines associated with $\epsilon _{%
\text{\textsc{a}}}$ and $\epsilon _{\text{\textsc{a+1}}}.$%
\begin{equation}
\begin{tabular}{lllllllll}
& $\alpha _{\text{\textsc{1}}}$ &  & $\alpha _{\text{\textsc{2}}}$ &  & $%
\alpha _{\text{\textsc{3}}}$ &  & $\alpha _{\text{\textsc{4}}}$ &  \\ 
$\textcolor{red}{|}$ & $\textcolor{red}{\bigcirc}$ & $\textcolor{red}{|}$ & $%
\textcolor{blue}{\bigcirc}$ & $\textcolor{blue}{|}$ & $\textcolor{red}{%
\bigcirc}$ & $\textcolor{blue}{|}$ & $\textcolor{blue}{\bigcirc}$ & $%
\textcolor{red}{|}$%
\end{tabular}%
\end{equation}%
For each pair of simple roots $\left( \alpha _{\text{\textsc{a}}},\alpha _{%
\text{\textsc{b}}}\right) $ with non vanishing intersection matrix $K_{\text{%
\textsc{ab}}}=\alpha _{\text{\textsc{a}}}.\alpha _{\text{\textsc{b}}}\neq 0$%
, we draw an arrow from the node $\alpha _{\text{\textsc{a}}}$ to the node $%
\alpha _{\text{\textsc{b}}};$ and we write the value $K_{\text{\textsc{ab}}}$
on the arrow. By hiding the vertical lines, we obtain the super- Dynkin
diagram of $sl\left( 3|2\right) $\ associated with the basis $\left(
\varepsilon _{1},\varepsilon _{2},\delta _{1},\delta _{2},\varepsilon
_{3}\right) $ as illustrated in the \textrm{Figure }\textbf{\ref{gl32}}. 
Notice that the ordering of $\left( \varepsilon _{1},\varepsilon _{2},\delta
_{1},\delta _{2},\varepsilon _{3}\right) $ is defined modulo the action of
the Weyl group $W_{sl_{3}}\times W_{sl_{2}}$ which permutes the basis
vectors without changing the $\mathbb{Z}_{2}$-grading. \newline
We end this description by noticing that this graphic representation applies
also to the highest weight $\lambda =\lambda _{\text{\textsc{a}}}\epsilon _{%
\text{\textsc{a}}}$ of modules of the Lie superalgebra $sl\left( m|n\right)
. $ Details regarding these graphs are given in the Appendix B.

\section{More on Chern-Simons with super- invariance}

In this section, we study the L-operators for $SL\left( m|n\right) $
supergroups by using CS theory in the presence of interacting 't Hooft and
Wilson super-lines. First, we revisit useful results regarding the building
of $\mathcal{L}_{sl_{m}}.$ We take this occasion to introduce a graphic
description to \textrm{imagine} all varieties of the $\mathcal{L}_{sl_{m}}$%
s; see the Figure \textbf{\ref{fig13}}. Then, we investigate the
generalisation of these results to supergroups $SL\left( m|n\right) .$ We
also give illustrating examples.

\subsection{From $SL\left( m\right) $ symmetry to super $SL\left( m|n\right) 
$}

Here, we consider CS theory living on $\mathbb{R}^{2}\times \mathbb{CP}^{1}$
with $SL\left( m\right) $ symmetry and gauge field action (\ref{ac}) in the
presence of crossing 't Hooft and Wilson lines. The 't Hooft line tH$%
_{\gamma _{0}}^{\mu }$ sits on the \emph{x-axis} of the topological plane $%
\mathbb{R}^{2}$ and the Wilson line W$_{\xi _{z}}^{\boldsymbol{R}}$ expands
along the vertical \emph{y-axis} as depicted by the Figure \textbf{\ref{fig1}%
.}

\subsubsection{L-operator for $SL\left( m\right) $ symmetry}

In the CS theory with $SL\left( m\right) $ gauge symmetry, the oscillator
realisation of the L-operator is given by eq(\ref{l}) namely $e^{X}z^{%
\mathbf{\mu }}e^{Y}$. We revisit below the explicit derivation of its
expression by using a projector operator language \cite{4J}.\  \  \ 

$\bullet $ \emph{Building L}$\left( z\right) $\newline
The explicit construction of the L-operator requires the knowledge of three
quantities: \newline
$\left( \mathbf{1}\right) $ the adjoint form of the coweight $\mathbf{\mu }$
which is the magnetic charge operator of tH$_{\gamma _{0}}^{\mathbf{\mu }}$. 
\newline
$\left( \mathbf{2}\right) $ the nilpotent matrix operators $X$ and $Y$
obeying the property $X^{k}=Y^{k}=0$ for some positive integer $k$. For $%
sl\left( m\right) $, this degree $k$ of nilpotency is $k=2$. As we will see
later on, this feature holds also for $sl(m|n)$.\newline
To describe the quantities $\left( \mathbf{\mu },X,Y\right) $, we begin by
recalling $\left( \mathbf{a}\right) $ the Levi-decomposition of $sl(m)$ with
respect to $\mu _{k}$ with label as $1\leq k\leq m-1.$ $\left( \mathbf{b}%
\right) $ the decomposition of its fundamental representation $\boldsymbol{m}%
.$ These two decompositions are given by%
\begin{equation}
\mathbf{\mu }_{k}:%
\begin{tabular}{lll}
$sl(m)$ & $\rightarrow $ & $\boldsymbol{l}_{\mu _{k}}\oplus \boldsymbol{n}%
_{+}\oplus \boldsymbol{n}_{-}$ \\ 
$\mathbf{\ }\boldsymbol{m}$ & $=$ & $\boldsymbol{k}_{1-\frac{k}{m}}\oplus
\left( \boldsymbol{m-k}\right) _{-\frac{k}{m}}$%
\end{tabular}
\label{slm}
\end{equation}%
In the first decomposition, the generators of Levi-subalgebra $\boldsymbol{l}%
_{\mu _{k}}$ and those of the nilpotent sub-algebras $\boldsymbol{n}_{\pm }$
are discriminated by the charges under $\mathbf{\mu }_{k}$; we have $\left[ 
\mathbf{\mu }_{k},\boldsymbol{l}_{\mu _{k}}\right] =0$ and $\left[ \mathbf{%
\mu }_{k},\boldsymbol{n}_{\pm }\right] =\pm \boldsymbol{n}_{\pm }$. In eq(%
\ref{slm}), the $\boldsymbol{l}_{\mu _{k}}$ and $\boldsymbol{n}_{\pm }$ are
is given by%
\begin{equation}
\begin{tabular}{lll}
$\boldsymbol{l}_{\mu _{k}}$ & $=$ & $sl(1)\oplus sl(k)\oplus sl(m-k)$ \\ 
$\boldsymbol{n}$ & $=$ & $\boldsymbol{k}\otimes \left( \boldsymbol{m-k}%
\right) $%
\end{tabular}%
\end{equation}%
with%
\begin{equation}
\begin{tabular}{lll}
$\dim \boldsymbol{l}_{\mu _{k}}$ & $=$ & $1+(k^{2}-1)+\left[ (m-k)^{2}-1%
\right] $ \\ 
$\dim \boldsymbol{n}$ & $=$ & $k\left( m-k\right) $%
\end{tabular}%
\end{equation}%
Regarding the decomposition of the fundamental representation $\boldsymbol{m}%
,$ it is given by the direct sum of representations of $sl(k)\oplus sl(m-k)$
namely 
\begin{equation}
\boldsymbol{k}_{1-\frac{k}{m}}\text{\qquad },\text{ \qquad }\left( 
\boldsymbol{m-k}\right) _{-\frac{k}{m}}
\end{equation}%
The lower label refers to the charges of $sl(1)$ generated by $\mathbf{\mu }%
_{k}.$ The values are constrained by the traceless property of $sl(m)$. From
the decomposition $\boldsymbol{m}=\boldsymbol{k}_{1-\frac{k}{m}}\oplus
\left( \boldsymbol{m-k}\right) _{-\frac{k}{m}},$ we learn two interesting
features \textrm{\cite{3C}}: \newline
$\left( \mathbf{i}\right) $ the $\mathbf{\mu }$ operator can be expressed in
terms of the orthogonal projectors $\Pi _{\boldsymbol{k}}$ and $\Pi _{%
\boldsymbol{m-k}}$ on the representations $\boldsymbol{k}_{1-\frac{k}{m}}$
and $\left( \boldsymbol{m-k}\right) _{-\frac{k}{m}}$ as follows 
\begin{equation}
\begin{tabular}{lllll}
$\mathbf{\mu }$ & $\mathbf{=}$ & $\frac{m-k}{m}\Pi _{\boldsymbol{k}}-\frac{k%
}{m}\Pi _{\boldsymbol{m-k}}$ & $\equiv $ & $q_{1}\Pi _{\boldsymbol{R}%
_{1}}+q_{2}\Pi _{\boldsymbol{R}_{2}}$ \\ 
$I_{id}$ & $\mathbf{=}$ & $\Pi _{\boldsymbol{k}}+\Pi _{\boldsymbol{m-k}}$ & 
& 
\end{tabular}%
\end{equation}%
where $\boldsymbol{R}_{i}$\ stand for the representations $\boldsymbol{k}_{1-%
\frac{k}{m}}$ of $sl(k)$ and $\left( \boldsymbol{m-k}\right) _{-\frac{k}{m}}$
of $sl(m-k);$ and where $\Pi _{\boldsymbol{R}_{1}}$ and $\Pi _{\boldsymbol{R}%
_{2}}$ are projectors satisfying $\Pi _{\boldsymbol{R}_{i}}.\Pi _{%
\boldsymbol{R}_{j}}=\delta _{ij}\Pi _{\boldsymbol{R}_{i}}$. The coefficients 
$q_{i}$ are given by 
\begin{equation}
q_{i}=Tr\left( \mathbf{\mu }\Pi _{\boldsymbol{R}_{i}}\right) .
\end{equation}%
$\left( \mathbf{ii}\right) $ the operators $z^{\mathbf{\mu }}$, $X$ and $Y$
involved in the calculation of $e^{X}z^{\mathbf{\mu }}e^{Y}$ can be also
expressed in terms of $\Pi _{\boldsymbol{k}}$ and $\Pi _{\boldsymbol{m-k}}$.
For example, we have%
\begin{equation}
z^{\mathbf{\mu }}=z^{\frac{m-k}{m}}\Pi _{\boldsymbol{k}}+z^{-\frac{k}{m}}\Pi
_{\boldsymbol{m-k}}
\end{equation}%
By using $I_{id}=\Pi _{\boldsymbol{k}}+\Pi _{\boldsymbol{m-k}}$ and $\Pi _{%
\boldsymbol{R}_{i}}.\Pi _{\boldsymbol{R}_{j}}=\delta _{ij}\Pi _{\boldsymbol{R%
}_{i}},$ we can split the matrix operators $X$ and $Y$ into four blocks like%
\begin{equation}
X_{ij}=\Pi _{\boldsymbol{R}_{i}}X\Pi _{\boldsymbol{R}_{j}}\qquad ,\qquad
Y_{ij}=\Pi _{\boldsymbol{R}_{i}}Y\Pi _{\boldsymbol{R}_{j}}
\end{equation}%
Substituting into $L=e^{X}z^{\mathbf{\mu }}e^{Y}$, we obtain the generic
expression of the L-matrix namely%
\begin{equation}
L_{ij}=\Pi _{\boldsymbol{R}_{i}}e^{X}\left( z^{\frac{m-k}{m}}\Pi _{%
\boldsymbol{k}}+z^{-\frac{k}{m}}\Pi _{\boldsymbol{m-k}}\right) e^{Y}\Pi _{%
\boldsymbol{R}_{j}}  \label{lz}
\end{equation}%
with $\boldsymbol{R}_{1}=\boldsymbol{k}_{1-\frac{k}{m}}$ and $\boldsymbol{R}%
_{2}=\left( \boldsymbol{m-k}\right) _{-\frac{k}{m}}.$ Moreover, using the
property $X^{2}=Y^{2}=0,$ we obtain after some straightforward calculations,
the following%
\begin{equation}
L=\left( 
\begin{array}{cc}
\Pi _{1}\left( z^{\frac{m-k}{m}}+z^{-\frac{k}{m}}XY\right) \Pi _{1} & z^{-%
\frac{k}{m}}\Pi _{1}X\Pi _{2} \\ 
z^{-\frac{k}{m}}\Pi _{2}Y\Pi _{1} & z^{-\frac{k}{m}}\Pi _{2}%
\end{array}%
\right)  \label{slmn}
\end{equation}%
with $XY=b^{ai}c_{ai}\Pi _{1}$ where $b^{ai}$ and $c_{ai}$\ are Darboux
coordinates of the phase space underlying the RLL equation of integrability 
\textrm{\cite{1D}},%
\begin{equation}
R_{rs}^{ik}\left( z-w\right) L_{j}^{r}\left( z\right) L_{l}^{s}\left(
w\right) =L_{r}^{i}\left( w\right) L_{s}^{k}\left( z\right)
R_{jl}^{rs}\left( z-w\right)  \label{511}
\end{equation}%
where $R_{rs}^{ik}\left( z\right) $\ is the usual R-matrix of Yang-Baxter
equation.\  \  \ 

\begin{figure}[tbph]
\begin{center}
\includegraphics[width=16cm]{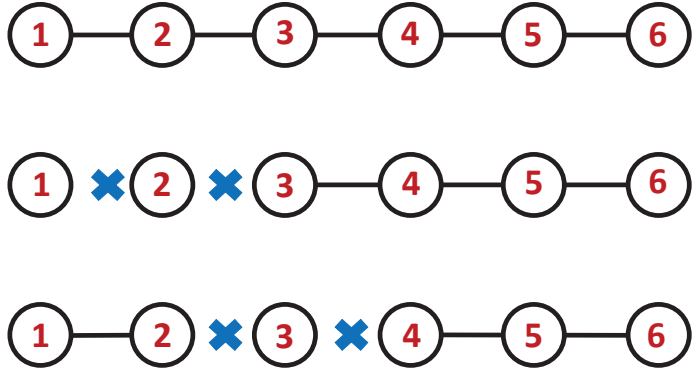}
\end{center}
\par
\vspace{-0.5cm}
\caption{Two Levi-decompositions of $sl_{7}$ in the D-language. The first
decomposition is given by cutting the second node corresponding to $\protect%
\alpha _{2}$. The second decomposition is given by cutting the third node.
Generally speaking, there are six ways to cut a node from $sl_{7}$.}
\label{fig13}
\end{figure}

$\bullet $\emph{\ Levi-decomposition\ in D- language}\newline
Here, we want to show that as far as the $sl\left( m\right) $ is concerned,
the Levi-decomposition with respect to $\mu _{k}$ is equivalent to cutting
the node labeled by the simple root $\alpha _{k}$ in the Dynkin diagram. We
state this correspondence as follows%
\begin{equation}
\begin{tabular}{lllllll}
$sl(m)$ \  \  \  & $:$ & $sl(m-k)$ & $\oplus $ & $sl(1)$ & $\oplus $ & $sl(k)$
\\ 
\ $\  \downarrow $ &  & \multicolumn{5}{l}{\  \  \  \  \  \  \  \  \  \ $\  \  \  \  \  \
\downarrow $} \\ 
$D_{m-1}$ & $:$ & $D_{m-k-1}$ & $\oplus $ & $D_{1}$ & $\oplus $ & $D_{k-1}$%
\end{tabular}
\label{57}
\end{equation}%
where the notation $D_{p-1}$ refers to the Dynkin diagram of $sl\left(
p\right) $ and where we have hidden the nilpotent sub-algebras $\boldsymbol{n%
}_{\pm }$; \textrm{see also the Figure }\textbf{\ref{fig13}}. Notice that $%
\boldsymbol{n}_{\pm }$ together with $sl(1)$ give the $sl\left( 2\right) $
associated with $D_{1}$. The correspondence (\ref{57}) is interesting for
two reasons.

\begin{description}
\item[$\left( \mathbf{1}\right) $] It indicates that the Levi- splitting (%
\ref{slm}) used in the oscillator realisation of the L-operator can be
nicely described by using the language of Dynkin diagram of $sl(m)$ (for
short D-language).

\item[$\left( \mathbf{2}\right) $] It offers a guiding algorithm to extend
the Levi-decomposition to Lie superalgebras, which to our knowledge, is
still an open problem \textrm{\cite{1KC,1KD}}. Because of this lack, we will
use this algorithm later on when we study the extension of the
Levi-decomposition to $sl(m|n).$ In this regard, it is interesting to notice
that in the context of supersymmetric gauge theory, it has been known that
the L-operator has an interpretation as a surface operator \cite{1R}$.$\ It
has been also known that the Levi-decomposition is relevant to the surface
operator; see, e.g. \cite{2R}$.$\ 
\end{description}

\  \  \newline
Recall that the Dynkin diagram of $sl(m)$ is given by a linear chain with $%
\left( m-1\right) $ nodes labeled by the $\left( m-1\right) $ simple roots $%
\alpha _{i}$; see the first graph of the Figure \textbf{\ref{fig13}}
describing $sl_{7}.$ \textrm{The nodes'} links are given by the intersection
matrix 
\begin{equation}
K_{ij}=\alpha _{i}.\alpha _{j}
\end{equation}%
which is just the Cartan matrix of $sl(m)$. In this graphic description, the
three terms $sl(m-k)\oplus sl(1)\oplus sl(k)$ making $\boldsymbol{l}_{\mu
_{k}}$ are nicely described in terms of pieces of the Dynkin diagram as
exhibited by (\ref{57}). The three pieces $D_{m-k-1}\oplus D_{1}\oplus
D_{k-1}$ are generated by cutting the node $\alpha _{k}$ with label as $%
2\leq k\leq m-2$; see the Figure \textbf{\ref{fig13}\ }\ for illustration. 

\  \  \newline
The case $k=1$ ( resp. $k=m-1$) concerns the cutting of boundary node $%
\alpha _{1}$ ( resp. $\alpha _{m-1}$): In this situation, we have the
following correspondence%
\begin{equation}
\begin{tabular}{lllll}
$sl(m)$ \  \  \  & $:$ & $\ sl(1)$ & $\oplus $ & $sl(m-1)$ \\ 
\ $\  \downarrow $ &  & \multicolumn{3}{l}{\  \  \  \  \  \  \ $\  \  \  \  \downarrow $%
} \\ 
$D_{m-1}$ & $:$ & \  \  \ $D_{1}$ & $\oplus $ & $D_{m-2}$%
\end{tabular}
\label{sm}
\end{equation}

\subsubsection{Extension to $SL\left( m|n\right) $ symmetry}

To extend the construction of the L-operator of $sl\left( m\right) $ to the
Lie superalgebra $sl(m|n),$ we use the relationship between $sl\left(
m\right) $ and $sl(m|n)$ algebras%
\begin{equation}
sl\left( m\right) \subset sl(m|n)_{\bar{0}}\subset sl(m|n)  \label{rs}
\end{equation}%
This embedding property holds also for the representations 
\begin{equation}
\mathbf{rep}_{sl_{m}}\subset \mathbf{rep}_{sl(m|n)_{\bar{0}}}\subset \mathbf{%
rep}_{sl(m|n)}
\end{equation}

$A)$ \emph{Field super-action}\newline
Here, the $sl(m)$ symmetry of the CS field action (\ref{ac}) is promoted to
the super $SL(m|n)$ and the usual trace (tr) is promoted to the super-trace
(str) \textrm{\cite{SG,SG1}}; that is:%
\begin{equation}
\begin{tabular}{lllllll}
$sl(m)$ & $\rightarrow $ & $sl(m|n)$ & \qquad ;\qquad & trace & $\rightarrow 
$ & super-trace%
\end{tabular}%
\end{equation}%
So, the generalised CS field action on $\mathbb{R}^{2}\times \mathbb{CP}^{1}$
invariant under $SL(m|n)$ is given by the supertrace of a Lagrangian like $%
\int dz\wedge \left[ str\mathcal{L}_{CS}\right] .$ This generalised action
reads as follows%
\begin{equation}
S_{4dCS}^{sl(m|n)}=\int_{\mathbb{R}^{2}\times \mathbb{CP}^{1}}dz\wedge str%
\left[ \mathcal{A}\wedge d\mathcal{A}+\frac{2}{3}\mathcal{A}\wedge \mathcal{A%
}\wedge \mathcal{A}\right]
\end{equation}%
In this generalisation, the Chern-Simons gauge field $\mathcal{A}$ is valued
in the Lie superalgebra $sl(m|n)$. It has the following expansion%
\begin{equation}
\mathcal{A}=\sum_{\text{\textsc{ab}}}A^{\text{\textsc{ab}}}\mathcal{E}_{%
\text{\textsc{ab}}}  \label{de}
\end{equation}%
where $\mathcal{E}_{\text{\textsc{ab}}}$ are the graded generators of $%
sl(m|n)$ obeying the graded commutation relations (\ref{gc}). In terms of
these super-generators, the super-trace of the Chern-Simons 3-form 
\begin{equation}
\tilde{\Omega}_{3}=\mathcal{A}\wedge d\mathcal{A}+\frac{2}{3}\mathcal{A}%
\wedge \mathcal{A}\wedge \mathcal{A}
\end{equation}%
is given by%
\begin{equation}
str\left( \tilde{\Omega}_{3}\right) =g_{\text{\textsc{abcd}}}A^{\text{%
\textsc{ab}}}dA^{\text{\textsc{cd}}}+\frac{2}{3}f_{\text{\textsc{abcdef}}}A^{%
\text{\textsc{ab}}}A^{\text{\textsc{cd}}}A^{\text{\textsc{ef}}}
\end{equation}%
where we have set%
\begin{equation}
g_{\text{\textsc{abcd}}}=str\left( \mathcal{E}_{\text{\textsc{ab}}}\mathcal{E%
}_{\text{\textsc{cd}}}\right) \qquad ,\qquad f_{\text{\textsc{abcdef}}%
}=str\left( \mathcal{E}_{\text{\textsc{ab}}}\mathcal{E}_{\text{\textsc{cd}}}%
\mathcal{E}_{\text{\textsc{ef}}}\right)
\end{equation}%
By using the notation (\ref{ep}), we can rewrite the development (\ref{de})
like, 
\begin{equation}
\mathcal{A}=A^{\text{ab}}\mathcal{E}_{\text{ab}}+A^{\prime \text{ij}}%
\mathcal{E}_{\text{ij}}^{\prime }+\tilde{A}^{\text{ai}}\mathcal{\tilde{E}}_{%
\text{ai}}+\tilde{A}^{\prime \text{ia}}\mathcal{\tilde{E}}_{\text{ia}%
}^{\prime }
\end{equation}%
where the 1-form potentials $A^{\text{ab}}$ and $A^{\prime \text{ij}}$ have
an even degree while the $\tilde{A}^{\text{ai}}$ and $\tilde{A}^{\prime 
\text{ia}}$ have an odd degree. The diagonal $A^{\text{ab}}$ and $A^{\prime 
\text{ij}}$ are respectively in the adjoints of $sl(m)$ and $sl(n).$ The
off-diagonal blocks $\tilde{A}^{\text{ai}}$ and $\tilde{A}^{\prime \text{ia}%
} $ are fermionic fields contained in the bi-fundamental $sl(m)\oplus sl(n)$%
. In the super-matrix representation, they are as follows%
\begin{equation}
\mathcal{A}=\left( 
\begin{array}{cc}
A^{\text{ab}}\mathcal{E}_{\text{ab}} & \tilde{A}^{\text{ai}}\mathcal{\tilde{E%
}}_{\text{ai}} \\ 
\tilde{A}^{\prime \text{ia}}\mathcal{\tilde{E}}_{\text{ia}}^{\prime } & 
A^{\prime \text{ij}}\mathcal{\tilde{E}}_{\text{ij}}%
\end{array}%
\right)
\end{equation}%
The 1-form gauge field $A^{\text{\textsc{ab}}}=str\left( \mathcal{E}^{\text{%
\textsc{ab}}}\mathcal{A}\right) $ splits explicitly like%
\begin{equation}
\begin{tabular}{lllll}
$A^{\text{ab}}$ & $=tr\left( \mathcal{E}^{\text{ab}}\mathcal{A}\right) $ & ,
& $A^{\prime \text{ij}}$ & $=-tr\left( \mathcal{E}^{\prime \text{ij}}%
\mathcal{A}\right) $ \\ 
$\tilde{A}^{\text{ai}}$ & $=tr\left( \mathcal{\tilde{E}}^{\text{ai}}\mathcal{%
A}\right) $ & , & $\tilde{A}^{\prime \text{ia}}$ & $=-tr\left( \mathcal{%
\tilde{E}}^{\prime \text{ia}}\mathcal{A}\right) $%
\end{tabular}%
\end{equation}%
In the distinguished basis of $gl(m|n)$ with even part $gl(m)\oplus gl(n)$,
the $A^{\text{ab}}$ is the gauge field of $gl(m)$ valued in the adjoint $%
\left( \boldsymbol{m,\bar{m}}\right) $\ and the $A^{\prime \text{ij}}$ is
the gauge field of $gl(n)$ valued in $\left( \boldsymbol{n,\bar{n}}\right) $%
. The fields $\tilde{A}^{\text{ai}}$ and $\tilde{A}^{\prime \text{ia}}$
describe topological gauge matter \textrm{\cite{SG2,SG3}} valued in the
bi-fundamentals $\left( \boldsymbol{m,\bar{n}}\right) $ and $\left( 
\boldsymbol{\bar{m},n}\right) $.\  \  \  \ 

\begin{figure}[tbph]
\begin{center}
\includegraphics[width=10cm]{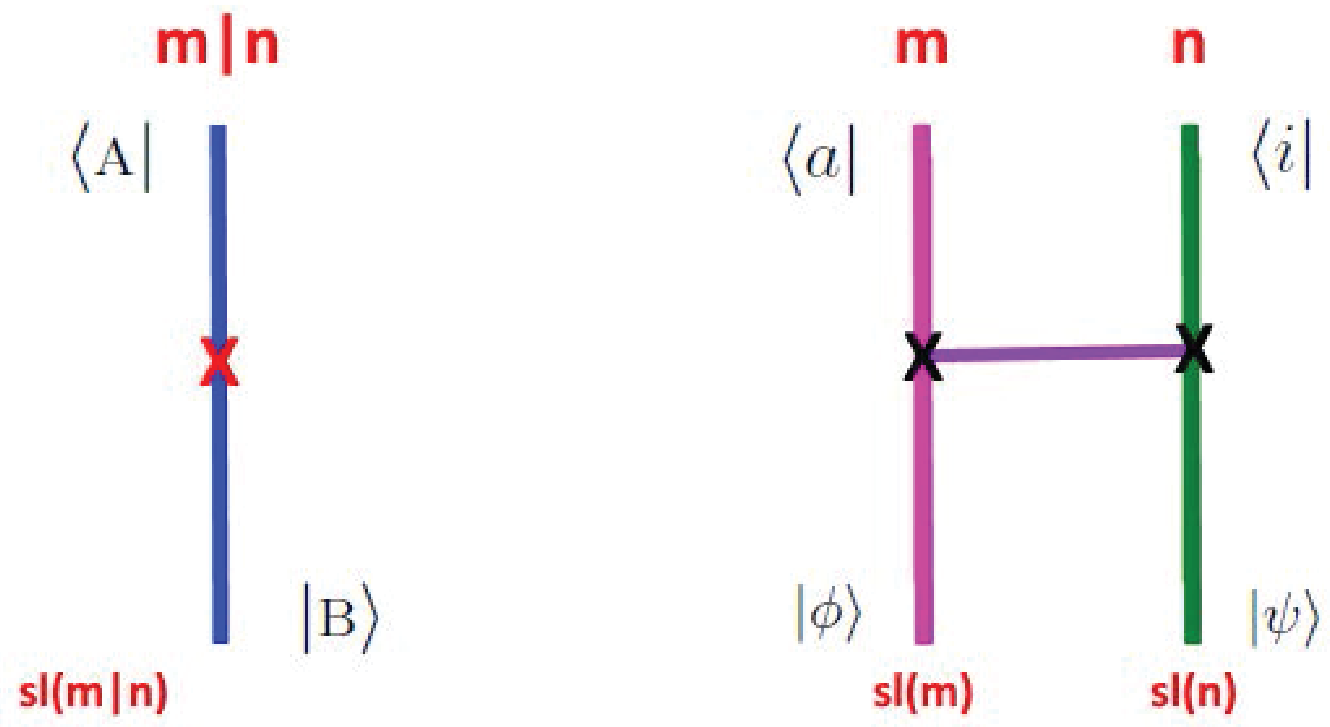}
\end{center}
\par
\vspace{-0.5cm}
\caption{A graphic representation of the Wilson superline $W_{\protect \xi %
_{z}}^{\boldsymbol{m|n}}$ charged under $sl(m|n)$. On \textrm{the }left, the
superline \textrm{is represented} in the topological plane. On \textrm{the}
right, the superline is\textrm{\ described }in terms of graded Wilson lines
related by a bridge \textrm{connecting} the two gradations $\mathbf{\bar{0}}$
and $\mathbf{\bar{1}}$. The states $\left \vert \text{\textsc{a}}%
\right
\rangle $ and $\left \vert \text{\textsc{b}}\right \rangle $ \textrm{%
can be bosonic and fermionic}.\textrm{\ The states }$\left \vert
a\right
\rangle $\textrm{\ are bosonic-like and} $\left \vert
i\right
\rangle $ are fermionic-like. The states $\left \vert \protect \phi %
\right
\rangle $ and $\left \vert \protect \psi \right \rangle $ can be
either bosonic or fermionic. }
\label{sl}
\end{figure}

\begin{figure}[tbph]
\begin{center}
\includegraphics[width=16cm]{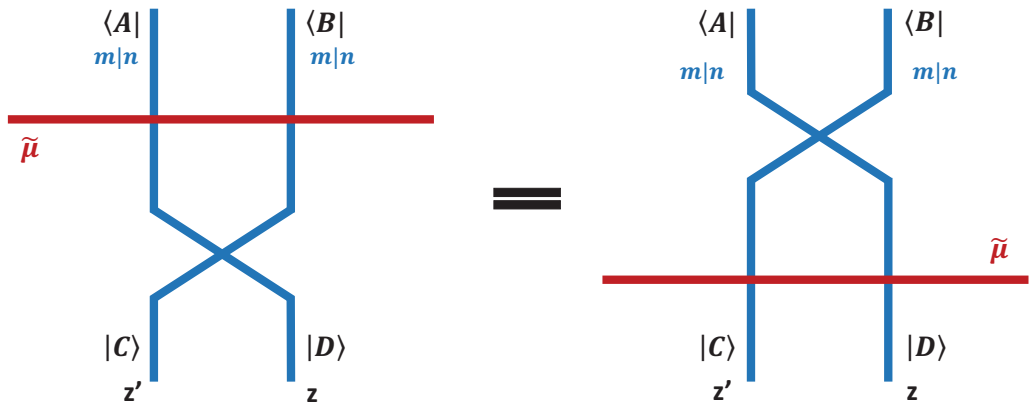}
\end{center}
\par
\vspace{-0.5cm}
\caption{A graphic representation of the RLL\ equation involving three
super-lines. Two "blue" interacting Wilson super-lines crossed by a "red "
't Hooft super-line. This super-equation contains the usual RLL\ equation of
bosonic like symmetries such as sl(m).}
\label{fig14}
\end{figure}

$B)$\emph{\ super-line operators}\newline
To extend the bosonic-like Wilson line W$_{\xi _{z}}^{\boldsymbol{m}}$ of
the CS gauge theory to the super- group $SL(m|n)$, we use the representation
language to think about this super-line as follows%
\begin{equation}
\begin{tabular}{l|l|l}
gauge symmetry & fund representation & \  \ Wilson line\  \  \\ \hline
$\  \ SL(m)$ & $\  \  \  \boldsymbol{R}=\boldsymbol{m}$ & $\  \  \  \text{W}_{\xi
_{z}}^{\boldsymbol{m}}$ \\ \hline
$\  \ SL(m|n)$ & $\  \  \  \boldsymbol{R}=\boldsymbol{m|n}$ & \  \  \  \ W$_{\xi
_{z}}^{\boldsymbol{m|n}}$%
\end{tabular}%
\end{equation}%
where the fundamental $\boldsymbol{m}$ of $sl\left( m\right) $ is promoted
to the fundamental $\boldsymbol{m|n}$ of $sl(m|n)$. In this picture, W$_{\xi
_{z}}^{\boldsymbol{m|n}}$ can be imagined as follows%
\begin{equation}
W_{\xi _{z}}^{\boldsymbol{m|n}}=str_{\boldsymbol{m|n}}\left[ P\exp \left(
\doint \nolimits_{\xi _{z}}\mathcal{A}\right) \right]  \label{524}
\end{equation}%
with $\mathcal{A}=A^{\text{\textsc{ab}}}\mathcal{E}_{\text{\textsc{ab}}}$ and%
\begin{equation}
str\left( 
\begin{array}{cc}
A & B \\ 
C & D%
\end{array}%
\right) =trA-trD
\end{equation}%
A diagrammatic representation of the Wilson superline W$_{\mathrm{\xi }%
_{z}}^{\boldsymbol{m|n}}$ charged under $SL(m|n)$ is given b the Figure 
\textbf{\ref{sl}}. 
Regarding the magnetically charged 't Hooft super-line, we think of it below
as tH$_{\mathrm{\gamma }_{\mathrm{0}}}^{\mathbf{\mu }}$having the same
extrinsic shape as in the bosonic CS theory, but with the intrinsic bosonic $%
SL\left( m\right) $ promoted to $SL\left( m|n\right) $. This definition
follows from eq(\ref{211}) of appendix A by extending the $g_{I}\left(
z\right) $ and $g_{II}\left( z\right) $ to supergroup elements $\tilde{g}%
_{I}\left( z\right) $ and $\tilde{g}_{II}\left( z\right) $ in $SL(m|n)$. In
other words, eq(\ref{211}) generalises as%
\begin{equation}
L_{sl_{\boldsymbol{m|n}}}=\tilde{g}_{I}.z^{\mathbf{\tilde{\mu}}}.\tilde{g}%
_{II}  \label{526}
\end{equation}%
with $\tilde{g}_{I}\left( z\right) $ and $\tilde{g}_{II}\left( z\right) $
belonging to $SL(m|n)$; and $\mathbb{\mu }$ generating the charge group $%
SL\left( 1\right) $ in the even part $SL(m|n)_{0}.$ Notice that using $%
sl(m|n)_{0}=s\left[ gl(m)\oplus gl(n)\right] $, the adjoint form of $\mathbf{%
\tilde{\mu}}$ has in general two contributions like%
\begin{equation}
\mathbf{\tilde{\mu}}=\mathbb{\mu }_{sl_{m}}+\mathbb{\mu }_{sl_{n}}\qquad
,\qquad \mathbb{\mu }_{sl_{m}}=\mathbf{\tilde{\mu}.}\Pi _{sl_{m}}\qquad
,\qquad \mathbb{\mu }_{sl_{n}}=\mathbf{\tilde{\mu}.}\Pi _{sl_{n}}
\end{equation}%
\textrm{where} $str(\mathbf{\tilde{\mu}})=0$ and $\Pi _{sl_{m}}\ $and $\Pi
_{sl_{n}}$ are orthogonal projectors on $sl(m)$ and $sl(n)$ respectively;
i.e: $\Pi _{sl_{m}}.\Pi _{sl_{n}}=0$. For an illustration; see for instance
eq(\ref{620}) given below. Notice that the super-traceless condition \textrm{%
reads} in terms of the usual trace like $tr(\mathbb{\mu }_{sl_{m}})-tr(%
\mathbb{\mu }_{sl_{n}})=0.$ By projecting $sl(m|n)_{0}$ down to $sl(m)$
disregarding the $sl(n)$ part, the super-trace condition $str(\mathbf{\tilde{%
\mu}})=0$ reduces to the familiar $tr(\mathbb{\mu }_{sl_{m}})=0.$ \newline
Except for the intrinsic properties we have described above, the extrinsic
features of the super-lines are quite similar to $sl\left( m\right) .$ In
particular, the positions of the two crossing super-lines in the topological
plane $\mathbb{R}^{2}$ and the holomorphic $\mathcal{C}$ are as in the
bosonic CS theory with $SL\left( m\right) $ gauge symmetry; see the Figure 
\textbf{\ref{fig1}}. In this regard, we expect that the extension of the YBE
and RLL equations (\ref{511}) to supergroups may be also derived from the
crossing of the super-lines. From the side of the superspin chains, these
algebraic equations were studied in literature; see for instance to \cite%
{YB,YB1,YB2,YB3,YB4,YB5,YB6} and references therein. From the gauge theory
side, the\ super-YBE\ and the super-RLL equations have not been yet
explored. In our formalism, the super-RLL equations are given by the diagram
of the Figure \textbf{\ref{fig14}.} 

\subsection{Decomposing super-Ds with one fermionic node}

Here, we give partial results regarding the extension of the decomposition (%
\ref{57}) concerning $sl(m)$ to the case of the Lie superalgebra $sl(m|n).$
We study three kinds of decompositions of DSDs of $sl(m|n).$ These
decompositions are labeled by an integer p constrained as $1\leq p\leq m$
and can be imagined in terms of the breaking pattern%
\begin{equation}
sl(m|n)\rightarrow sl(p|0)\oplus sl(m-p|n)  \label{pa}
\end{equation}%
The three kinds of decomposition patterns concern the following intervals of
the label p:

$\bullet $ the particular case $p=1$.

$\bullet $ the generic case $2\leq p\leq m-1.$

$\bullet $ the special case $p=m$.\newline
This discrimination for the values of p is for convenience; they can be
described in a compact way. \textrm{Notice that the above decomposition can
be also applied for the pattern} 
\begin{equation}
sl(m|n)\rightarrow sl(0|q)\oplus sl(m|n-q)
\end{equation}%
with $1\leq q\leq n$. \textrm{We omit the details of this case; the results
can be read from (\ref{pa}).}

\subsubsection{Cutting the left node in the super\textbf{\ }$\hat{D}_{m+n-1}$%
}

The decomposition of the DSDs denoted\footnote{%
\ The rank of $sl(m|n)$ is $m+n-1;$ its DSDs have $m+n-1$ nodes. To
distinguish these super-diagrams from the bosonic $D_{m-1}$ and $D_{n-1}$
ones of $sl\left( m\right) $ and $sl\left( n\right) $, we denote them as $%
\hat{D}_{m+n-1}$.} below like\textbf{\ }$\hat{D}_{m+n-1}$ generalises the
correspondence (\ref{sm}). It is illustrated on the Figure \textbf{\ref%
{fig12} }describing two examples of typical decompositions of $\hat{D}_{6}$: 
$\left( i\right) $ a bosonic decomposition corresponding to cutting the
second node labeled by the bosonic root $\alpha _{2}.$ $\left( ii\right) $ a
fermionic decomposition corresponding to cutting the fourth node labeled by
the fermionic root $\alpha _{4}.$%
\begin{figure}[tbph]
\begin{center}
\includegraphics[width=16cm]{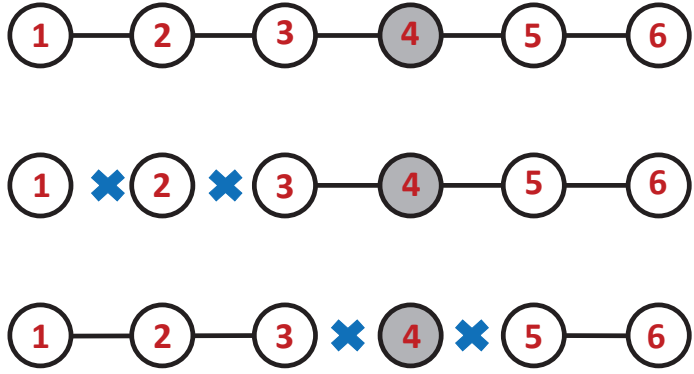}
\end{center}
\par
\vspace{-0.5cm}
\caption{Two decompositions of a DSD of the distinguished Lie superalgebra A$%
_{5|1}$. The first decomposition is given by cutting the second node
corresponding to the simple root $\protect \alpha _{2}$. The other
decomposition is given by cutting the fermionic (fourth) node.}
\label{fig12}
\end{figure}

\  \  \newline
By cutting the left node of the DSD, that is the node labeled by $\alpha
_{1} $ with positive length $\alpha _{1}^{2}=2,$ the super- Dynkin diagram $%
\hat{D}_{m+n-1}$ breaks into two pieces $D_{1}$ and $\hat{D}_{m+n-2}$ as
given by the following correspondence%
\begin{equation}
\begin{tabular}{lllll}
$sl(m|n)$ \  \  \  & $:$ & $\ sl(1)$ & $\oplus $ & $sl(m-1|n)$ \\ 
\ $\  \  \downarrow $ &  & \multicolumn{3}{l}{\  \  \  \  \  \  \ $\  \  \  \  \
\downarrow $} \\ 
$\hat{D}_{m+n-1}$ & $:$ & \  \  \ $D_{1}$ & $\oplus $ & $\hat{D}_{m+n-2}$%
\end{tabular}
\label{ab}
\end{equation}%
Notice that by setting $n=0,$ we recover the bosonic $sl(m)$ case (\ref{sm}%
). \textrm{Here,} the $\hat{D}_{m+n-1}$ refers to the \emph{distinguished}
DSDs of $sl(m|n)$ having $m+n-1$ nodes; one of them is a fermionic; it is
labeled by the odd simple root $\tilde{\alpha}_{m}.$ All the other $m+n-2$
nodes are bosonic simple roots as shown \textrm{in} the following table,%
\begin{equation}
\begin{tabular}{|l|l|l|l|l|}
\hline
{\small simple root } & $\  \  \alpha _{1}$ & $\left \{ \alpha _{a}\right \} _{%
{\small 1<a<m-1}}$ & $\tilde{\alpha}=\tilde{\alpha}_{m}$ & $\left \{ \alpha
_{i}^{\prime }\right \} _{{\small 1\leq i<n}}$ \\ \hline
{\small \  \  \  \ degree} & {\small even} & $\ ${\small even} & {\small \  \ odd%
} & $\ ${\small even\  \  \ } \\ \hline
{\small \  \ length }$\alpha ^{2}$ & ${\small \  \  \ 2}$ & ${\small \  \  \ 2}$
& ${\small \  \  \  \  \ 0}$ & ${\small \  \ -2}$ \\ \hline
\end{tabular}%
\end{equation}%
By cutting the node labeled by $\alpha _{1}$, the root system $\Phi
_{sl(m|n)}$ splits into two subsets. The first concerns the root sub-system
of $sl(m-1|n)$ containing $\left( m+n-2\right) \left( m+n-1\right) $ roots
dispatched as follows%
\begin{equation}
\begin{tabular}{|l|l|l|l|}
\hline
{\small root} & $\  \  \  \  \alpha _{ab}$ & $\  \  \  \  \  \  \alpha _{ij}^{\prime }$
& $\  \  \pm \tilde{\alpha}_{ai}$ \\ \hline
{\small value} & $\pm {\small \{ \varepsilon }_{a}{\small -\varepsilon }_{b}%
{\small \}}_{{\small 1\leq a<b<m}}$ & $\pm {\small \{ \delta }_{i}{\small %
-\delta }_{j}{\small \}}_{{\small 1\leq i<j\leq n}}$ & $\pm \{ \varepsilon
_{a}-\delta _{i}\}_{{\small 1\leq a,b<m}}^{{\small 1\leq i\leq n}}${\small \
\  \  \ } \\ \hline
{\small number} & $\left( m-1\right) \left( m-2\right) $ \  & $\  \ n\left(
n-1\right) $ \  \  & $\  \  \  \  \  \ 2\left( m-1\right) n$ \\ \hline
{\small degree} & {\small \  \  \  \ even} & {\small \  \  \  \ even} & {\small \
\  \  \  \  \  \  \  \  \ odd \  \ } \\ \hline
\end{tabular}%
\end{equation}%
This root sub-system $\left \{ \beta =\sum n_{i}\alpha _{i}\right \} $ is a
subset of $\Phi _{sl_{m|n}};$ it has no dependence in $\alpha _{1}$ as it
has been removed. This property can be stated like%
\begin{equation}
\frac{\partial \beta }{\partial \alpha _{1}}=0\quad ,\quad \beta \in \Phi
_{sl_{m|n}}
\end{equation}%
The second subset contains $2\left( m-1\right) +2n$ roots $\beta =\sum
n_{i}\alpha _{i};$ it is a subset of $\Phi _{sl_{m|n}}$ with $\partial \beta
/\partial \alpha _{1}\neq 0$. These roots are distributed as follows 
\begin{equation}
\begin{tabular}{|l|l|l|}
\hline
{\small root} & $\  \left \{ \pm \alpha _{1b}\right \} $ \  \  \  \  & $\  \  \  \
\  \  \pm \tilde{\alpha}_{1i}$ \\ \hline
{\small value} & ${\small \  \pm \{ \varepsilon }_{1}{\small -\varepsilon }%
_{b}{\small \}}_{{\small 1<b\leq m}}$ & $\pm {\small \{ \varepsilon }_{1}%
{\small -\delta }_{i}{\small \}}_{{\small 1\leq i\leq n}}${\small \  \  \  \ }
\\ \hline
{\small number} & $\  \ 2\left( m-1\right) $ \  & $\  \  \  \  \  \  \ 2n$ \\ \hline
{\small degree} & {\small \  \  \  \  \  \ even} & {\small \  \  \  \  \  \  \ odd \  \ }
\\ \hline
\end{tabular}
\label{nn}
\end{equation}%
The decomposition (\ref{ab}) can be checked by calculating the dimensions
and the ranks of \textrm{the} $sl(m|n)$ pieces resulting from the breaking 
\begin{equation}
sl(m|n)=\boldsymbol{l}_{1}\oplus \boldsymbol{N}_{1}^{+}\oplus \boldsymbol{N}%
_{1}^{-}
\end{equation}%
where%
\begin{equation}
\begin{tabular}{lll}
$\  \  \  \boldsymbol{l}_{1}$ & $=$ & $s\left[ gl(1|0)\oplus gl(m-1|n)\right] $
\\ 
$\  \  \  \boldsymbol{l}_{1}$ & $=$ & $\left( \boldsymbol{l}_{1}\right) _{\bar{0%
}}\oplus \left( \boldsymbol{l}_{1}\right) _{\bar{1}}$%
\end{tabular}%
\end{equation}%
and where the nilpotent $\boldsymbol{N}_{1}^{\pm }$ are in the
bifundamentals of $gl(1|0)\oplus gl(m-1|n)$. From this splitting, we learn $%
\dim \left( \boldsymbol{l}_{1}\right) _{\bar{0}}=\left( m+n-1\right) ^{2}$
and $\dim \left( \boldsymbol{l}_{1}\right) _{\bar{1}}=2\left( m-1\right)
+2n. $ Recall that the Lie superalgebra $gl(m-1|n)$ has dimension $\left(
m+n-1\right) ^{2}$ and the super-traceless $sl(m-1|n)$ has dimension $\left(
m+n-1\right) ^{2}-1$; it decomposes like%
\begin{equation}
\begin{tabular}{lll}
$gl(m-1|n)$ & $=$ & $gl(m-1|n)_{\bar{0}}\oplus gl(m-1|n)_{\bar{1}}$ \\ 
$gl(m-1|n)_{\bar{0}}$ & $=$ & $gl(m-1)\oplus gl(n)$%
\end{tabular}%
\end{equation}%
such that the even part of $\boldsymbol{l}_{1}$ is given by the
super-traceless%
\begin{equation}
\left( \boldsymbol{l}_{1}\right) _{\bar{0}}=s\left[ gl(1)\oplus
gl(m-1)\oplus gl(n)\right]
\end{equation}%
Then, the nilpotent $\boldsymbol{N}_{1}^{\pm }$ are given by \textrm{the}
direct sum of representations $(\boldsymbol{r}_{x}^{gl_{{\small m-1}}},%
\boldsymbol{r}_{y}^{gl_{{\small n}}})$ of the subalgebra $\left( \boldsymbol{%
l}_{1}\right) _{\bar{0}}$ with $x,$ $y$ referring to the charge of $gl(1)$.
From eq(\ref{nn}), we learn 
\begin{equation}
\begin{tabular}{lll}
$\boldsymbol{N}_{1}^{+}$ & $=$ & $({\small [}\boldsymbol{m-1}{\small ]},%
\boldsymbol{\bar{1}})\oplus (\boldsymbol{1},\boldsymbol{\bar{n}})$ \\ 
$\boldsymbol{N}_{1}^{-}$ & $=$ & $(\overline{\boldsymbol{m-1}},\boldsymbol{1}%
_{0})\oplus (\boldsymbol{1},\boldsymbol{n})$%
\end{tabular}
\label{p1}
\end{equation}%
The $\left( m-1\right) +n$ generators $X_{\text{\textsc{a}}}=\left( X_{a},%
\mathcal{X}_{i}\right) $ of \textrm{the} nilpotent algebra $\boldsymbol{N}%
_{1}^{+}$ and the $\left( m-1\right) +n$ generators $Y_{\text{\textsc{a}}%
}=\left( Y_{a},\mathcal{Y}_{i}\right) $ of $\boldsymbol{N}_{1}^{-}$ are
realised using the kets $\left \vert \text{\textsc{a}}\right \rangle $ and
bra $\left \langle \text{\textsc{a}}\right \vert $ as follows%
\begin{equation}
\begin{tabular}{lllll|l}
\multicolumn{5}{l|}{\  \  \  \  \  \  \  \  \  \  \  \  \  \ {\small generators}} & 
{\small degree} \\ \hline
$X_{a}$ & $=$ & $\left \vert 1\right \rangle \left \langle 1+a\right \vert $
& , & $a=1,...,m-1$ & {\small even} \\ 
$Y_{a}$ & $=$ & $\left \vert 1+a\right \rangle \left \langle 1\right \vert $
& , & $a=1,...,m-1$ & {\small even} \\ 
$\mathcal{X}_{i}$ & $=$ & $\left \vert 1\right \rangle \left \langle
m+i\right \vert $ & , & $i=1,...,n$ & {\small odd} \\ 
$\mathcal{Y}_{i}$ & $=$ & $\left \vert m+i\right \rangle \left \langle
1\right \vert $ & , & $i=1,...,n$ & {\small odd} \\ \hline
\end{tabular}
\label{x1y}
\end{equation}%
They are nilpotent since \textrm{we have} $X_{a}X_{b}=\mathcal{X}_{i}%
\mathcal{X}_{j}=X_{a}\mathcal{X}_{i}=0$ and the same for the $Y$'s. These
properties are interesting for the calculation of the super- Lax operators.

\subsubsection{Cutting an internal node $\protect \alpha _{p}$ with $1<p<m-1$}

In this generic case, the Lie superalgebra $sl(m|n)$ decomposes like $%
\boldsymbol{l}_{p}\oplus \boldsymbol{N}_{p}^{+}\oplus \boldsymbol{N}_{p}^{-}$
with the sub-superalgebra $\boldsymbol{l}_{p}$ as%
\begin{equation}
\boldsymbol{l}_{p}=s\left[ gl(p)\oplus gl(1)\oplus gl(m-p|n)\right]
\end{equation}%
and the nilpotent $\boldsymbol{N}_{p}^{\pm }$ given by the bi-fundamentals
of $gl(p)\oplus gl(m-p|n)$ with $\pm 1$ charges under $gl(1).$ Being a
superalgebra, the $\boldsymbol{l}_{p}$ decomposes in turns like%
\begin{equation}
\begin{tabular}{lll}
$\  \  \boldsymbol{l}_{p}$ & $=$ & $\left( \boldsymbol{l}_{p}\right) _{\bar{0}%
}\oplus \left( \boldsymbol{l}_{p}\right) _{\bar{1}}$ \\ 
$\left( \boldsymbol{l}_{p}\right) _{\bar{0}}$ & $=$ & $gl(p)\oplus
gl(1)\oplus gl(m-p)\oplus gl(n)$ \\ 
$\left( \boldsymbol{l}_{p}\right) _{\bar{1}}$ & $=$ & $gl(m-p|n)_{\bar{1}}$%
\end{tabular}
\label{slp}
\end{equation}%
The decomposition of $sl(m|n)\subset gl(m|n)$ and the associated super-
diagram $\hat{D}_{m+n-1}$ generalise the correspondence (\ref{57}). It given
by%
\begin{equation}
\begin{tabular}{lllllll}
$sl(m|n)$ \  \  \  & $\rightarrow $ \  \  \  \  & $sl(p|0)$ & $\oplus $ & $sl(1)$
& $\oplus $ & $sl(m-p|n)$ \\ 
\ $\  \  \downarrow $ &  & \multicolumn{5}{l}{\  \  \  \  \  \  \  \  \  \ $\  \  \  \  \  \
\  \  \downarrow $} \\ 
$\hat{D}_{m+n-1}$ & $\rightarrow $ & $D_{p-1}$ & $\oplus $ & $D_{1}$ & $%
\oplus $ & $\hat{D}_{m+n-p-1}$%
\end{tabular}
\label{ssd}
\end{equation}%
where we have hidden the nilpotent $\boldsymbol{N}_{p}^{\pm }$. In this
generic $1<p<m,$ the simple roots of $sl(m|n)$ are dispatched as follows,%
\begin{equation}
\begin{tabular}{|l|l|l|l|l|l|}
\hline
{\small simple root } & $\left \{ \alpha _{a}\right \} _{{\small 1<a<p}}$ & $%
\alpha _{p}$ & $\left \{ \alpha _{a}\right \} _{p{\small <a<m}}$ & $\tilde{%
\alpha}=\tilde{\alpha}_{m}$ & $\left \{ \alpha _{i}^{\prime }\right \} _{%
{\small 1\leq i<n}}$ \\ \hline
\  \  \  \ {\small degree} & even & even & $\ $even & odd & $\ $even\  \  \  \\ 
\hline
\end{tabular}%
\end{equation}%
By cutting the p-th node of the DSD of $sl(m|n)$ labeled by the simple root $%
\alpha _{p}$, the super $\hat{D}_{m+n-1}$ breaks into three pieces like $%
D_{p-1}\oplus D_{1}\oplus \hat{D}_{m+n-p-1}$. Then, the root system $\Phi
_{sl_{m|n}}$ splits into three subsets as commented below.

$\bullet $ \emph{case }$\hat{D}_{m+n-p-1}$\newline
The first subset concerns the roots of $sl(m-p|n)$ containing ${\small %
(m+n-p)(m+n-p-1)}$ elements dispatched as follows%
\begin{equation}
\begin{tabular}{|l|l|l|l|}
\hline
{\small root} & $\  \  \  \  \  \pm \alpha _{ab}$ & $\  \  \pm \alpha _{ij}^{\prime
}$ & $\pm \tilde{\alpha}_{ai}$ \\ \hline
{\small value} & ${\small \pm \{ \varepsilon }_{a}{\small -\varepsilon }_{b}%
{\small \}}_{p{\small <a<b\leq m}}$ & ${\small \pm \{ \delta }_{i}{\small %
-\delta }_{j}{\small \}}_{{\small 1\leq i<j\leq n}}$ & $\pm {\small \{
\varepsilon }_{a}{\small -\delta }_{i}{\small \}}_{p{\small <a\leq m}}^{%
{\small 1\leq i\leq n}}${\small \  \  \  \ } \\ \hline
{\small number} & $\left( m-p\right) \left( m-p-1\right) $ \  & $\  \ n\left(
n-1\right) $ \  \  & $\ 2\left( m-p\right) n$ \\ \hline
{\small degree} & {\small \  \  \  \ even} & {\small \  \  \  \ even} & {\small \
\  \  \ odd \  \ } \\ \hline
\end{tabular}
\label{nb1}
\end{equation}

$\bullet $ \emph{case }$D_{p-1}$\newline
The second subset concerns the roots of $sl(p);$ it contains $p\left(
p-1\right) $ even roots given by%
\begin{equation}
\pm \alpha _{ab}=\pm \left \{ \varepsilon _{a}-\varepsilon _{b}\right \} _{%
{\small 1\leq a<b\leq p}}  \label{nb2}
\end{equation}

$\bullet $ \emph{case of bi-fundamentals}\newline
The third subset of roots regards the bi-fundamentals $\boldsymbol{N}%
_{p}^{\pm }$; it contains $2p\left( m-p\right) $ even roots and $2pn$ odd
ones as shown on the following table%
\begin{equation}
\begin{tabular}{|l|l|l|}
\hline
{\small root} & $\  \  \pm \alpha _{ab}$ \  \  \  \  & $\  \  \  \  \pm \tilde{\alpha}%
_{ai}$ \\ \hline
{\small value} & ${\small \pm \{ \varepsilon }_{a}{\small -\varepsilon }_{b}%
{\small \}}_{{\small 1\leq a\leq p}}^{{\small p<b\leq m}}$ & ${\small \pm \{
\varepsilon }_{a}{\small -\delta }_{i}{\small \}}_{{\small 1\leq a\leq p}}^{%
{\small 1\leq i\leq n}}$ \  \  \  \\ \hline
{\small number} & $\  \  \ 2p\left( m-p\right) $ \  & $\  \  \  \  \  \  \ 2pn$ \\ 
\hline
{\small degree} & {\small \  \  \  \  \  \ even} & {\small \  \  \  \  \  \  \ odd \  \ }
\\ \hline
\end{tabular}
\label{nb3}
\end{equation}%
Notice that by adding the numbers of the roots in (\ref{nb1}) and (\ref{nb2}%
) as well as (\ref{nb3}), we obtain the desired equality 
\begin{equation}
(m+n-p)(m+n-p-1)+p\left( p-1\right) +2p\left( m-p\right) +2pn=\left(
m+n-1\right) \left( m+n\right)
\end{equation}%
Notice also that the algebraic structure of the nilpotent $\boldsymbol{N}%
_{p}^{\pm }$ can be described by using the bosonic-like symmetry $\left( 
\boldsymbol{l}_{p}\right) _{\bar{0}}$ given by (\ref{slp}) namely $%
gl(p)\oplus gl(m-p)\oplus gl(n)$ where we have hidden $gl(1)$ as it is an
abelian charge operator. We have%
\begin{equation}
\begin{tabular}{lll}
$\boldsymbol{N}_{p}^{+}$ & $=$ & $\left( \boldsymbol{p},\overline{%
\boldsymbol{m-p}},\boldsymbol{1}\right) \oplus (\boldsymbol{p},\boldsymbol{1}%
,\boldsymbol{\bar{n}})$ \\ 
$\boldsymbol{N}_{p}^{-}$ & $=$ & $\left( \boldsymbol{\bar{p}},\boldsymbol{m-p%
},\boldsymbol{1}\right) \oplus (\boldsymbol{\bar{p}},\boldsymbol{1},%
\boldsymbol{n})$%
\end{tabular}
\label{p2}
\end{equation}%
The $p\left( m-p\right) +pn$ generators $X_{\text{\textsc{a}}}=\left( X_{a%
\dot{b}},\mathcal{X}_{ai}\right) $ of \textrm{the} nilpotent $\boldsymbol{N}%
_{p}^{+}$ and the $p\left( m-p\right) +pn$ generators $Y_{\text{\textsc{a}}%
}=\left( Y_{a\dot{b}},\mathcal{Y}_{ai}\right) $ of $\boldsymbol{N}_{1}^{-}$
are realised by using the super- kets $\left \vert \text{\textsc{a}}%
\right
\rangle $ and super- bra $\left \langle \text{\textsc{a}}%
\right
\vert $ as follows%
\begin{equation}
\begin{tabular}{lllllll|l}
\multicolumn{7}{l|}{\  \  \  \  \  \  \  \  \  \  \  \  \  \ {\small generators}} & 
{\small degree} \\ \hline
$X_{a\dot{b}}$ & $=$ & $\left \vert a\right \rangle \left \langle p+b\right
\vert $ & , & ${\small a=1,...,p}$ & {\small ,} & ${\small \dot{b}=1,...,m-p}
$ & {\small even} \\ 
$Y_{a\dot{b}}$ & $=$ & $\left \vert p+b\right \rangle \left \langle a\right
\vert $ & , & ${\small a=1,...,p}$ & {\small ,} & ${\small \dot{b}=1,...,m-p}
$ & {\small even} \\ 
$\mathcal{X}_{ai}$ & $=$ & $\left \vert a\right \rangle \left \langle
m+i\right \vert $ & , & ${\small a=1,...,p}$ & {\small ,} & ${\small %
i=1,...,n}$ & {\small odd} \\ 
$\mathcal{Y}_{ai}$ & $=$ & $\left \vert m+i\right \rangle \left \langle
a\right \vert $ & , & ${\small a=1,...,p}$ & {\small ,} & ${\small i=1,...,n}
$ & {\small odd} \\ \hline
\end{tabular}
\label{x2y}
\end{equation}%
They are nilpotent since $X_{a\dot{b}}X_{c\dot{d}}=\mathcal{X}_{ai}\mathcal{X%
}_{bj}=X_{a\dot{b}}\mathcal{X}_{ai}=0$ and the same for the Y's.

\subsubsection{Cutting the fermionic node $\protect \alpha _{m}$}

In this case, the Lie superalgebra $sl(m|n)$ decomposes like $\boldsymbol{l}%
_{m}\oplus \boldsymbol{N}_{m}^{+}\oplus \boldsymbol{N}_{m}^{-}$ with%
\begin{equation}
\begin{tabular}{lllll}
$\left( \boldsymbol{l}_{m}\right) _{\bar{0}}$ & $=$ & $sl(m|n)_{\bar{0}}$ & $%
=$ & $s\left[ gl(m)\oplus gl(n)\right] $ \\ 
& $\equiv $ & \multicolumn{3}{l}{$sl(m)\oplus sl(n)\oplus sl(1)$}%
\end{tabular}
\label{p3}
\end{equation}%
and \textrm{the} odd part $\left( \boldsymbol{l}_{m}\right) _{\bar{1}}$
given by%
\begin{equation}
\boldsymbol{N}_{m}^{+}=sl(m|n)_{+1},\qquad \boldsymbol{N}%
_{m}^{-}=sl(m|n)_{-1}
\end{equation}%
The $\boldsymbol{N}_{m}^{\pm }$ are in the \textrm{bi-fundamentals} of $%
gl(m)\oplus gl(n)$ with $\pm 1$ charges under $sl(1).$ The $\boldsymbol{N}%
_{m}^{+}$ is given by $(\boldsymbol{m},\boldsymbol{\bar{n}})$ and the $%
\boldsymbol{N}_{m}^{-}$ is given by $(\boldsymbol{\bar{m}},\boldsymbol{n})$
with generators 
\begin{equation}
\begin{tabular}{lllllll|l}
\multicolumn{7}{l|}{\  \  \  \  \  \  \  \  \  \  \  \  \  \ {\small generators}} & 
{\small degree} \\ \hline
$\mathcal{X}_{ai}$ & $=$ & $\left \vert a\right \rangle \left \langle
m+i\right \vert $ & , & ${\small a=1,...,m}$ & {\small ,} & ${\small %
i=1,...,n}$ & {\small odd} \\ 
$\mathcal{Y}_{ai}$ & $=$ & $\left \vert m+i\right \rangle \left \langle
a\right \vert $ & , & ${\small a=1,...,m}$ & {\small ,} & ${\small i=1,...,n}
$ & {\small odd} \\ \hline
\end{tabular}
\label{x3y}
\end{equation}%
\begin{equation*}
\text{ \  \ }
\end{equation*}%
\textrm{Here as well, the} generators are nilpotent because $\mathcal{X}_{ai}%
\mathcal{X}_{cj}=0$ and the same \textrm{goes} for the Y's. The novelty for
this case is that we have only fermionic generators.\newline
The decomposition of $sl(m|n)$ and its super- diagram $\hat{D}_{m+n-1}$ is a
very special case in the sense \textrm{that} it corresponds to cutting the
unique fermionic node of the distinguished super- diagram%
\begin{equation}
\begin{tabular}{lllllll}
$sl(m|n)$ \  \  \  & $:$ & $sl(m|0)$ & $\oplus $ & $sl(1)$ & $\oplus $ & $%
sl(0|n)$ \\ 
\ $\  \  \  \downarrow $ &  & \multicolumn{5}{l}{\  \  \  \  \  \  \  \  \  \ $\  \  \  \  \
\  \  \  \  \  \  \downarrow $} \\ 
$\hat{D}_{m+n-1}$ & $:$ & $D_{m-1}$ & $\oplus $ & $\hat{D}_{1}$ & $\oplus $
& $D_{n-1}$%
\end{tabular}%
\end{equation}%
where we have hidden the nilpotent $\boldsymbol{N}_{m}^{\pm }$. \textrm{%
Strictly speaking,} the diagram $\hat{D}_{1}$ has one fermionic node
corresponding to the unique simple root of $sl(1|1)$ which is fermionic.

\section{L-operators for supergroup $SL\left( m|n\right) $}

In this section, we focus on the distinguished\ DSD\emph{\ }and construct
the super- Lax operators by using the cutting algorithm studied in \textrm{%
the} previous section. We give two types of L-operators: The first type has
mixed bosonic and fermionic phase space variables; see eqs\textrm{(\ref{mn})}
and (\ref{lab}). The second type is purely fermionic; it corresponds to the $%
\mathbb{Z}_{2}$-gradation of $SL\left( m|n\right) ;$ see eqs(\ref{LL}) and (%
\ref{AB}).

\subsection{L-operators with bosonic and fermionic variables}

Here, we construct the super- Lax operator for Chern-Simons theory with $%
SL(m|n)$ gauge symmetry with $m\neq n$. This is a family of super-line
operators associated with the decompositions of the distinguished $sl(m|n)$
given by eq(\ref{p1}), (\ref{p2})\ and (\ref{p3}). The L-operator factorises
as 
\begin{equation}
\mathcal{L}=e^{\Psi }z^{\mathbf{\mu }}e^{\Phi }  \label{pf}
\end{equation}%
with $\mathbf{\mu }$ generating $sl\left( 1\right) $ and $\Psi ,\Phi $
belonging to the nilpotent $N_{\pm }$ sub-superalgebras. The L-operator
describes the coupling between a 't Hooft super-line tH$_{\mathrm{\gamma }%
_{0}}^{_{\mathbf{\mu }}}$ with magnetic charge $\mathbf{\mu }$ and a Wilson
super-line W$_{\mathrm{\xi }_{z}}^{\boldsymbol{m|n}}.$

\subsubsection{More on the decompositions (\protect \ref{p1}) and (\protect
\ref{p2})}

We start by recalling that the $sl(m|n)$ \textrm{decomposes} as $\boldsymbol{%
l}_{\mu _{p}}^{0}\oplus \boldsymbol{N}_{p}^{+}\oplus \boldsymbol{N}_{p}^{-}$
with \textrm{graded} sub-superalgebra $\boldsymbol{l}_{\mu _{p}}^{0}$ 
\textrm{equal to} $(\boldsymbol{l}_{\mu _{p}}^{0})_{\bar{0}}\oplus (%
\boldsymbol{l}_{\mu _{p}}^{0})_{\bar{1}}$ \textrm{such that}%
\begin{equation}
\begin{tabular}{lll}
$(\boldsymbol{l}_{\mu _{p}}^{0})_{\bar{0}}$ & $=$ & $s\left[ gl(p)\oplus
gl(1)\oplus gl(m-p|n)\right] $ \\ 
$(\boldsymbol{l}_{\mu _{p}}^{0})_{\bar{1}}$ & $=$ & $sl(m-p|n)_{\bar{1}}$%
\end{tabular}
\label{mu}
\end{equation}%
\textrm{and} $\boldsymbol{N}_{p}^{\pm }$ \textrm{as} given by (\ref{p1}) and
(\ref{p2}). The decomposition of the fundamental $\boldsymbol{m|n}$
representation of $sl(m|n)$ with respect to $\mu _{p}$ is given by%
\begin{equation}
\begin{tabular}{lll}
$\boldsymbol{m}$ & $=$ & $\boldsymbol{p}_{\frac{m-p-n}{m-n}}\oplus \left( 
\boldsymbol{m-p}\right) _{-\frac{p}{m-n}}$ \\ 
$\boldsymbol{n}$ & $=$ & $\boldsymbol{n}_{-\frac{p}{m-n}}$%
\end{tabular}
\label{pm}
\end{equation}%
where the lower labels refer to the $sl\left( 1\right) $ charges. These
charges are fixed by the vanishing condition of the super-trace of the
representation $\mathbf{m|n}$ reading like,%
\begin{equation}
px_{1}+\left( m-p\right) x_{2}=nx_{2}  \label{x1x2}
\end{equation}%
and solved for $m\neq n$ as $x_{1}=\frac{m-p-n}{m-n}$ and $x_{2}=-\frac{p}{%
m-n}$. Notice that the\textrm{\ special case }$m=n$\textrm{\ needs a
separate construction as it corresponds to the second family of Lie
superalgebras listed in the table (\ref{tabg}). Notice also that }the $%
sl\left( 1\right) $ charges allow to construct the generator $\mathbf{\mu }%
_{p}$ in terms of projectors on three representations: \newline
$\left( \mathbf{1}\right) $ the projector $\Pi _{1}$ on the fundamental
representation $\boldsymbol{p}$ of $sl(p).$ \newline
$\left( \mathbf{2}\right) \ $the projector $\Pi _{2}$ on the representation $%
\left( \boldsymbol{m-p}\right) $ of $sl(m-p).$ \newline
$\left( \mathbf{3}\right) $\ the projector $\Pi _{3}$ on the representation $%
\boldsymbol{n}$ of $sl(n).$ \newline
So, we have%
\begin{equation}
\begin{tabular}{lll}
$\mathbf{\mu }_{p}$ & $=$ & $\frac{m-p-n}{m-n}\Pi _{1}-\frac{p}{m-n}\Pi _{2}-%
\frac{p}{m-n}\Pi _{3}$ \\ 
$z^{\mathbf{\mu }_{p}}$ & $=$ & $z^{\frac{m-p-n}{m-n}}\Pi _{1}+z^{-\frac{p}{%
m-n}}\Pi _{2}+z^{-\frac{p}{m-n}}\Pi _{3}$%
\end{tabular}
\label{mp}
\end{equation}%
and%
\begin{equation}
\lbrack \mathbf{\mu }_{p},\Psi ]=\Psi \qquad ,\qquad \lbrack \mathbf{\mu }%
_{p},\Phi ]=-\Phi  \label{66}
\end{equation}%
Observe in passing that $\mathbf{\mu }_{p}$ can be also expressed like $%
\left( 1-\frac{p}{m-n}\right) \Pi _{1}-\frac{p}{m-n}\left( \Pi _{2}+\Pi
_{3}\right) ;$ this feature will be exploited in the appendix C to rederive
the result of [29].

\subsubsection{The L-operator associated with (\protect \ref{mu})}

To calculate the L-operator associated with the decomposition (\ref{mu}),
notice that the graded matrix operators $\Psi $ and $\Phi $ in (\ref{pf})
satisfy (\ref{66}) and can be split into two contributions: $\left( i\right) 
$ an even contribution $\left. \Psi \right \vert _{even}=X$ and $\left. \Phi
\right \vert _{even}=Y$. $\left( ii\right) $ an odd contribution $\left.
\Psi \right \vert _{odd}=\mathcal{X}$ and $\left. \Phi \right \vert _{odd}=%
\mathcal{Y}$. So, we have%
\begin{equation}
\Psi =X+\mathcal{X}\qquad ,\qquad \Phi =Y+\mathcal{Y}  \label{xx}
\end{equation}%
The $X$ and $Y$ are generated by the bosonic generators $X_{a\dot{b}}$ and $%
Y^{a\dot{b}}$; they read as follows%
\begin{equation}
X=\sum_{a=1}^{p}\sum_{\dot{b}=p+1}^{m}\mathrm{b}^{a\dot{b}}X_{a\dot{b}%
}\qquad ,\qquad Y=\sum_{a=1}^{p}\sum_{\dot{b}=p+1}^{m}\mathrm{c}_{\dot{b}%
a}Y^{\dot{b}a}  \label{xy}
\end{equation}%
where $\mathrm{b}^{a\dot{b}}$ and $\mathrm{c}_{\dot{b}a}$ are bosonic-like
Darboux coordinates. The $\mathcal{X}$ and $\mathcal{Y}$ are generated by
fermionic generators $\mathcal{X}_{ai}$ and $\mathcal{Y}^{ia}$; they are
given by%
\begin{equation}
\mathcal{X}=\sum_{a=1}^{p}\sum_{i=1}^{n}\mathrm{\beta }^{ai}\mathcal{X}%
_{ai}\qquad ,\qquad \mathcal{Y}=\sum_{a=1}^{p}\sum_{i=1}^{n}\mathrm{\gamma }%
_{ia}\mathcal{Y}^{ia}  \label{yx}
\end{equation}%
where $\mathrm{\beta }^{ai}$ and $\mathrm{\gamma }_{ia}$ are fermionic-like
phase space variables.\newline
The explicit expression of $X_{a\dot{b}}$ and $Y^{\dot{b}a}$ as well as
those of $\mathcal{X}_{ai}$ and $\mathcal{Y}^{ia}$ are given by (\ref{x2y}).
They satisfy the useful features%
\begin{equation}
\begin{tabular}{lllllllllll}
$\Pi _{1}X$ & $=X$ & $,$ & $X\Pi _{1}$ & $=0$ & $,$ & $\Pi _{2}X$ & $=0$ & $%
, $ & $X\Pi _{2}$ & $=X$ \\ 
$\Pi _{2}Y$ & $=Y$ & $,$ & $Y\Pi _{2}$ & $=0$ & $,$ & $\Pi _{1}Y$ & $=0$ & $%
, $ & $Y\Pi _{1}$ & $=Y$%
\end{tabular}
\label{pp}
\end{equation}%
and%
\begin{equation}
\begin{tabular}{lllllllllll}
$\Pi _{1}\mathcal{X}$ & $=\mathcal{X}$ & $,$ & $\mathcal{X}\Pi _{1}$ & $=0$
& $,$ & $\Pi _{2}\mathcal{X}$ & $=0$ & $,$ & $\mathcal{X}\Pi _{2}$ & $=%
\mathcal{X}$ \\ 
$\Pi _{2}\mathcal{Y}$ & $=\mathcal{Y}$ & $,$ & $\mathcal{Y}\Pi _{2}$ & $=0$
& $,$ & $\Pi _{1}\mathcal{Y}$ & $=0$ & $,$ & $\mathcal{Y}\Pi _{1}$ & $=%
\mathcal{Y}$%
\end{tabular}
\label{pq}
\end{equation}%
as well as%
\begin{equation}
X\Pi _{3}=\Pi _{3}Y=0,\qquad \mathcal{X}\Pi _{3}=\mathcal{X},\qquad \Pi _{3}%
\mathcal{Y}=\mathcal{Y}  \label{pr}
\end{equation}%
These relations indicate that $\Psi \Pi _{1}=0$ and $\Psi \Pi _{2}=\Psi $
while $\Pi _{2}\Phi =\Phi $ and $\Phi \Pi _{2}=0$. Notice also that these
matrices satisfy $X^{2}=Y^{2}=0$ and $\mathcal{X}^{2}=\mathcal{Y}^{2}=0$ as
well as $X\mathcal{X}=X\mathcal{Y}=0$ and $Y\mathcal{X}=Y\mathcal{Y}=0$. By
using $\Psi =X+\mathcal{X}$ and $\Phi =Y+\mathcal{Y}$, we also have $\Psi
^{2}=0$ and $\Phi ^{2}=0$. So, the super - Lax operator $\mathcal{L}=e^{\Psi
}z^{\mathbf{\mu }}e^{\Phi }$ reads as follows%
\begin{equation}
\mathcal{L}=z^{\mathbf{\mu }_{p}}+z^{\mathbf{\mu }_{p}}\Phi +\Psi z^{\mathbf{%
\mu }_{p}}+\Psi z^{\mathbf{\mu }_{p}}\Phi  \label{zx}
\end{equation}%
Substituting (\ref{mp}), we can put the above relation into the form%
\begin{equation}
\begin{tabular}{lll}
$\mathcal{L}$ & $=$ & $z^{\mathbf{\mu }_{p}}+z^{\frac{m-p-n}{m-n}}\Pi
_{1}\Phi +z^{-\frac{p}{m-n}}\Pi _{2}\Phi +z^{-\frac{p}{m-n}}\Pi _{3}\Phi $
\\ 
&  & $+z^{\frac{m-p-n}{m-n}}\Psi \Pi _{1}+z^{-\frac{p}{m-n}}\Psi \Pi
_{2}+z^{-\frac{p}{m-n}}\Psi \Pi _{3}$ \\ 
&  & $+z^{\frac{m-p-n}{m-n}}\Psi \Pi _{1}\Phi +z^{-\frac{p}{m-n}}\Psi \Pi
_{2}\Phi +z^{-\frac{p}{m-n}}\Psi \Pi _{3}\Phi $%
\end{tabular}
\label{zy}
\end{equation}%
Then, using the properties (\ref{pp}-\ref{pr}), we end up with%
\begin{equation}
\mathcal{L}=\left( 
\begin{array}{ccc}
z^{\frac{m-p-n}{m-n}}\Pi _{1}+z^{-\frac{p}{m-n}}\Pi _{1}\Psi \Phi \Pi _{1} & 
z^{-\frac{p}{m-n}}\Pi _{1}\Psi \Pi _{2} & z^{-\frac{p}{m-n}}\Pi _{1}\Psi \Pi
_{3} \\ 
z^{-\frac{p}{m-n}}\Pi _{2}\Phi \Pi _{1} & z^{-\frac{p}{m-n}}\Pi _{2} & 0 \\ 
z^{-\frac{p}{m-n}}\Pi _{3}\Phi \Pi _{1} & 0 & z^{-\frac{p}{m-n}}\Pi _{3}%
\end{array}%
\right) \text{ }  \label{mn}
\end{equation}%
having the remarkable $\Pi _{1}\Psi \Phi \Pi _{1}$ mixing bosons and
fermions like $\mathrm{b}^{b\dot{b}}\mathrm{c}_{\dot{b}a}+\mathrm{\beta }%
^{bi}\mathrm{\gamma }_{ia}$. The quadratic term $\mathrm{b}^{b\dot{b}}%
\mathrm{c}_{\dot{b}a}$ can be put in correspondence with the energy operator
of $p\left( m-p\right) $ free bosonic harmonic oscillators. However, the
term $\mathrm{\beta }^{bk}\mathrm{\gamma }_{ka}$ describes the energy
operator of $pn$ free fermionic harmonic oscillators. Below, we shed more
light on this aspect by investigating the quantum version of eq\textrm{(\ref%
{mn}).}

\subsubsection{Quantum Lax operator\emph{\ }$\mathcal{\hat{L}}$}

To get more insight into the classical Lax super-operator (\ref{mn}) and in
order to compare with known results obtained in the literature of integrable
superspin chain using the Yangian algebra\textrm{\ }$\mathcal{Y}\left(
sl\left( m|n\right) \right) $\textrm{, }we investigate here the quantum%
\textrm{\ }$\mathcal{\hat{L}}$\textrm{\ }associated with the classical%
\textrm{\ }$\mathcal{L}$\textrm{\ }(\ref{mn}). To that purpose, we proceed
in four steps as described below:\textrm{\ }\newline
$\left( \mathbf{1}\right) $\textrm{\ }We start from eq(\ref{mn}) and
substitute\textrm{\ }$X$ and $Y$\textrm{\ }as well as\textrm{\ }$\mathcal{X}$
and $\mathcal{Y}$ by their expressions in terms of the classical
oscillators. Putting eqs(\ref{xy})\textrm{\ }and\textrm{\ }(\ref{yx}) into (%
\ref{mn}), we obtain a $3\times 3$ block graded matrix of the form%
\begin{equation}
\mathcal{L}_{A}^{B}=\left( 
\begin{array}{ccc}
\mathcal{L}_{a}^{b} & \mathcal{L}_{a}^{\dot{b}} & \mathcal{L}_{a}^{j} \\ 
\mathcal{L}_{\dot{a}}^{b} & \mathcal{L}_{\dot{a}}^{\dot{b}} & \mathcal{L}_{%
\dot{a}}^{j} \\ 
\mathcal{L}_{i}^{a} & \mathcal{L}_{i}^{\dot{b}} & \mathcal{L}_{i}^{j}%
\end{array}%
\right)
\end{equation}%
with entries as follows\textrm{\ }%
\begin{equation}
\mathcal{L}_{A}^{B}=\left( 
\begin{array}{ccc}
z^{\frac{m-p-n}{m-n}}\delta _{a}^{b}+z^{-\frac{p}{m-n}}\mathrm{b}^{b\dot{b}}%
\mathrm{c}_{\dot{b}a}\mathbf{+}z^{-\frac{p}{m-n}}\mathrm{\beta }^{bk}\mathrm{%
\gamma }_{ka} & z^{-\frac{p}{m-n}}\delta _{ac}\mathrm{b}^{c\dot{b}} & z^{-%
\frac{p}{m-n}}\delta _{ac}\mathrm{\beta }^{cj} \\ 
z^{-\frac{p}{m-n}}\mathrm{c}_{\dot{a}c}\delta ^{cb} & z^{-\frac{p}{m-n}%
}\delta _{\dot{a}}^{\dot{b}} & 0 \\ 
z^{-\frac{p}{m-n}}\mathrm{\gamma }_{ic}\delta ^{cb} & 0 & z^{-\frac{p}{m-n}%
}\delta _{i}^{j}%
\end{array}%
\right)  \label{La}
\end{equation}%
$\left( \mathbf{2}\right) $ In the above expression of the Lax operator (\ref%
{La}), the products $\mathrm{b}^{b\dot{b}}\mathrm{c}_{\dot{b}a}$ and $%
\mathrm{\beta }^{bk}\mathrm{\gamma }_{ka}$ are classical; they can be
respectively thought of as%
\begin{equation}
\begin{tabular}{lll}
$\mathrm{b}^{b\dot{b}}\mathrm{c}_{\dot{b}a}$ & $=$ & $\frac{1}{2}\left( 
\mathrm{b}^{b\dot{b}}\mathrm{c}_{\dot{b}a}+\mathrm{c}_{\dot{b}a}\mathrm{b}^{b%
\dot{b}}\right) $ \\ 
$\mathrm{\beta }^{bk}\mathrm{\gamma }_{ka}$ & $=$ & $\frac{1}{2}\left( 
\mathrm{\beta }^{bk}\mathrm{\gamma }_{ka}-\mathrm{\gamma }_{ka}\mathrm{\beta 
}^{bk}\right) $%
\end{tabular}
\label{bg}
\end{equation}%
with $\left( i\right) $ bosonic $\mathrm{b}^{b\dot{b}}$ and $\mathrm{c}_{%
\dot{b}a}$ tensors represented by the following $p\times \left( m-p\right) $
and $\left( m-p\right) \times p$ rectangular matrices%
\begin{equation}
\mathrm{b}^{b\dot{b}}=\left( 
\begin{array}{ccc}
\mathrm{b}^{1\dot{1}} & \cdots & \mathrm{b}^{1\dot{p}} \\ 
\vdots & \ddots & \vdots \\ 
\mathrm{b}^{p\dot{1}} & \cdots & \mathrm{b}^{p\dot{p}}%
\end{array}%
\right) ,\qquad \mathrm{c}_{\dot{b}a}=\left( 
\begin{array}{ccc}
\mathrm{c}_{\dot{1}1} & \cdots & \mathrm{c}_{\dot{1}p} \\ 
\vdots & \ddots & \vdots \\ 
\mathrm{c}_{\dot{p}1} & \cdots & \mathrm{c}_{\dot{p}p}%
\end{array}%
\right)  \label{19}
\end{equation}%
with $\dot{p}=m-p$; and $\left( ii\right) $ fermionic $\mathrm{\beta }^{bk}$
and $\mathrm{\gamma }_{ka}$ tensors represented by the $p\times n$ and $%
n\times p$ rectangular matrices%
\begin{equation}
\mathrm{\beta }^{bk}=\left( 
\begin{array}{ccc}
\mathrm{\beta }^{11} & \cdots & \mathrm{\beta }^{1n} \\ 
\vdots & \ddots & \vdots \\ 
\mathrm{\beta }^{p1} & \cdots & \mathrm{\beta }^{pn}%
\end{array}%
\right) ,\qquad \mathrm{\gamma }_{ka}=\left( 
\begin{array}{ccc}
\mathrm{\gamma }_{11} & \cdots & \mathrm{\gamma }_{1p} \\ 
\vdots & \ddots & \vdots \\ 
\mathrm{\gamma }_{n1} & \cdots & \mathrm{\gamma }_{np}%
\end{array}%
\right)  \label{20}
\end{equation}%
$\left( \mathbf{3}\right) $ \textrm{At the }quantum level, the bosonic $%
\mathrm{b}^{a\dot{b}}$ and $\mathrm{c}_{\dot{a}b}$ as well as the fermionic $%
\mathrm{\beta }^{ai}$ and $\mathrm{\gamma }_{ia}$ are promoted to the
creation $\mathrm{\hat{b}}^{a\dot{b}}$ and annihilation $\mathrm{\hat{c}}_{%
\dot{a}b}$ operators as well as the creation $\mathrm{\hat{\beta}}^{ai}$ and
the annihilation $\mathrm{\hat{\gamma}}_{ia}$. In this regard, notice that
in the $su(m|n)$ unitary theory, these creation and annihilation operators
are related like $\mathrm{\hat{b}}^{a\dot{b}}=(\mathrm{\hat{c}}_{\dot{b}%
a})^{\dagger }$ and $\mathrm{\hat{\beta}}^{ai}=(\mathrm{\hat{\gamma}}%
_{ia})^{\dagger }$. In the 4D super Chern-Simons theory, the unitary gauge
symmetry is complexified like $sl(m|n).$ Notice also that the usual
classical Poisson bracket of the $\mathbb{Z}_{2}$- graded phase space
variables are replaced in quantum mechanics by the following graded
canonical commutation relations 
\begin{equation}
\begin{tabular}{lll}
$\lbrack \mathrm{\hat{c}}_{\dot{c}d},\mathrm{\hat{b}}^{a\dot{b}}]$ & $=$ & $%
\delta _{\dot{c}}^{\dot{b}}\delta _{d}^{a}$ \\ 
$\lbrack \mathrm{\hat{b}}^{a\dot{b}},\mathrm{\hat{b}}^{d\dot{c}}]$ & $=$ & $%
0 $ \\ 
$\left[ \mathrm{\hat{c}}_{\dot{b}a},\mathrm{\hat{c}}_{\dot{c}d}\right] $ & $%
= $ & $0$%
\end{tabular}%
\qquad ,\qquad 
\begin{tabular}{lll}
$\{ \mathrm{\hat{\gamma}}_{jb},\mathrm{\hat{\beta}}^{ai}\}$ & $=$ & $\delta
_{b}^{a}\delta _{j}^{i}$ \\ 
$\{ \mathrm{\hat{\beta}}^{ai},\mathrm{\hat{\beta}}^{bj}\}$ & $=$ & $0$ \\ 
$\{ \mathrm{\hat{\gamma}}_{ia},\mathrm{\hat{\gamma}}_{jb}\}$ & $=$ & $0$%
\end{tabular}
\label{21}
\end{equation}%
and $[\mathrm{\hat{c}}_{\dot{c}d},\mathrm{\hat{\beta}}^{ai}]=[\mathrm{\hat{c}%
}_{\dot{c}d},\mathrm{\hat{\gamma}}_{jb}]=0$ as well as $[\mathrm{\hat{b}}^{d%
\dot{c}},\mathrm{\hat{\beta}}^{ai}]=[\mathrm{\hat{b}}^{d\dot{c}},\mathrm{%
\hat{\gamma}}_{jb}]=0$. \newline
$\left( \mathbf{4}\right) $ Under the substitution $\left( b,c\right)
\rightarrow (\hat{b},\hat{c})$ and $\left( \beta ,\gamma \right) \rightarrow
(\hat{\beta},\hat{\gamma}),$ the classical eq(\ref{bg}) gets promoted to
operators as follows%
\begin{equation}
\begin{tabular}{lll}
$\frac{1}{2}\left( \mathrm{\hat{b}}^{b\dot{d}}\mathrm{\hat{c}}_{\dot{d}a}+%
\mathrm{\hat{c}}_{\dot{d}a}\mathrm{\hat{b}}^{b\dot{d}}\right) $ & $=$ & $%
\mathrm{\hat{b}}^{b\dot{d}}\mathrm{\hat{c}}_{\dot{d}a}+\frac{p\left(
m-p\right) }{2}\delta _{a}^{b}$ \\ 
$\frac{1}{2}\left( \mathrm{\hat{\beta}}^{bk}\mathrm{\hat{\gamma}}_{ka}-%
\mathrm{\hat{\gamma}}_{ka}\mathrm{\hat{\beta}}^{bk}\right) $ & $=$ & $%
\mathrm{\hat{\beta}}^{bk}\mathrm{\hat{\gamma}}_{ka}-\frac{pn}{2}\delta
_{a}^{b}$%
\end{tabular}%
\end{equation}%
Substituting these quantum expressions into (\ref{La}), we obtain the
explicit oscillator realisation of the quantum Lax operator namely%
\begin{equation}
\mathcal{\hat{L}}_{A}^{B}=\left( 
\begin{array}{ccc}
z^{\frac{m-p-n}{m-n}}\delta _{a}^{b}+z^{-\frac{p}{m-n}}\hat{K}_{a}^{b} & z^{-%
\frac{p}{m-n}}\delta _{ac}\mathrm{\hat{b}}^{c\dot{b}} & z^{-\frac{p}{m-n}%
}\delta _{ac}\mathrm{\hat{\beta}}^{cj} \\ 
z^{-\frac{p}{m-n}}\mathrm{\hat{c}}_{\dot{a}c}\delta ^{cb} & z^{-\frac{p}{m-n}%
}\delta _{\dot{a}}^{\dot{b}} & 0 \\ 
z^{-\frac{p}{m-n}}\mathrm{\hat{\gamma}}_{ic}\delta ^{cb} & 0 & z^{-\frac{p}{%
m-n}}\delta _{i}^{j}%
\end{array}%
\right)  \label{lab}
\end{equation}%
with%
\begin{equation}
\hat{K}_{a}^{b}=\mathrm{\hat{b}}^{b\dot{d}}\mathrm{\hat{c}}_{\dot{d}a}+%
\mathrm{\hat{\beta}}^{bk}\mathrm{\hat{\gamma}}_{ka}+\left[ \frac{p\left(
m-p\right) }{2}-\frac{pn}{2}\right] \delta _{a}^{b}
\end{equation}%
This result, obtained from the 4D super Chern-Simons theory, can be compared
with the quantum Lax operator \textrm{(2.20) in} \cite{1G} \textrm{calculated%
} using the super Yangian $Y\left( su\left( m|n\right) \right) $ \textrm{%
representation}.\  \textrm{Actually, we can multiply this graded matrix by
the quantity }$z^{-\frac{p}{m-n}}$\textrm{\ to obtain}%
\begin{equation}
\mathcal{\hat{L}}_{A}^{B}=\left( 
\begin{array}{ccc}
z\delta _{a}^{b}+\hat{K}_{a}^{b} & \delta _{ac}\mathrm{\hat{b}}^{c\dot{b}} & 
\delta _{ac}\mathrm{\hat{\beta}}^{cj} \\ 
\mathrm{\hat{c}}_{\dot{a}c}\delta ^{cb} & \delta _{\dot{a}}^{\dot{b}} & 0 \\ 
\mathrm{\hat{\gamma}}_{ic}\delta ^{cb} & 0 & \delta _{i}^{j}%
\end{array}%
\right)
\end{equation}%
\textrm{Note that the multiplication by a function of the spectral parameter
does not affect the RLL equation \cite{1D,1H}. More details concerning the
comparison between these results and those obtained in [29] are reported in
appendix C. }\newline
\textrm{Thanks to these results, we can }rely on the consistency of the CS
theory approach \textrm{based on the} decomposition of Lie superalgebras to
state that the study performed here for one fermionic node can be
straightforwardly extended to $\left( i\right) $\ the case of several
fermionic nodes of $sl(m|n);$\ and\ to $\left( ii\right) $ the other
classical Lie superalgebras of table (\ref{tabg}) such as the
orthosymplectic $osp\left( m|2n\right) $\ spin chain.

\subsection{Pure fermionic L-operator}

This is an interesting decomposition of the Lie superalgebra $sl(m|n)$ with
distinguished DSD. It corresponds to cutting of the unique fermionic node $%
\tilde{\alpha}_{m}$ of the distinguished Dynkin super- graph. This
decomposition coincides with the usual $\mathbb{Z}_{2}$-gradation of the Lie
superalgebra 
\begin{equation}
sl(m|n)=sl(m|n)_{\bar{0}}\oplus sl(m|n)_{\bar{1}}  \label{lmn}
\end{equation}%
Here, we begin by calculating the classical Lax matrix $L$\ associated with
cutting $\tilde{\alpha}_{m}$\ in $\hat{D}_{m+n-1}$,\textrm{\ then we
investigate its quantum version }$\hat{L}$\textrm{.}

\subsubsection{Constructing the classical Lax matrix}

In the pure fermionic case, the Lie superalgebra decomposes as in (\ref{p3}%
). Besides $\left( \boldsymbol{l}_{m}\right) _{\bar{0}}=s\left[ gl(m)\oplus
gl(n)\right] $ and the nilpotent $\boldsymbol{N}_{+}=sl(m-1|n)_{+1}$ and $%
\boldsymbol{N}_{-}=sl(m-1|n)_{-1},$ we need the decomposition of the
fundamental $\boldsymbol{m|n}$ representation of $sl(m|n).$ We have 
\begin{equation}
\boldsymbol{m|n}=\left( \boldsymbol{m}_{x},\boldsymbol{n}_{y}\right)
\end{equation}%
with lower labels $x$ and $y$ referring to the $sl\left( 1\right) $ charges.
These charges are determined by the vanishing condition of the supertrace
namely $mx-ny=0$ which is solved as follows 
\begin{equation}
x=\frac{-n}{m-n}\qquad ,\qquad y=\frac{-m}{m-n}
\end{equation}%
Notice that this solution corresponds just to setting $p=m$ in (\ref{x1x2}).
These charges allow to construct the generator of the charge operator $%
\mathbf{\mu }$ in terms of two projectors $\Pi _{1}$ and $\Pi _{2}$ on the
representations $\boldsymbol{m}$ of $sl(m)$ and $\boldsymbol{n}$ of $sl(n).$
Using the kets $\left \vert a\right \rangle $ of even degree generating the $%
\boldsymbol{m}$ and the kets $\left \vert m+i\right \rangle $ of odd degree
generating $\boldsymbol{n},$ we have%
\begin{equation}
\Pi _{1}=\sum_{a=1}^{m}\left \vert a\right \rangle \left \langle a\right
\vert \qquad ,\qquad \Pi _{2}=\sum_{i=1}^{n}\left \vert m+i\right \rangle
\left \langle m+i\right \vert
\end{equation}%
where $\Pi _{1}$ is the projector on $sl\left( m\right) $ sector in the even
part $sl\left( m|n\right) _{\bar{0}}$ of the Lie superalgebra and $\Pi _{2}$
is the projector on its $sl\left( n\right) $ sector. The generator of $%
gl\left( 1\right) $ is then given by%
\begin{equation}
\mathbf{\mu }=\frac{n}{n-m}\Pi _{1}+\frac{m}{n-m}\Pi _{2}  \label{620}
\end{equation}%
The $\Psi $ and $\Phi $ nilpotent matrices in (\ref{pf}) read as%
\begin{equation}
\Psi =\sum_{a=1}^{m}\sum_{i=1}^{n}\mathrm{\beta }^{ai}\mathcal{X}_{ai}\qquad
,\qquad \Phi =\sum_{a=1}^{m}\sum_{i=1}^{n}\mathrm{\gamma }_{ia}\mathcal{Y}%
^{ia}  \label{pg}
\end{equation}%
where%
\begin{equation}
\mathcal{X}_{ai}=\left \vert a\right \rangle \left \langle m+i\right \vert
\qquad ,\qquad \mathcal{Y}^{ia}=\left \vert m+i\right \rangle \left \langle
a\right \vert
\end{equation}%
with $\mathrm{\beta }^{ai}$ and $\mathrm{\gamma }_{ia}$ describing $mn$
fermionic phase space coordinates. This realisation satisfies some
properties, in particular 
\begin{equation}
\begin{tabular}{lllll}
$\Pi _{1}\Psi $ & $=\Psi $ & $\qquad ,\qquad $ & $\Pi _{2}\Psi $ & $=0$ \\ 
$\Pi _{2}\Phi $ & $=\Phi $ & $\qquad ,\qquad $ & $\Pi _{1}\Phi $ & $=0$ \\ 
$\Psi \Pi _{2}$ & $=\Psi $ & $\qquad ,\qquad $ & $\Psi \Pi _{1}$ & $=0$ \\ 
$\Phi \Pi _{1}$ & $=\Phi $ & $\qquad ,\qquad $ & $\Phi \Pi _{2}$ & $=0$%
\end{tabular}%
\end{equation}%
showing that 
\begin{equation}
\left[ \mathbf{\mu },\Psi \right] =\Psi \qquad ,\qquad \left[ \mathbf{\mu }%
,\Phi \right] =-\Phi
\end{equation}%
Moreover, using the properties $\Psi ^{2}=\Phi ^{2}=0,$ we have $e^{\Psi
}=1+\Psi $ and $e^{\Phi }=1+\Phi .$ Putting back into the L-operator (\ref%
{pf}), we obtain 
\begin{equation}
\begin{tabular}{lll}
$\mathcal{L}$ & $=$ & $z^{\frac{n}{n-m}}\Pi _{1}+z^{\frac{m}{n-m}}\Pi _{2}+$
\\ 
&  & $z^{\frac{n}{n-m}}\Pi _{1}\Phi +z^{\frac{m}{n-m}}\Pi _{2}\Phi +$ \\ 
&  & $z^{\frac{n}{n-m}}\Psi \Pi _{1}+z^{\frac{m}{n-m}}\Psi \Pi _{2}+$ \\ 
&  & $z^{\frac{n}{n-m}}\Psi \Pi _{1}\Phi +z^{\frac{m}{n-m}}\Psi \Pi _{2}\Phi 
$%
\end{tabular}%
\end{equation}%
Substituting $\Pi _{1}\Phi =0$ and $\Psi \Pi _{1}=0$ as well as $\Phi \Pi
_{2}=0$ and $\Pi _{2}\Psi =0$, the above expression reduces to%
\begin{equation}
\begin{tabular}{lll}
$\mathcal{L}$ & $=$ & $z^{\frac{n}{n-m}}\Pi _{1}+z^{\frac{m}{n-m}}\left(
\Psi \Phi \right) \Pi _{1}+$ \\ 
&  & $+z^{\frac{m}{n-m}}\Pi _{2}\Phi +z^{\frac{m}{n-m}}\Psi \Pi _{2}+$ \\ 
&  & $+z^{\frac{m}{n-m}}\Pi _{2}$%
\end{tabular}%
\end{equation}%
It reads in super matrix language as follows%
\begin{equation}
\mathcal{L}=\left( 
\begin{array}{cc}
z^{\frac{n}{n-m}}\Pi _{1}+z^{\frac{m}{n-m}}\Pi _{1}\Psi \Phi \Pi _{1} & z^{%
\frac{m}{n-m}}\Pi _{1}\Psi \Pi _{2} \\ 
z^{\frac{m}{n-m}}\Pi _{2}\Phi \Pi _{1} & z^{\frac{m}{n-m}}\Pi _{2}%
\end{array}%
\right)  \label{LL}
\end{equation}%
where $\Pi _{1}\left( \Psi \Phi \right) \Pi _{1}$ is given by $\sum_{k}%
\mathrm{\beta }^{bk}\mathrm{\gamma }_{ka}.$ Notice that the term $\mathrm{%
\beta }^{bk}\mathrm{\gamma }_{ka}$ can be put in correspondence with the
energy of $mn$ free fermionic harmonic oscillators as described below.

\subsubsection{Quantum version of eq(\protect \ref{LL})}

To derive the quantum version\textrm{\ }$\mathcal{\hat{L}}$\textrm{\ }%
associated with the classical (\ref{LL}) and its properties, we use the
correspondence between the phase space variables and the quantum
oscillators. We determine\textrm{\ }$\mathcal{\hat{L}}$\textrm{\ }by
repeating the analysis that we have done in the sub-subsection 6.1.2 to the
fermionic oscillators. To that purpose, we perform this derivation by
following four steps as follows. \newline
$\left( \mathbf{1}\right) $ We substitute the $\Psi $ and $\Phi $\ in (\ref%
{LL}) by their expressions in terms of the classical fermionic oscillators $%
\mathrm{\beta }^{bk}$ and $\mathrm{\gamma }_{ka}$. By putting eq\textrm{(\ref%
{pg})} into (\textrm{\ref{LL}}), we obtain the following $2\times 2$ block
matrix%
\begin{equation}
\mathcal{L}_{B}^{A}=\left( 
\begin{array}{cc}
\mathcal{L}_{b}^{a} & \mathcal{L}_{j}^{a} \\ 
\mathcal{L}_{i}^{b} & \mathcal{L}_{ij}%
\end{array}%
\right)
\end{equation}%
\textrm{\ }with entries as follows%
\begin{equation}
\mathcal{L}_{A}^{B}=\left( 
\begin{array}{cc}
z^{\frac{n}{n-m}}\delta _{a}^{b}+z^{\frac{m}{n-m}}\mathrm{\beta }^{bk}%
\mathrm{\gamma }_{ka} & z^{\frac{m}{n-m}}\delta _{ab}\mathrm{\beta }^{bj} \\ 
z^{\frac{m}{n-m}}\mathrm{\gamma }_{ic}\delta ^{cb} & z^{\frac{m}{n-m}}\delta
_{i}^{j}%
\end{array}%
\right)  \label{LH}
\end{equation}%
$\left( \mathbf{2}\right) $ We replace the product $\mathrm{\beta }^{bk}%
\mathrm{\gamma }_{ka}$ in the above classical Lax matrix (\ref{LH}) by the
following equivalent expression where $\mathrm{\beta }^{bk}$ and $\mathrm{%
\gamma }_{ka}$ are treated on equal footing,%
\begin{equation}
\mathrm{\beta }^{bk}\mathrm{\gamma }_{ka}=\frac{1}{2}\left( \mathrm{\beta }%
^{bk}\mathrm{\gamma }_{ka}-\mathrm{\gamma }_{ka}\mathrm{\beta }^{bk}\right)
\label{gg}
\end{equation}%
with $\mathrm{\beta }^{bk}$ and $\mathrm{\gamma }_{ka}$ given by the
following $m\times n$ and $n\times m$ rectangular matrices 
\begin{equation}
\mathrm{\beta }^{bj}=\left( 
\begin{array}{ccc}
\mathrm{\beta }^{11} & \cdots & \mathrm{\beta }^{1n} \\ 
\vdots & \ddots & \vdots \\ 
\mathrm{\beta }^{m1} & \cdots & \mathrm{\beta }^{mn}%
\end{array}%
\right) ,\qquad \mathrm{\gamma }_{ia}=\left( 
\begin{array}{ccc}
\mathrm{\gamma }_{11} & \cdots & \mathrm{\gamma }_{1m} \\ 
\vdots & \ddots & \vdots \\ 
\mathrm{\gamma }_{n1} & \cdots & \mathrm{\gamma }_{nm}%
\end{array}%
\right)
\end{equation}%
$\left( \mathbf{3}\right) $ \textrm{At the qu}antum level, the classical
fermionic oscillators $\mathrm{\beta }^{bj}$ and $\mathrm{\gamma }_{ia}$ are
promoted to the creation $\mathrm{\hat{\beta}}^{bj}$ and the annihilation $%
\mathrm{\hat{\gamma}}_{ia}$ operators satisfying the following graded
canonical commutation relations 
\begin{equation}
\begin{tabular}{lll}
$\{ \mathrm{\hat{\gamma}}_{ia},\mathrm{\hat{\beta}}^{bj}\}$ & $=$ & $\delta
_{a}^{b}\delta _{i}^{j}$ \\ 
$\{ \mathrm{\hat{\beta}}^{ai},\mathrm{\hat{\beta}}^{bj}\}$ & $=$ & $0$ \\ 
$\{ \mathrm{\hat{\gamma}}_{ia},\mathrm{\hat{\gamma}}_{jb}\}$ & $=$ & $0$%
\end{tabular}
\label{gb}
\end{equation}%
As noticed before regarding the $su(m|n)$ unitary theory, we have the
relation $\mathrm{\hat{\beta}}^{ai}=(\mathrm{\hat{\gamma}}_{ia})^{\dagger }.$
By using this quantum extension, the classical $\mathbf{(}\beta \gamma
-\gamma \beta \mathbf{)}/2$ gets promoted in turns to the quantum operator $%
\mathbf{(}\hat{\beta}\hat{\gamma}-\hat{\gamma}\hat{\beta}\mathbf{)}/2.$
Then, using (\ref{gb}), we can also express $\mathbf{(}\hat{\beta}\hat{\gamma%
}-\hat{\gamma}\hat{\beta}\mathbf{)}/2$ as a normal ordered operator with the
creation operators $\hat{\beta}$ put on the left like $\mathrm{\beta }^{bk}%
\mathrm{\gamma }_{ka}-\left( mn/2\right) \delta _{a}^{b}$. So, eq(\ref{gg})
gets replaced by the following normal ordered quantum quantity%
\begin{equation}
\frac{1}{2}\left( \mathrm{\hat{\beta}}^{bk}\mathrm{\hat{\gamma}}_{ka}-%
\mathrm{\hat{\gamma}}_{ka}\mathrm{\hat{\beta}}^{bk}\right) =\mathrm{\hat{%
\beta}}^{bk}\mathrm{\hat{\gamma}}_{ka}-\frac{mn}{2}\delta _{a}^{b}
\end{equation}%
$\left( \mathbf{3}\right) $ Substituting the above quantum relation into eq(%
\ref{LH}), we obtain the quantum Lax operator $\mathcal{\hat{L}}$ given by%
\begin{equation}
\mathcal{\hat{L}}_{A}^{B}=\left( 
\begin{array}{cc}
z^{\frac{n}{n-m}}\delta _{a}^{b}+z^{\frac{m}{n-m}}\left( \mathrm{\hat{\beta}}%
^{bk}\mathrm{\hat{\gamma}}_{ka}-\frac{mn}{2}\delta _{a}^{b}\right) & z^{%
\frac{m}{n-m}}\delta _{ab}\mathrm{\hat{\beta}}^{bj} \\ 
z^{\frac{m}{n-m}}\mathrm{\hat{\gamma}}_{ic}\delta ^{cb} & z^{\frac{m}{n-m}%
}\delta _{i}^{j}%
\end{array}%
\right)  \label{AB}
\end{equation}%
By multiplying this relation by $z^{-\frac{m}{n-m}}$, the above graded
matrix becomes%
\begin{equation}
\mathcal{\hat{L}}_{A}^{B}=\left( 
\begin{array}{cc}
z\delta _{a}^{b}+\left( \mathrm{\hat{\beta}}^{bk}\mathrm{\hat{\gamma}}_{ka}-%
\frac{mn}{2}\delta _{a}^{b}\right) & \delta _{ab}\mathrm{\hat{\beta}}^{bj}
\\ 
\mathrm{\hat{\gamma}}_{ic}\delta ^{cb} & \delta _{i}^{j}%
\end{array}%
\right)
\end{equation}

\section{Conclusion and comments}

In this paper, we investigated the 4D Chern-Simons theory with gauge
symmetry given by the $SL(m|n)$ super-group family ($m\neq n$) and
constructed the super- Lax operator solving the RLL equations of the
superspin chain. We described the Wilson and 't Hooft super-lines for the $%
SL(m|n)$\ symmetry and explored their interaction and their implementation
in the extended 4D CS super- gauge theory. We also developed a DSDs
algorithm for the distinguished basis of $sl(m|n)$\ to generalise the Levi-
decomposition of Lie algebras to the Lie superalgebras. Our findings agree
with partial results obtained in literature on integrable superspin chains.
The solutions for $SL(m|n)$\ are of two types: a generic one having a
mixture between bosonic and fermionic oscillators, and a special purely
fermionic type corresponding to the $\mathbb{Z}_{2}$-gradation of $sl(m|n)$. 
\newline
To perform this study, we started by revisiting the explicit derivation of
the expression of the L-operator in 4D CS theory with bosonic gauge symmetry
by following the method of Costello-Gaiotto-Yagi used in \textrm{\cite{1D}}.
We also investigated the holomorphy of $L\left( z\right) $ and described
properties of interacting Wilson and 't Hooft lines. We showed how the Dirac
singularity of the magnetic 't Hooft line lead to an exact description of
the oscillator Lax operator for the XXX spin chains with bosonic symmetry.%
\newline
Then, we worked out the differential equation $\mathcal{D}L=0$ solved by the
CGY realisation of the L-operator. We also gave a link of this differential
equation with the usual time evolution equation of the Lax operator. We used
the algebraic structure of $\mathcal{D}L=0$ to motivate the generalisation
of the L-operator to supergroups. As illustration, we considered two
particular symmetries: $\left( i\right) $ the bosonic $GL\left( 2\right) $
as a simple representative of $GL\left( m\right) $. $\left( ii\right) $ the
supergroup $GL\left( 1|1\right) $ as a representative of $GL(m|n).$ \newline
After that, we investigated the general case of 4D CS with supergroups by
focussing on the $sl(m|n)$ family. As there is no known extension for the
Levi-theorem concerning the decomposition of Lie superalgebras, we developed
an algorithm to circumvent this lack. This algorithm, which uses the Dynkin
diagram language, has been checked in the case of bosonic Lie algebras to be
just a rephrasing \textrm{of} the Levi-theorem. The extension to Lie
superalgebras is somehow subtle because a given Lie superalgebra has in
general several representative DSDs. In this context, recall that a bosonic
finite dimensional Lie algebra has one Dynkin diagram. But this is not true
for Lie superalgebras as described in section 4. As a first step towards the
construction of the Lax operators for classical gauge supergroups, we
focused our attention on the particular family of distinguished $sl(m|n).$
For this family, we showed that the Levi-theorem extends naturally as
detailed in section 5. We used this result to derive the various types of
super- Lax operators for the distinguished DSDs containing one fermionic
node. \newline
We hope to return to complete this investigation by performing three more
steps in the study of L-operators of Lie superalgebras. First, enlarge the
construction \textrm{to} other classical Lie superalgebras like $A(m|m)$, $%
B(m|n)$, $C(m+1)$ and $D(m|n)$. Second, extend the present $sl(m|n)$ study
to DSDs with two fermionic nodes and more. Third, use the so-called
Gauge/Bethe correspondence to work out D-brane realisations of the superspin
chains in type II strings.

\section{Appendices}

Here we provide complementary materials that are useful for this
investigation. We give two appendices A and B. In appendix A, we revisit the
derivation of the L-operators in 4D CS with bosonic gauge symmetries and
their properties. In section B, we describe the Verma modules of $sl(m|n).$

\subsection{Appendix A: L-operators in 4D CS theory}

First, we study the link between Dirac singularity of monopoles and the Lax
operator obtained in \textrm{\cite{1D}}. Then, we revisit the\ explicit%
\textrm{\ }derivation of the minuscule Lax operators by using
Levi-decomposition of gauge symmetries.

\subsubsection{From Dirac singularity to the L-operator}

Following \textrm{\cite{1D}}, a similar analysis to the Yang-Mills theory
monopoles holds for the 4D-CS theory in the presence of a 't Hooft line with
magnetic charge given by the coweight $\mu $. In this case, the special
behaviour of the singular gauge field implies dividing the region
surrounding the 't Hooft line into the two intersecting regions $U_{I}$ and $%
U_{2}$ with line intersection $U_{I}\cap U_{II}=\mathrm{\gamma }_{0}$. By
choosing the 't Hooft line $\mathrm{\gamma }_{0}$ as sitting on the x-axis $%
(y=0)$ of the topological plane $\mathbb{R}^{2}$ and at $z=0$ of the
holomorphic line, we have 
\begin{equation}
\begin{tabular}{lll}
$U_{I}$ & $=$ & $\left \{ y\leq 0,z=0\right \} $ \\ 
$U_{II}$ & $=$ & $\left \{ y\geq 0,z=0\right \} $%
\end{tabular}%
\end{equation}%
On the region $U_{I},$ we have a trivialised gauge field described by a $G$%
-valued holomorphic function that needs to be regular at $z=0$, say a
holomorphic gauge transformation $\mathfrak{g}_{I}(z)\in G_{[[z]]}$. The
same behaviour is valid for the region $U_{II}$\ where we have $\mathfrak{g}%
_{II}(z)\in G_{[[z]]}.$ These trivial bundles are glued by a transition
function (isomorphism) on the region $U_{I}\cap U_{II},$ it serves as a
parallel transport of the gauge field from the region $y<0$ to the region $%
y>0$ near the line (say in the disc $y=0,\left \vert z\right \vert \leq
\varepsilon $). This parallel transport is given by the local Dirac
singularity \textrm{\cite{1I}}%
\begin{equation}
\mathfrak{g}_{0}\left( z;\mu \right) =z^{\mu }\in G_{((z))}
\end{equation}%
In \cite{1D}, the observable $L(z;\mu )$ is given by the parallel transport
of the gauge field bundle sourced by the magnetically charged 't Hooft line
of magnetic charge\ $\mu $ from $y\ll 0$ to $y\gg 0.$ It reads as,%
\begin{equation}
L(z;\mu )=\mathcal{P}Exp\left[ \int_{y}\mathcal{A}_{y}(z)\right]
\end{equation}%
Because of the singular behaviour of the gauge configuration described
above, the line operator $L(z)$ near $z\simeq 0$ takes the general form%
\begin{equation}
L(z;\mu )=g_{I}\left( z\right) z^{\mu }g_{II}\left( z\right)  \label{211}
\end{equation}%
and belongs to the moduli space $G_{[[z]]}\backslash G_{((z))}/G_{[[z]]}$.
Notice that because of the topological nature of the Dirac monopole (a Dirac
string stretching between two end states), we also need to consider another
't Hooft line with the opposite magnetic charge $-\mu $ at $z=\infty .$ In
the region near $z\simeq \infty ,$ the gauge configuration is treated in the
same way as in the neighbourhood of $z\simeq 0.$ The corresponding parallel
transport takes the form%
\begin{equation}
G_{[[z^{-1}]]}z^{-\mu }G_{[[z^{-1}]]}
\end{equation}%
with gauge transformations in $G_{[[z^{-1}]]}$ going to the identity $I_{id}$
when $z=\infty .$ Consequently, the parallel transport from $y\ll 0$ to $%
y\gg 0$ of the gauge field, sourced by the 't Hooft lines having the charge $%
\mu $\ at $z=0$\ and $-\mu $\ at $z=\infty ,$ is given by the holomorphic
line operator,%
\begin{equation}
L(z;\mu )=A\left( z\right) z^{\mu }B\left( z\right)  \label{L}
\end{equation}%
It is characterized by zeroes and poles at $z=0$ and $z=\infty $ manifesting
the singularities implied by the two 't Hooft lines at zero and infinity.

\subsubsection{Minuscule L-operator}

\begin{figure}[tbph]
\begin{center}
\includegraphics[width=14cm]{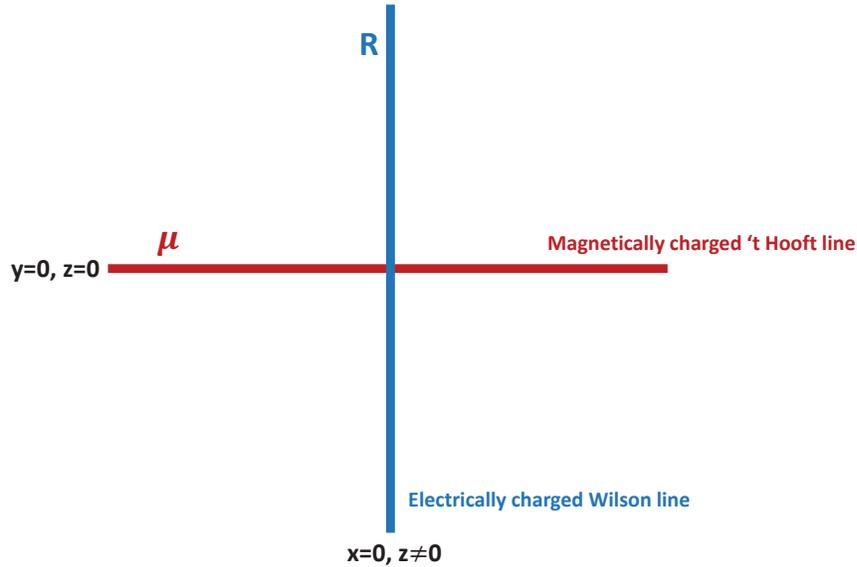}
\end{center}
\par
\vspace{-0.5cm}
\caption{A magnetically charged 't Hooft line crossing an electrical charged
Wilson line. The two lines expands in the topological plane $\mathbb{R}^{2}$
of the 4D Chern-Simons theory.}
\label{fig1}
\end{figure}

Below, we focus on the special family of 't Hooft defects given by the
minuscule 't Hooft lines. They are characterized by magnetic charges given
by the minuscule coweights $\mu $ of the gauge symmetry group $G$. For this
family, the L-operator (\ref{L}) has interesting properties due to the Levi-
decomposition of the Lie algebra $\boldsymbol{g}$ with respect to $\mu $.
Indeed, if $\mu $ is a minuscule coweight in the Cartan of $\boldsymbol{g}$,
it can be decomposed into three sectors 
\begin{equation}
\boldsymbol{g}=\boldsymbol{n}_{+}\oplus \boldsymbol{l}_{\mu }\oplus 
\boldsymbol{n}_{-}\qquad ,\qquad e^{\boldsymbol{g}}=e^{\boldsymbol{n}_{+}}e^{%
\boldsymbol{l}_{\mu }}e^{\boldsymbol{n}_{-}}  \label{gl}
\end{equation}%
with%
\begin{equation}
\left[ \mathbf{\mu },\boldsymbol{n}_{\pm }\right] =\pm \boldsymbol{n}_{\pm
}\qquad ,\qquad \left[ \mathbf{\mu },\boldsymbol{l}_{\mu }\right] =0
\end{equation}%
The L-operator for a minuscule 't Hooft line with charge $\mu $ at $z=0$ and 
$-\mu $\ at $z=\infty $ reads as in (\ref{L}) such that $A(z)$ and $B(z)$
are factorised as follows%
\begin{equation}
\begin{tabular}{lllllll}
$A(z)$ & $=$ & $e^{a_{+}(z)}A^{0}(z)e^{a_{-}(z)}$ & , & $A^{0}(z)$ & $=$ & $%
e^{a^{0}(z)}$ \\ 
$B(z)$ & $=$ & $e^{b_{+}(z)}B^{0}(z)e^{b_{-}(z)}$ & , & $B^{0}(z)$ & $=$ & $%
e^{b^{0}(z)}$%
\end{tabular}
\label{A}
\end{equation}%
Here, the functions $a_{+}$ and $b_{+}$ are valued in $\boldsymbol{n}_{+}$,
the $a_{0}$ and $b_{0}$ valued in $\boldsymbol{l}_{\mu }$ and the $a_{-}$
and $b_{-}$ in $\boldsymbol{n}_{-}.$ For $z\sim 0,$ these functions have the
typical expansion 
\begin{equation}
\digamma \left( z\right) =\sum_{n\geq 0}z^{n}\digamma _{n}=\digamma
_{0}+z\digamma _{1}+...
\end{equation}%
while for $z\sim \infty ,$ we have the development 
\begin{equation}
\digamma \left( z\right) =\sum_{n\geq 1}z^{-n}\digamma _{-n}=\frac{1}{z}%
\digamma _{-1}+\frac{1}{z^{2}}\digamma _{-2}+...
\end{equation}%
Now we turn to establish the expression (\ref{l}) of the L-operator.\newline
We start from (\ref{L}) by focussing on the singularity at $z=0.$
Substituting(\ref{A}), we obtain%
\begin{equation}
L(z;\mu )=\left[ e^{a^{+}(z)}A^{0}(z)e^{a^{-}(z)}\right] z^{\mu }\left[
e^{b^{+}(z)}B^{0}(z)e^{b^{-}(z)}\right]
\end{equation}%
Then, using the actions of the minuscule coweight on $b^{+}$ and $a^{-},$
taking into account that $A^{0}$ and $B^{0}$ commute with $\mu $, we can
bring the above expression to the following form 
\begin{equation}
L(z;\mu )=\left[ e^{a^{+}(z)+zb^{+}(z)}\right] M_{0}z^{\mu }\left[
e^{za^{-}(z)+b^{-}(z)}\right]  \label{regular}
\end{equation}%
Using the regularity of $a^{\pm }(z)$ and $b^{\pm }(z)$ at $z=0,$ we can
absorb the term $zb^{+}(z)$ into $a^{+}(z)$ and $za^{-}(z)$ into $b^{-}(z).$
So, the above expression reduces to%
\begin{equation}
L(z;\mu )=e^{a^{+}(z)}M_{0}z^{\mu }e^{b^{-}(z)}  \label{F1}
\end{equation}%
\textrm{\ }A\textrm{\ }similar treatment for the singular L-operator $%
L=Cz^{\mu }D$ at $z=\infty $ yields the following factorization 
\begin{equation}
L(z;\mu )=e^{zd^{+}(z)}\tilde{M}_{0}z^{\mu }e^{zc^{-}(z)}  \label{F2}
\end{equation}%
Equating the two eqs(\ref{F1}-\ref{F2}), we end up with the three following
constraint relations 
\begin{equation}
a^{+}(z)=zd^{+}(z)\quad ,\quad b^{-}(z)=zc^{-}(z)\quad ,\quad M_{0}(z)=%
\tilde{M}_{0}(z)
\end{equation}%
Because of the expansion properties%
\begin{equation}
\begin{tabular}{lll}
$a^{+}(z)$ & $=$ & $a_{0}^{+}+za_{1}^{+}+...$ \\ 
$zd^{+}(z)$ & $=$ & $d_{-1}^{+}+\frac{1}{z}d_{-2}^{+}...$%
\end{tabular}%
\end{equation}%
it follows that the solution of $a^{+}(z)=zd^{+}(z)$ is given by $%
a^{+}(z)=a_{0}^{+}$ and $zd^{+}(z)=a_{0}^{+}$. The same expansion features
hold for the second constraint $b^{-}(z)=zc^{-}(z);$ thus leading to $%
b^{-}(z)=b_{0}^{-}$ and $zc^{-}(z)=b_{0}^{-}.$ Regarding the third $M_{0}(z)=%
\tilde{M}_{0}(z)$, we have 
\begin{equation}
\begin{tabular}{lll}
$M_{0}(z)$ & $=$ & $m_{0}+zm_{1}+...$ \\ 
$\tilde{M}_{0}(z)$ & $=$ & $I_{id}+\frac{1}{z}m_{-1}(z)+.....$%
\end{tabular}%
\end{equation}%
leading to $M_{0}=\tilde{M}_{0}=I_{id}.$ Substituting \textrm{this} solution
back into the L-operator, we end up with the following expression%
\begin{equation}
L(z;\mu )=e^{X}z^{\mu }e^{Y}  \label{2}
\end{equation}%
where we have set $X=a_{0}^{+}$ and $Y=b_{0}^{-}.$ Moreover, seen that $X$
is valued in the nilpotent algebra $\boldsymbol{n}_{+}$ and $Y$ in the
nilpotent $\boldsymbol{n}_{-}$, they can be expanded like 
\begin{equation}
X=\sum_{i=1}^{\dim \boldsymbol{n}_{+}}b^{i}X_{i}\qquad ,\qquad
Y=\sum_{i=1}^{\dim \boldsymbol{n}_{-}}c_{i}Y^{i}  \label{exp}
\end{equation}%
The $X_{i}$'s and $Y^{i}$'s are the generators of $\boldsymbol{n}_{+}$ and $%
\boldsymbol{n}_{-}.$ The coefficients $b^{i}$ and $c_{i}$ are interpreted as
the Darboux coordinates of the phase space of the L-operator. Eq(\ref{2}) is
precisely the form of $L$ given by eq(\ref{l}). At quantum level, we also
have the following typical commutation relations of bosonic harmonic
oscillators%
\begin{equation}
\left[ \hat{c}_{k},\hat{b}^{i}\right] =\delta _{k}^{i}\qquad ,\qquad \left[ 
\hat{b}^{i},\hat{b}^{k}\right] =\left[ \hat{c}_{i},\hat{c}_{k}\right] =0
\end{equation}%
Notice that the typical quadratic relation $\sum b^{i}c_{i}$ that appears in
our calculations as the trace $Tr\left( XY\right) $ is put in correspondence
with the usual quantum oscillator hamiltonian $\sum (a_{i}^{\dagger
}a_{i}+1/2).$\newline
We end this section by giving a comment regarding the evaluation of the
L-operator between two quantum states as follows%
\begin{equation}
L_{\psi \phi }=\left \langle \psi |L|\phi \right \rangle
\end{equation}%
In this expression, the particle states $\psi $ and $\phi $ have internal
degrees of freedom described by a representation $\boldsymbol{R}$ of the
gauge symmetry $G$. They are respectively interpreted as incoming and
out-going states propagating along a Wilson line W$_{\mathrm{\xi }_{z}}^{%
\boldsymbol{R}}$ crossing the 't Hooft line tH$_{\mathrm{\gamma }_{0}}^{%
\mathbf{\mu }_{R}}$. For an illustration see the Figure \textbf{\ref{fig1}}. 

\subsection{Appendix B: Verma modules of $gl(m|n)$}

The content of this appendix complements the study given in section 4.
Representations of $gl(m|n)$ in $\mathbb{Z}_{2}$-graded vector \textrm{space}
$V$ are Lie superalgebra homomorphisms $\varrho :gl(m|n)\rightarrow End(V)$
where the generators $\varrho \left( \mathcal{E}_{\text{\textsc{AB}}}\right) 
$ belonging to End($V$ ) obey the graded commutators (\ref{gc}). Below, we
focus on the highest weight representations of $gl(m|n)$.

\subsubsection{Highest weight representations}

We begin by recalling that as for bosonic-like Lie algebras, a Verma module $%
M\left( \lambda \right) :gl(m|n)\rightarrow End(V_{\lambda })$ is
characterised by a highest weight vector $\lambda .$ By using the unit
weight vector basis $\epsilon _{\text{\textsc{a}}},$ this highest weight can
be expanded as follows \textrm{\cite{1H}} 
\begin{equation}
\lambda =\sum_{\text{\textsc{a=1}}}^{m+n}\lambda _{\text{\textsc{a}}%
}\epsilon _{\text{\textsc{a}}}
\end{equation}%
where generally speaking the components $\lambda _{\text{\textsc{a}}}\in 
\mathbb{C}$. Below, we restrict to Verma modules with integral highest
weights having integers $\lambda _{\text{\textsc{a}}}$ ordered like $\lambda
_{\text{\textsc{a}}}\geq \lambda _{\text{\textsc{a+1}}}$ and moreover as 
\begin{equation}
\lambda _{1}\geq \lambda _{2}\geq ...\geq \lambda _{m}\geq 0\geq \lambda
_{m+1}\geq ...\geq \lambda _{m+n}
\end{equation}%
In practice, the highest weight representation $M\left( \lambda \right) $
can be built out of a highest-weight vector $\left \vert \Omega _{\lambda
}\right \rangle $ (say the vacuum state) by acting on it by the $\mathcal{E}%
_{\text{\textsc{ab}}}$ generators of the superalgebra $End(V_{\lambda }).$
The $\left \vert \Omega _{\lambda }\right \rangle $ is an eigenstate of the
diagonal operators $\mathcal{E}_{\text{\textsc{aa}}}$, and is annihilated by
the step operators $\mathcal{E}_{\text{\textsc{ab}}}$ with \textsc{a%
\TEXTsymbol{<}b},%
\begin{equation}
\begin{tabular}{lllll}
$\mathcal{E}_{\text{\textsc{aa}}}\left \vert \Omega _{\lambda }\right
\rangle $ & $=$ & $\lambda _{\text{\textsc{a}}}\left \vert \Omega _{\lambda
}\right \rangle $ & , & $1\leq $\textsc{a}$\leq m+n$ \\ 
$\mathcal{E}_{\text{\textsc{ab}}}\left \vert \Omega _{\lambda }\right
\rangle $ & $=$ & $0$ & , & $\beta _{\text{\textsc{ab}}}\text{ positive roots%
\textsc{, a\TEXTsymbol{<}b}}$%
\end{tabular}%
\end{equation}%
Notice that the step operators $\mathcal{E}_{\text{\textsc{ab}}}$ are just
the annihilation operators $\mathcal{E}_{+\beta _{\text{\textsc{ab}}}}$
associated with the positive roots $\beta _{\text{\textsc{ab}}}$. The other
vectors in the $V_{\lambda }$ - module are obtained by acting on $%
\left
\vert \Omega _{\lambda }\right \rangle $ by the creation operators as
follows%
\begin{equation}
\left \vert n_{1},...,n_{p}\right \rangle =\mathcal{E}_{-\mathrm{\gamma }%
_{1}}^{n_{1}}.....\mathcal{E}_{-\mathrm{\gamma }_{p}}^{n_{p}}\left \vert
\Omega _{\lambda }\right \rangle \quad with\qquad n_{l}\in \left \{ 
\begin{array}{ccc}
\mathbb{N}\text{\  \  \  \ } & for & \deg n_{l}=0 \\ 
\left \{ 0,1\right \} & for & \deg n_{l}=1%
\end{array}%
\right.  \label{st}
\end{equation}%
Here, the $\mathrm{\gamma }_{l}$'s stand for the positive roots $\beta _{%
\text{\textsc{ab}}}$ and the step operators $\mathcal{E}_{-\mathrm{\gamma }%
_{l}}^{n_{l}}$'s are the creation operators (lowering operators). The $%
\mathrm{\gamma }_{l}$'s expand in terms of the simple roots $\alpha _{\text{%
\textsc{a}}}=\epsilon _{\text{\textsc{a}}}-\epsilon _{\text{\textsc{a+1}}}$
as follows%
\begin{equation}
\mathrm{\gamma }_{l}=\sum_{\text{\textsc{a}}}n_{l\text{\textsc{a}}}\alpha _{%
\text{\textsc{a}}}
\end{equation}%
with $n_{l\text{\textsc{a}}}$ some positive integers. Notice that two states 
$\left \vert n_{1},...,n_{p}\right \rangle $ and $\left \vert n_{1}^{\prime
},...,n_{p}^{\prime }\right \rangle $ in $V_{\lambda }$ are identified if
they are related by the super-commutation relations (\ref{gc}). Moreover,
seen that the lowering operator $\mathcal{E}_{-\beta _{\text{\textsc{ab}}}}$
changes the highest weight $\lambda $ by the roots $-\beta _{\text{\textsc{ab%
}}}=-\left( \epsilon _{\text{\textsc{a}}}-\epsilon _{\text{\textsc{b}}%
}\right) $ (with \textsc{a\TEXTsymbol{<}b}) we can determine the weight $%
\eta \left( \lambda \right) $ of the state 
\begin{equation}
\left \vert \omega _{\eta \left( \lambda \right) }\right \rangle \equiv
\left \vert n_{1},...,n_{p}\right \rangle
\end{equation}%
By using the simple roots $\alpha _{\text{\textsc{a}}}$ and the
decomposition $\epsilon _{\text{\textsc{b}}}-\epsilon _{\text{\textsc{a}}%
}=\alpha _{\text{\textsc{b}}}+...+\alpha _{\text{\textsc{a-1}}},$ the weight 
$\eta $ of the state $\left \vert \omega _{\eta \left( \lambda \right)
}\right \rangle $ has the form%
\begin{equation}
\eta =\lambda -\sum_{\text{\textsc{a=1}}}^{m+n-1}M_{\text{\textsc{a}}}\alpha
_{\text{\textsc{a}}}\qquad ,\qquad M_{\text{\textsc{a}}}\in \mathbb{N}
\end{equation}%
We end this description by noticing that the Verma modules $M\left( \lambda
\right) $ of the Lie superalgebra $gl\left( m|n\right) $ are infinite
dimensional. However, for the particular case $\left( m|n\right) =\left(
1|1\right) $, we have only one lowering operator $\mathcal{E}_{+\alpha }$
obeying the nilpotency property $\mathcal{E}_{+\alpha }^{2}=0$. As such, eq(%
\ref{st}) reduces to%
\begin{equation}
\left \vert l\right \rangle =\mathcal{E}_{-\mathrm{\alpha }}^{l}\left \vert
\Omega _{\lambda }\right \rangle \qquad ,\qquad l=0,1
\end{equation}%
with%
\begin{equation}
\begin{tabular}{lll}
$\mathcal{E}_{\text{\textsc{11}}}\left \vert \Omega _{\lambda }\right
\rangle $ & $=$ & $\lambda _{\text{\textsc{1}}}\left \vert \Omega _{\lambda
}\right \rangle $ \\ 
$\mathcal{E}_{\text{\textsc{22}}}\left \vert \Omega _{\lambda }\right
\rangle $ & $=$ & $\lambda _{\text{\textsc{2}}}\left \vert \Omega _{\lambda
}\right \rangle $ \\ 
$\mathcal{E}_{+\alpha }\left \vert \Omega _{\lambda }\right \rangle $ & $=$
& $0$%
\end{tabular}%
\end{equation}%
Recall that $gl\left( 1|1\right) $ has four generators given by the two
diagonal $\mathcal{E}_{\text{\textsc{11}}},\mathcal{E}_{\text{\textsc{22}}}$
and two odd step operators $\mathcal{E}_{\pm \alpha }$ corresponding to the
roots $\pm \alpha =\pm \left( \varepsilon -\delta \right) $. A highest
weight $\lambda $ of $gl\left( 1|1\right) $ expands as $\lambda =\lambda
_{1}\varepsilon +\lambda _{2}\delta $ and the Verma module $M\left( \lambda
\right) $ associated with this $\lambda $ is generated by the two states
namely%
\begin{equation}
\left \vert \Omega _{\lambda }\right \rangle \qquad ,\qquad \mathcal{E}_{-%
\mathrm{\alpha }}\left \vert \Omega _{\lambda }\right \rangle
\end{equation}

\subsubsection{Dynkin and Weight super- diagrams}

Knowing the simple roots $\alpha _{\text{\textsc{a}}}=\epsilon _{\text{%
\textsc{a}}}-\epsilon _{\text{\textsc{a+1}}}$ of the Lie superalgebra $%
sl\left( m|n\right) $ and the highest weight $\lambda =\lambda _{\text{%
\textsc{a}}}\epsilon _{\text{\textsc{a}}}$ as well as the descendent $\eta
=\lambda -M_{\text{\textsc{a}}}\alpha _{\text{\textsc{a}}}$ of a module $%
V_{\lambda },$ we can draw the content of the Dynkin graph of sl$\left(
m|n\right) $ and the weight diagram of $V_{\lambda }$ in terms of quiver
graphs. As roots and weights are expressed in terms of the unit weight
vectors $\epsilon _{\text{\textsc{a}}}$, it is interesting to begin by
representing the $\epsilon _{\text{\textsc{a}}}$'s. These $\epsilon _{\text{%
\textsc{a}}}$'s are represented by a vertical line\emph{.} However, because
of the two possible degrees of $\epsilon _{\text{\textsc{a}}}$, the vertical
lines should be distinguished; they have different colors depending of the
grading and are taken as: \newline
$\left( i\right) $ red color for $\deg $\textsc{a=0}; that is for the real
weight $\varepsilon _{a}$. \newline
$\left( ii\right) $ blue color for $\deg $\textsc{a=1}, that is for the pure
imaginary weight $\delta _{i}$. \newline
So, we have the following building blocks for the $\epsilon _{\text{\textsc{a%
}}}$s, 
\begin{equation}
\begin{array}{ccccccc}
\varepsilon _{a} & : & \textcolor{red}{|} & \qquad ,\qquad & \delta _{i} & :
& \textcolor{blue}{|}%
\end{array}%
\end{equation}%
where we have used the splitting $\epsilon _{\text{\textsc{a}}}=\left(
\varepsilon _{a},\delta _{i}\right) $. \newline
Using these vertical lines, the ordered basis set $\left( \epsilon
_{1},...,\epsilon _{m+n}\right) $ is then represented graphically by $m$ red
vertical lines and n vertical blue lines placed in the order specified by
the choice of the $\mathbb{Z}_{2}$-grading. For the example $gl\left(
3|2\right) $ with basis set $\left( \varepsilon _{1},\varepsilon _{2},\delta
_{1},\delta _{2},\varepsilon _{3}\right) ,$ we have the following graph%
\begin{equation}
\begin{tabular}{lllll}
$\varepsilon _{1}$ & $\varepsilon _{2}$ & $\delta _{1}$ & $\delta _{2}$ & $%
\varepsilon _{3}$ \\ 
$\  \textcolor{red}{|}$ & $\textcolor{red}{|}$ & $\textcolor{blue}{|}$ & $%
\textcolor{blue}{|}$ & $\textcolor{red}{|}$%
\end{tabular}
\label{428}
\end{equation}%
The next step to do is to represent the roots and the weights. Simple roots $%
\alpha _{\text{\textsc{a}}}=\epsilon _{\text{\textsc{a}}}-\epsilon _{\text{%
\textsc{a+1}}}$ are represented by circle nodes $\bigcirc $ between each
pair of adjacent vertical lines associated with $\epsilon _{\text{\textsc{a}}%
}$ and $\epsilon _{\text{\textsc{a+1}}}.$ For the previous example namely $%
sl\left( 3|2\right) $ with basis set $\left( \varepsilon _{1},\varepsilon
_{2},\delta _{1},\delta _{2},\varepsilon _{3}\right) ,$ we have%
\begin{equation}
\begin{tabular}{lllllllll}
& $\alpha _{\text{\textsc{1}}}$ &  & $\alpha _{\text{\textsc{2}}}$ &  & $%
\alpha _{\text{\textsc{3}}}$ &  & $\alpha _{\text{\textsc{4}}}$ &  \\ 
$\textcolor{red}{|}$ & $\textcolor{red}{\bigcirc}$ & $\textcolor{red}{|}$ & $%
\textcolor{blue}{\bigcirc}$ & $\textcolor{blue}{|}$ & $\textcolor{red}{%
\bigcirc}$ & $\textcolor{blue}{|}$ & $\textcolor{blue}{\bigcirc}$ & $%
\textcolor{red}{|}$%
\end{tabular}%
\end{equation}

\begin{figure}[tbph]
\begin{center}
\includegraphics[width=14cm]{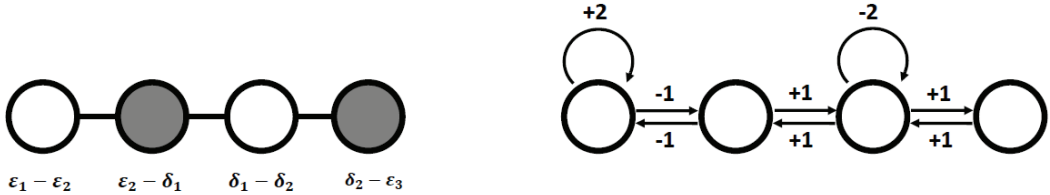}
\end{center}
\par
\vspace{-0.5cm}
\caption{Building the Dynkin diagram of Lie superalgebra gl$\left(
m|n\right) .$ Here we give the example the Dynkin diagram of gl$\left(
3|2\right) $ with weight basis ordered as $\left( \protect \varepsilon _{1},%
\protect \varepsilon _{2},\protect \delta _{1},\protect \delta _{2},\protect%
\varepsilon _{3}\right) .$ This graph is borrowed from \textrm{\protect \cite%
{1H}}.}
\label{fig 11}
\end{figure}

$\bullet $ \emph{Super-} \emph{Dynkin diagram}\newline
For each pair of simple roots $\left( \alpha _{\text{\textsc{a}}},\alpha _{%
\text{\textsc{b}}}\right) $ with non vanishing intersection matrix $K_{\text{%
\textsc{ab}}}=\alpha _{\text{\textsc{a}}}.\alpha _{\text{\textsc{b}}}\neq 0$%
, we draw an arrow from the \textsc{a}-th node to the \textsc{b}-th node.
The $K_{\text{\textsc{ab}}}$ is an integer and written on the arrow. By
hiding the vertical lines, we obtain the super- Dynkin diagram of $sl\left(
m|n\right) $\ with the specified basis $\left( \epsilon _{1},...,\epsilon
_{m+n}\right) $. In the\ Figure\textrm{\ }\textbf{\ref{fig 11}}, we give the
super- Dynkin diagram of $sl\left( 3|2\right) $ with weight basis as $\left(
\varepsilon _{1},\varepsilon _{2},\delta _{1},\delta _{2},\varepsilon
_{3}\right) $. 
Notice that the ordering is defined modulo the action of the Weyl group $%
W_{m}\times W_{n}$ which permutes the basis vectors $\left( \epsilon _{\text{%
\textsc{1}}},...,\epsilon _{\text{\textsc{m+n}}}\right) $ without changing
the $\mathbb{Z}_{2}$-grading. For instance, the choice $\left( \varepsilon
_{2},\varepsilon _{1},\delta _{1},\delta _{2},\varepsilon _{3}\right) $
leads to the same Dynkin diagram as the one given by the Figure \textbf{\ref%
{fig 11}}.

\begin{figure}[tbph]
\begin{center}
\includegraphics[width=12cm]{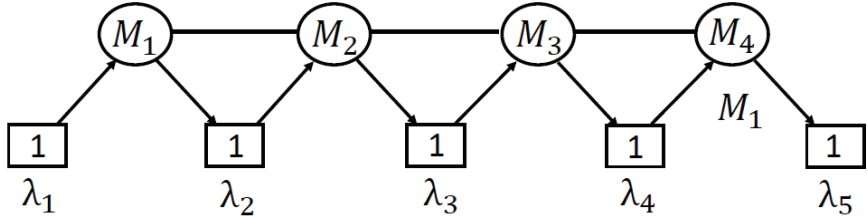}
\end{center}
\par
\vspace{-0.5cm}
\caption{Building the weight diagram of representation of Lie superalgebra sl%
$\left( m|n\right) .$ Here the diagram of the weight $\protect \eta =\protect%
\lambda -M_{\text{\textsc{a}}}\protect \alpha _{\text{\textsc{a}}}$ in the
Lie superalgebra gl$\left( 3|2\right) $with $M_{\text{\textsc{a}}}$-integers
as $\left( M_{1},M_{2},M_{3},M_{4}\right) =(2,3,2,1).$ This graph is
borrowed from \textrm{\protect \cite{1H}}. For generalisations and more
information, we refer to this interesting study.}
\label{wh}
\end{figure}

$\bullet $ \emph{Super-} \emph{weight diagrams}\newline
To represent the highest weight $\lambda =\lambda _{\text{\textsc{a}}%
}\epsilon _{\text{\textsc{a}}}$ of modules of the Lie superalgebra $gl\left(
m|n\right) $, we first draw the (red and blue) vertical lines representing $%
\epsilon _{\text{\textsc{a}}}$ as in (\ref{428}). \newline
Then, for each vertical line representing $\epsilon _{\text{\textsc{a}}},$
we implement the coefficient $\lambda _{\text{\textsc{a}}}$ by drawing a
diagonal line ending on the vertical $\epsilon _{\text{\textsc{a}}}$ and
write $\lambda _{\text{\textsc{a}}}$ as in the Figure \textbf{\ref{wh}}
illustrating highest weights 
\begin{equation}
\lambda =\lambda _{\text{\textsc{1}}}\epsilon _{\text{\textsc{1}}}+\lambda _{%
\text{\textsc{2}}}\epsilon _{\text{\textsc{2}}}+\lambda _{\text{\textsc{3}}%
}\epsilon _{\text{\textsc{3}}}+\lambda _{\text{\textsc{4}}}\epsilon _{\text{%
\textsc{4}}}+\lambda _{\text{\textsc{5}}}\epsilon _{\text{\textsc{5}}}
\end{equation}%
in the Lie superalgebra $gl\left( 3|2\right) .$ \newline
To represent the weights $\eta =\lambda -M_{\text{\textsc{a}}}\alpha _{\text{%
\textsc{a}}}$ of the descendent states (\ref{st}), we draw $M_{\text{\textsc{%
a}}}$ horizontal line segments between the \textsc{a}-th and $\left( \text{%
\textsc{a}}+1\right) $-st vertical lines. For the example of $gl\left(
3|2\right) $ with basis set $\left( \varepsilon _{1},\varepsilon _{2},\delta
_{1},\delta _{2},\varepsilon _{3}\right) $ and $\left(
M_{1},M_{2},M_{3},M_{4}\right) =(2,3,2,1);$ that is%
\begin{equation}
\eta =\lambda -2\alpha _{\text{\textsc{1}}}-3\alpha _{\text{\textsc{2}}%
}-2\alpha _{\text{\textsc{3}}}-\alpha _{\text{\textsc{4}}}
\end{equation}%
we have the weight diagram the Figure \textbf{\ref{wh}}. 

\subsection{Appendix C: Derivation of eq(2.20) of ref.[29]}

In this appendix, we give the explicit derivation \textrm{of the Lax
operator of eq(2.20) in} ref.\textrm{[29]} obtained by \emph{Frassek,
Lukowski, Meneghelli, Staudacher} (FLMS solution). This solution \textrm{was
obtained} by using the Yangian formalism; but here \textrm{we show that we
can }derive it from the Chern-Simons theory with gauge super group family $%
SL(m|n)$ with $m\neq n$. For a recent description of these two formalisms
(Yangian and Chern-Simons) applied to the bosonic like symmetries; see \cite%
{osp}.\newline
We begin by recalling that \textrm{the FLMS solution was constructed in [29]}
for the Lie superalgebra $su(m|n)$,\textrm{\ which naturally extends} to its
complexification $sl(m|n)$ that we are treating here. The L-operator
obtained by FLMS has been presented as 2$\times 2$ matrix with entries given
by matrix blocks that we present as follows 
\begin{equation}
\mathcal{L}_{FLMS}=\left( 
\begin{array}{cc}
L_{xy} & L_{x\dot{y}} \\ 
L_{\dot{x}y} & L_{\dot{x}\dot{y}}%
\end{array}%
\right)  \label{flmn}
\end{equation}%
where $x,\dot{x},y,\dot{y}$ are labels and where $L_{xy},L_{x\dot{y}},L_{%
\dot{x}y},L_{\dot{x}\dot{y}}$ are given by \textrm{eq(2.20) in} \textrm{%
[29]; see also eq(\ref{la}) derived below}. \textrm{So, in order to} recover 
\textrm{this} solution from our analysis, we start from \textrm{eq(6.5)} of
our paper that w\textrm{e can rewrite in condensed form as} 
\begin{equation}
\begin{tabular}{lll}
$\mathbf{\mu }_{p}$ & $=$ & $\frac{m-p-n}{m-n}\mathcal{P}_{1}-\frac{p}{m-n}%
\mathcal{P}_{2}$ \\ 
$z^{\mathbf{\mu }_{p}}$ & $=$ & $z^{\frac{m-p-n}{m-n}}\mathcal{P}_{1}+z^{-%
\frac{p}{m-n}}\mathcal{P}_{2}$%
\end{tabular}
\label{zm}
\end{equation}%
\textrm{Here, }we have set $\mathcal{P}_{1}=\Pi _{1}$ and $\mathcal{P}%
_{2}=\Pi _{2}+\Pi _{3}$ which are also projectors that satisfy the usual
relations $\mathcal{P}_{k}.\mathcal{P}_{l}=\delta _{kl}\mathcal{P}_{k}$. The
use of $\mathcal{P}_{1}$ and $\mathcal{P}_{2}$ instead of $\Pi _{1}$,$\Pi
_{2},\Pi _{3}$ is to recover the 2$\times $2 representation (\ref{flmn}).
Using the bra-ket language and our label notations, we have $\mathcal{P}%
_{1}=\sum_{a=1}^{p}\left \vert a\right \rangle \left \langle a\right \vert $
and $\mathcal{P}_{2}=\sum_{A=p+1}^{m+n}\left \vert A\right \rangle
\left
\langle A\right \vert $ with matrix representations as follows%
\begin{equation}
\mathcal{P}_{1}=\left( 
\begin{array}{cc}
I_{{\small p\times p}} & 0_{{\small p\times Q}} \\ 
0_{{\small Q\times p}} & 0_{{\small Q\times Q}}%
\end{array}%
\right) ,\qquad \mathcal{P}_{2}=\left( 
\begin{array}{cc}
0_{{\small p\times p}} & 0_{{\small p\times Q}} \\ 
0_{{\small Q}\times p} & I_{{\small Q}\times {\small Q}}%
\end{array}%
\right)
\end{equation}%
where we have set ${\small Q=m+n-p}$. These projectors satisfy the usual%
\textrm{\ identity resolution,} namely $\mathcal{P}_{1}+\mathcal{P}%
_{2}=I_{\left( m+n\right) \times \left( m+n\right) }$.\newline
Putting the expression (\ref{zm}) of $z^{\mathbf{\mu }_{p}}$ into the super
L-operator $\mathcal{L}=e^{\Psi }z^{\mathbf{\mu }_{p}}e^{\Phi }$ given by eq(%
\ref{pf}), we end up with eqs(\ref{zx}-\ref{zy}) that \textrm{read} in terms
of the projectors $\mathcal{P}_{1}$ and $\mathcal{P}_{2}$ as follows 
\begin{equation}
\begin{tabular}{lll}
$\mathcal{L}$ & $=$ & $z^{\mathbf{\mu }_{p}}+z_{1}^{\frac{m-p-n}{m-n}}%
\mathcal{P}_{1}\Phi +z^{-\frac{p}{m-n}}\mathcal{P}_{2}\Phi $ \\ 
&  & $+z^{\frac{m-p-n}{m-n}}\Psi \mathcal{P}_{1}+z^{-\frac{p}{m-n}}\Psi 
\mathcal{P}_{2}$ \\ 
&  & $+z^{\frac{m-p-n}{m-n}}\Psi \mathcal{P}_{1}\Phi +z^{-\frac{p}{m-n}}\Psi 
\mathcal{P}_{2}\Phi $%
\end{tabular}
\label{C1}
\end{equation}%
In this expression, $\Psi $ and\ $\Phi $ are valued in the nilpotent
sub-superalgebras $\boldsymbol{N}_{+}$ and $\boldsymbol{N}_{-}$; they are
given by (\ref{xx}-\ref{yx}). For convenience, we rewrite them in terms of
super labels $\left( a,A\right) $ and $\left( b,B\right) $ as follows 
\begin{equation}
\Psi =\sum_{a=1}^{p}\sum_{A=p+1}^{m+n}B^{aA}\mathbb{X}_{aA}\qquad ,\qquad
\Phi =\sum_{b=1}^{p}\sum_{B=p+1}^{m+n}C_{Bb}\mathbb{Y}^{Bb}
\end{equation}%
where $\mathbb{X}_{aA}$ and $\mathbb{Y}^{bA}$ are respectively the
generators of the \textrm{nilpotents} $\boldsymbol{N}_{+}$ and $\boldsymbol{N%
}_{-}$. These graded generators are realised in terms of the canonical super
states as follows 
\begin{equation}
\mathbb{X}_{aA}=\left \vert a\right \rangle \left \langle A\right \vert
\qquad ,\qquad \mathbb{Y}^{Bb}=\left \vert B\right \rangle \left \langle
b\right \vert  \label{exx}
\end{equation}%
The coefficients $B^{aA}$ and $C_{Bb}$ are super Darboux coordinates of the
phase space of the 't Hooft super line; their canonical quantization,
denoted like $\hat{B}^{aA}$ and $\hat{C}_{Bb},$ describe the associated
quantum super oscillators. In matrix notation, the $B^{aA}$ and $C_{Bb}$
have the following form%
\begin{equation}
B^{aA}=\left( 
\begin{array}{ccc}
B^{1(p+1)} & \cdots & B^{1(m+n)} \\ 
\vdots & \ddots & \vdots \\ 
B^{p(p+1)} & \cdots & B^{p(m+n)}%
\end{array}%
\right) \qquad ,\qquad C_{Bb}=\left( 
\begin{array}{ccc}
C_{\left( p+1\right) 1} & \cdots & C_{\left( p+1\right) p} \\ 
\vdots & \ddots & \vdots \\ 
C_{\left( m+n\right) 1} & \cdots & C_{\left( m+n\right) p}%
\end{array}%
\right)
\end{equation}%
and similarly for the $\hat{B}^{aA}$ and $\hat{C}_{Bb}$ operators. For later
use, notice that the product $\Psi \Phi $ reads as $\sum B^{aA}\mathbb{X}%
_{aA}\mathbb{Y}^{Bb}C_{Bb}$; by substituting the generators \textrm{with
their expressions (\ref{exx})}, we obtain $\mathbb{X}_{aA}\mathbb{Y}%
^{Bb}=\delta _{A}^{B}\left \vert a\right \rangle \left \langle b\right \vert
,$ and consequently 
\begin{equation}
\Psi \Phi =\sum_{D}B^{aD}C_{Db}\left \vert a\right \rangle \left \langle
b\right \vert  \label{fp}
\end{equation}%
As far \textrm{as} this classical quantity is concerned, notice the three
following interesting features: \newline
$\left( \mathbf{1}\right) $ the product $\Psi \Phi $ can be expanded as $%
\sum_{D=p+1}^{m}b^{aD}c_{Db}+\sum_{D=m}^{m+n}\beta ^{aD}\gamma _{Db},$ where
the bosonic $b^{aD}$ and $c_{Db}$ as well as the fermionic $\beta ^{aD}$ and 
$\gamma _{Db}$ are as in eqs(\ref{19}-\ref{20}). \newline
$\left( \mathbf{2}\right) $ Classically speaking, the quadratic product $%
\Psi \Phi $ (\ref{fp}) can be also presented as follows%
\begin{equation}
\Psi \Phi =\frac{1}{2}\sum_{D}\left( B^{aD}C_{Db}+\left( -\right) ^{\left
\vert D\right \vert }C_{Db}B^{aD}\right) \left \vert a\right \rangle \left
\langle b\right \vert
\end{equation}%
just because the normal ordering is not required classically. The number $%
\left \vert D\right \vert $ refers \textrm{here} to the $\mathbb{Z}_{2}$%
-grading $0,1$. At the quantum level, the $\Psi $ and $\Phi $ are promoted
to the operators $\hat{\Psi}$ and $\hat{\Phi}$; as such, the above product $%
\Psi \Phi $ must be replaced by the operator $\hat{\Psi}\hat{\Phi}$ which is
given by the expansion%
\begin{equation}
\hat{\Psi}\hat{\Phi}=\frac{1}{2}\sum_{D}\left( \hat{B}^{aD}\hat{C}%
_{Db}+\left( -\right) ^{\left \vert D\right \vert }\hat{C}_{Db}\hat{B}%
^{aD}\right) \left \vert a\right \rangle \left \langle b\right \vert
\end{equation}%
Here, the graded commutators between the super oscillators $\hat{B}^{aA}$
and $\hat{C}_{Bb}$ are defined as usual by the super commutator $[\hat{C}%
_{Bb},\hat{B}^{aA}\}=\delta _{b}^{a}\delta _{B}^{A},$ which is a condensed
form of eqs(\ref{21}).\newline
$\left( \mathbf{3}\right) $ From these super commutators, we learn that $%
\hat{C}_{Bb}\hat{B}^{aA}$ is given by $\delta _{b}^{a}\delta _{B}^{A}+\left(
-\right) ^{\left \vert A\right \vert \times \left \vert B\right \vert }\hat{B%
}^{aA}\hat{C}_{Bb}.$ Using this result, we can express $\left( -\right)
^{\left \vert D\right \vert }\hat{C}_{Db}\hat{B}^{aD}$ as follows 
\begin{equation}
\left( -\right) ^{\left \vert D\right \vert }\hat{C}_{Db}\hat{B}^{aD}=\hat{B}%
^{aD}\hat{C}_{Db}+\left( -\right) ^{\left \vert D\right \vert }\delta
_{D}^{D}\delta _{b}^{a}
\end{equation}%
thus leading to%
\begin{equation}
\hat{\Psi}\hat{\Phi}=\sum \limits_{D}\left( \hat{B}^{aD}\hat{C}_{Db}+\frac{1%
}{2}\left( -\right) ^{\left \vert D\right \vert }\delta _{D}^{D}\delta
_{b}^{a}\right) \left \vert a\right \rangle \left \langle b\right \vert
\label{qm}
\end{equation}%
with $\frac{1}{2}\sum \limits_{D}\left( -\right) ^{\left \vert D\right \vert
}\delta _{D}^{D}$ given by%
\begin{equation}
\frac{1}{2}\left( \sum \limits_{p+1}^{m}\left( -\right) ^{\left \vert
D\right \vert }\delta _{D}^{D}+\sum \limits_{m}^{m+n}\left( -\right) ^{\left
\vert D\right \vert }\delta _{D}^{D}\right) =\frac{1}{2}\left( m-p\right) -%
\frac{1}{2}n
\end{equation}%
Returning to the explicit calculation of (\ref{C1}), we use the\textrm{\
useful }properties $\Psi ^{2}=0$ and $\Phi ^{2}=0$, as well as%
\begin{equation}
\begin{tabular}{lllll}
$\Psi \mathcal{P}_{1}=0$ & , & $\mathcal{P}_{1}\Psi =\Psi $ & , & $\Psi 
\mathcal{P}_{2}=\Psi $ \\ 
$\mathcal{P}_{1}\Phi =0$ & , & $\Phi \mathcal{P}_{1}=\Phi $ & , & $\mathcal{P%
}_{2}\Phi =\Phi $%
\end{tabular}
\label{C2}
\end{equation}%
So, eq(\ref{C1}) reduces to%
\begin{equation}
\begin{tabular}{lll}
$\mathcal{L}$ & $=$ & $z^{\mathbf{\mu }_{p}}+z^{-\frac{p}{m-n}}\Psi \Phi $
\\ 
&  & $+z^{-\frac{p}{m-n}}\Psi \mathcal{P}_{2}+z^{-\frac{p}{m-n}}\mathcal{P}%
_{2}\Phi $%
\end{tabular}%
\end{equation}%
Using the properties (\ref{C2}), we obtain an expression in terms of the
projectors $\mathcal{P}_{1}$ and $\mathcal{P}_{2}$ as well as $\mathcal{P}%
_{1}\Psi \Phi \mathcal{P}_{1},$ $\mathcal{P}_{1}\Psi \mathcal{P}_{2}$ and $%
\mathcal{P}_{2}\Phi \mathcal{P}_{1}$ that we present as follows 
\begin{equation}
\mathcal{L}=\left( 
\begin{array}{cc}
z^{1-\frac{p}{m-n}}\mathcal{P}_{1}+z^{-\frac{p}{m-n}}\mathcal{P}_{1}\Psi
\Phi \mathcal{P}_{1} & z^{-\frac{p}{m-n}}\mathcal{P}_{1}\Psi \mathcal{P}_{2}
\\ 
z^{-\frac{p}{m-n}}\mathcal{P}_{2}\Phi \mathcal{P}_{1} & z^{-\frac{p}{m-n}}%
\mathcal{P}_{2}%
\end{array}%
\right)
\end{equation}%
By multiplying by $z^{\frac{p}{m-n}}$ due to known properties of $\mathcal{L}
$ as commented in the main text, we end up with the remarkable expression 
\begin{equation}
\mathcal{L}=\left( 
\begin{array}{cc}
z\mathcal{P}_{1}+\mathcal{P}_{1}\Psi \Phi \mathcal{P}_{1} & \mathcal{P}%
_{1}\Psi \mathcal{P}_{2} \\ 
\mathcal{P}_{2}\Phi \mathcal{P}_{1} & \mathcal{P}_{2}%
\end{array}%
\right)  \label{Lo}
\end{equation}%
Quantum mechanically, eq(\ref{Lo}) is promoted to the hatted L-operator%
\begin{equation}
\mathcal{\hat{L}}=\left( 
\begin{array}{cc}
z\mathcal{P}_{1}+\mathcal{P}_{1}\hat{\Psi}\hat{\Phi}\mathcal{P}_{1} & 
\mathcal{P}_{1}\hat{\Psi}\mathcal{P}_{2} \\ 
\mathcal{P}_{2}\hat{\Phi}\mathcal{P}_{1} & \mathcal{P}_{2}%
\end{array}%
\right)  \label{la}
\end{equation}%
with $\hat{\Psi}\hat{\Phi}$ given by (\ref{qm}) which is precisely the%
\textrm{\ FLMS solution obtained in [29]}. \newline
We end this appendix by noticing that the present analysis can be extended
to the families of Lie superalgebras listed in the table (\ref{tabg}). This
generalisation can be achieved by extending the bosonic construction done in 
\cite{osp} to supergroups including fermions. Progress in this direction
will be reported in a future occasion.

\  \

\end{document}